\documentclass[12pt]{article}
\usepackage{ssi}
\usepackage{times}

\usepackage{graphicx}
\def\citex#1{[\citenum{#1}]}

\def\OMIT#1{{}}
\def\lsim{\mathrel{\!\mathpalette\vereq<}\!}

\def\vereq#1#2{\lower3.5pt\vbox{\baselineskip1.5pt \lineskip1.5pt
\ialign{$#1\hfill##\hfil$\crcr#2\crcr\sim\crcr}}}

\def\lqcd{\Lambda_{\rm QCD}}

\def\d{{\rm d}}
\def\ov{\overline}
\def\ds{\displaystyle}
\def\bra#1{\langle#1|}
\def\ket#1{|#1\rangle}

\def\GeV{{\rm GeV}}
\def\MeV{{\rm MeV}}

\newcommand{\Bbar}{\,\overline{\!B}}
\newcommand{\Dbar}{\,\overline{\!D}}
\newcommand{\Kbar}{\,\overline{\!K}}
\def\B0bar{\Bbar{}^0}
\def\D0bar{\Dbar{}^0}
\def\K0bar{\Kbar{}^0}

\def\FD{{\cal F}}
\def\FDs{\FD_*}
\def\FDt{\FD_{(*)}}

\newcommand{\nn}{\nonumber}
\newcommand{\beq}{\begin{equation}}
\newcommand{\eeq}{\end{equation}}
\newcommand{\beqa}{\begin{eqnarray}}
\newcommand{\eeqa}{\end{eqnarray}}

\begin{document}

\hfill\vbox{\hbox{LBNL--51835}\vspace*{-4pt}\hbox{hep-ph/0302031}}

\title{\boldmath Introduction to Heavy Meson Decays\\ and $CP$ Asymmetries}

\author{Zoltan Ligeti\\
Ernest Orlando Lawrence Berkeley National Laboratory\\
University of California, Berkeley, CA 94720}

\maketitle

\baselineskip 16pt
\begin{abstract}%
These lectures are intended to provide an introduction to heavy meson decays
and $CP$ violation.  The first lecture contains a brief review of the standard
model and how the CKM matrix and $CP$ violation arise, mixing and $CP$
violation in neutral meson systems, and explanation of the cleanliness of the
$\sin 2\beta$ measurement.  The second lecture deals with the heavy quark
limit, some applications of heavy quark symmetry and the operator product
expansion for exclusive and inclusive semileptonic $B$ decays.  The third
lecture concerns with theoretically clean $CP$ violation measurements that may
become possible in the future, and some developments toward a better
understanding of nonleptonic $B$ decays.  The conclusions include a subjective
best buy list for the near future.

\vfil
\begin{center}
\it
Lectures given at the\\[2pt]
30th SLAC Summer Institute on Particle Physics: Secrets of the B Meson\\[2pt]
SLAC, Menlo Park, California, August 5--16, 2002
\end{center}
\end{abstract}

\pagestyle{plain}
\renewcommand{\thepage}{\roman{page}}
\baselineskip 19pt
\tableofcontents
\baselineskip 16pt
\newpage

\renewcommand{\thepage}{\arabic{page}}
\setcounter{page}{1}

\section{\boldmath Introduction to Flavor Physics: Standard Model Review,
Mixing and $CP$ Violation in Neutral Mesons}

\subsection{Motivation}

Flavor physics is the study of interactions that distinguish between the
generations.  In the standard model (SM), flavor physics in the quark sector
and $CP$ violation in flavor changing processes arise from the
Cabibbo-Kobayashi-Maskawa (CKM) quark mixing matrix.  The goal of the $B$
physics program is to precisely test this part of the theory.  In the last
decade we tested the SM description of electroweak gauge interactions with an
accuracy that is an order of magnitude (or even more) better than before.  In
the coming years tests of the flavor sector and our ability to probe for flavor
physics and $CP$ violation beyond the SM may improve in a similar manner.  

In contrast to the hierarchy problem of electroweak symmetry breaking, there is
no similarly robust argument that new flavor physics must appear near the
electroweak scale.  Nevertheless, the flavor sector provides severe constraints
for model building, and many extensions of the SM do involve new flavor physics
near the electroweak scale which may be observable at the $B$ factories. 
Flavor physics also played an important role in the development of the SM:
(i)~the smallness of $K^0-\K0bar$ mixing led to the GIM mechanism and a
calculation of the charm mass before it was discovered; (ii)~$CP$ violation led
to the KM proposal that there should be three generations before any third
generation fermions were discovered; and (iii)~the large $B^0-\B0bar$ mixing
was the first evidence for a very large top quark mass.

To test the SM in low energy experiments, such as $B$ decays, the main obstacle
is that strong interactions become nonperturbative at low energies.  The scale
dependence of the QCD coupling constant is
\beq
\alpha_s(\mu) = \displaystyle {\alpha_s(M) \over \ds
  1 + {\alpha_s\over 2\pi}\, \beta_0  \ln{\mu\over M}} + \ldots \,.
\eeq
This implies that at high energies (short distances) perturbation theory is a
useful tool.  However, at low energies (long distances) QCD becomes
nonperturbative, and it is very hard and often impossible to do reliable
calculations.  There are two scenarios in which making precise predictions is
still possible: (i) using extra symmetries of QCD (such as chiral or heavy
quark symmetry); or (ii) certain processes are determined by short distance
physics.  For example, the measurement of $\sin 2\beta$ from $B\to \psi K_S$ is
theoretically clean because of $CP$ invariance of the strong interaction, while
inclusive $B$ decays are calculable with small model dependence because they
are short distance dominated.  These will be explained later in detail. 
Sometimes it is also possible to combine different measurements with the help
of symmetries to eliminate uncalculable hadronic physics; this is the case, for
example, in $K\to \pi\nu\bar\nu$, which is theoretically clean because the form
factors that enter this decay are related by symmetries to those measured in
semileptonic kaon decay.

These lectures fall short of providing a complete introduction to flavor
physics and $CP$ violation, for which there are several excellent books and
reviews~\cite{book1,book2,book3,babook,tevbook,lhcrep,yossi,adam}.  Rather, I
tried to sample topics that illustrate the richness of the field, both in terms
of the theoretical methods and the breadth of the interesting measurements. 
Some omissions might be justified as other lectures covered historical aspects
of the field~\cite{sanda}, lattice QCD~\cite{lattice}, physics beyond the
standard model~\cite{np}, and the experimental status and prospects in flavor
physics~\cite{tb,dbm,is,rt,fkw}.  Unfortunately, the list of references is also
far from complete.  This writeup follows closely the actual slides shown at the
SLAC Summer Institute.

\subsection{Standard model --- bits and pieces}

To define the standard model, we need to specify the gauge symmetry, the
particle content, and the pattern of symmetry breaking.  The SM gauge group is
\beq\label{groups}
SU(3)_c \times SU(2)_L \times U(1)_Y \,.
\eeq
Of this, $SU(3)_c$ is the gauge symmetry of the strong interaction, while 
$SU(2)_L \times U(1)_Y$ corresponding to the electroweak theory.  The particle
content is defined as three generations of the following representations
\beqa\label{reps}
&&Q_L(3,2)_{1/6}\,,\qquad u_R(3,1)_{2/3}\,,\qquad d_R(3,1)_{-1/3}\,, \nn\\
&&L_L(1,2)_{-1/2}\,,\qquad \ell_R(1,1)_{-1}\,,
\eeqa
where $Q_L$ and $L_L$ are left-handed quark and lepton fields, and $u_R$,
$d_R$, and $\ell_R$ are right-handed up-type quarks, down-type quarks, and
charged leptons, respectively.  The quantum numbers in Eq.~(\ref{reps}) are
given in the same order as the gauge groups in Eq.~(\ref{groups}).  Finally the
electroweak symmetry, $SU(2)_L \times U(1)_Y$, is broken to electromagnetism,
$U(1)_{\rm EM}$, by the vacuum expectation value (VEV) of the Higgs field,
$\phi(1,2)_{1/2}$,
\beq\label{higgs}
\langle\phi\rangle = \pmatrix{0\cr v/\sqrt2} ,
\eeq
where $v \approx 246\,\GeV$.  Once these ingredients of the SM are specified,
in principle all particle physics phenomena are determined in terms of 18
parameters, of which 10 correspond to the quark sector (6 masses and 4 CKM
parameters).

Some of the most important questions about the SM are the origin of electroweak
and flavor symmetry breaking.  Electroweak symmetry is spontaneously broken by
the dimensionful VEV in Eq.~(\ref{higgs}), but it is not known yet whether
there is an elementary scalar Higgs particle corresponding to $\phi$.  What we
do know, essentially because $v$ is dimensionful, is that the mass of the Higgs
(or whatever physics is associated with electroweak symmetry breaking) cannot
be much above the TeV scale, since in the absence of new particles, scattering
of $W$ bosons would violate unitarity and become strong around a TeV.  In
contrast, there is no similar argument that flavor symmetry breaking has to do
with physics at the TeV scale.  If the quarks were massless then the SM would
have a global $U(3)_Q \times U(3)_u \times U(3)_d$ symmetry, since the three
generations of left handed quark doublets and right handed singlets would be
indistinguishable.  This symmetry is broken by dimensionless quantities (the
Yukawa couplings that give mass to the quarks, see Eq.~(\ref{Lyuk}) below) to
$U(1)_B$, where $B$ is baryon number, and so we do not know what scale is
associated with flavor symmetry breaking. (For the leptons it is not even known
yet whether lepton number is conserved; see the discussion below.)  One may
nevertheless hope that these scales are related, since electroweak and flavor
symmetry breaking are connected in many new physics scenarios.  There may be
new flavor physics associated with the TeV scale, which could have observable
consequences, most probably for flavor changing neutral current processes
and/or for $CP$ violation.

The most important question in flavor physics is to test whether the SM (i.e.,
only virtual quarks, $W$, and $Z$ interacting through CKM matrix in tree and
loop diagrams) explain all flavor changing interactions.  To be able to answer
this question, we need experimental precision, which is being provided by the
$B$ factories, and theoretical precision, which can only be achieved in a
limited set of processes.  Thus, the key processes in this program are those
which can teach us about high energy physics with small hadronic uncertainties.

The SM so far agrees with all observed phenomena.  Testing the flavor sector as
precisely as possible is motivated by the facts that (i) almost all extensions
of the SM contain new sources of $CP$ and flavor violation; (ii) the flavor
sector is a major constraint for model building, and may distinguish between
new physics models; (iii) the observed baryon asymmetry of the Universe
requires $CP$ violation beyond the SM.  If the scale of new flavor physics is
much above the electroweak scale then there will be no observable effects in
$B$ decays, and the $B$ factories will make precise SM measurements.  However,
if there is new flavor physics near the electroweak scale then sizable
deviations from the SM predictions are possible, and we could get detailed
information on new physics.  So the point is not only to measure CKM elements,
but to overconstrain the SM predictions by as many ``redundant" measurements as
possible.

\subsubsection[Flavor and $CP$ violation in the SM]{\boldmath Flavor and $CP$
violation in the SM}

The SM is the most general renormalizable theory consistent with the gauge
symmetry and particle content in Eqs.~(\ref{groups}) and (\ref{reps}).  Its
Lagrangian has three parts.  (The discussion in this section follows
Ref.~\citex{yossi}.)  The kinetic terms are
\beq\label{Lkin}
{\cal L}_{\rm kin} = 
  -\frac14 \sum_{\mbox{\footnotesize groups}} (F_{\mu\nu}^a)^2 
  + \sum_{\mbox{\footnotesize rep's}} \ov\psi\, i D\!\!\!\!\slash\,\, \psi\,,
\eeq
where $D_\mu = \partial_\mu + ig_s G^a_\mu L^a + ig W^b_\mu T^b + ig' B_\mu Y$.
Here $L_a$ are the $SU(3)$ generators (0 for singlets, and the Gell-Mann
matrices, $\lambda_a/2$, for triplets), $T_b$ are the $SU(2)_L$ generators (0
for singlets, and the Pauli matrices, $\tau_a/2$, for doublets), and $Y$ are
the $U(1)_Y$ charges.  The $(F_{\mu\nu}^a)^2$ terms are always $CP$
conserving.  Throughout these lectures we neglect a possible $(\theta_{\rm QCD}
/ 16\pi^2) F_{\mu\nu} \widetilde F^{\mu\nu}$ term in the QCD Lagrangian, which
violates $CP$.  The constraints on the electron and neutron electric dipole
moments imply that the effects of $\theta_{\rm QCD}$ in flavor changing
processes are many orders of magnitude below the sensitivity of any proposed
experiment (see Ref.~\citex{strongCP} for details).  The Higgs terms,
\beq\label{Lhiggs}
{\cal L}_{\rm Higgs} = |D_\mu \phi|^2 + \mu^2 \phi^\dagger\phi
  - \lambda (\phi^\dagger\phi)^2\,, \qquad v^2=\mu^2/\lambda\,,
\eeq
cannot violate $CP$ if there is only one Higgs doublet.  With an extended Higgs
sector, $CP$ violation would be possible.  Finally, the Yukawa couplings are
given by
\beq\label{Lyuk}
{\cal L}_Y = - Y_{ij}^d\, \ov{Q_{Li}^I}\, \phi\, d_{Rj}^I
  - Y_{ij}^u\, \ov{Q_{Li}^I}\, \widetilde\phi\, u_{Rj}^I
  - Y_{ij}^\ell\, \ov{L_{Li}^I}\, \phi\, \ell_{Rj}^I + {\rm h.c.}\,, \qquad
\widetilde\phi = {\textstyle\pmatrix{0 & 1 \cr -1 & 0}} \phi^*\,,
\eeq
where $i,j$ label the three generations, and the superscripts $I$ denote that
the quark fields in the interaction basis.  To see that $CP$ violation is
related to unremovable phases of Yukawa couplings note that the terms
\beq\label{y1}
Y_{ij}\, \ov{\psi_{Li}}\, \phi\, \psi_{Rj}
  + Y_{ij}^*\, \ov{\psi_{Rj}}\, \phi^\dagger\, \psi_{Li} \,,
\eeq
become under $CP$ transformation
\beq\label{y2}
Y_{ij}\, \ov{\psi_{Rj}}\, \phi^\dagger\, \psi_{Li}
  + Y_{ij}^*\, \ov{\psi_{Li}}\, \phi\, \psi_{Rj} \,.
\eeq
Eqs.~(\ref{y1}) and (\ref{y2}) are identical if and only if a basis for the
quark fields can be chosen such that $Y_{ij} = Y_{ij}^*$, i.e., that $Y_{ij}$
are real.

After spontaneous symmetry breaking, the Yukawa couplings in Eq.~(\ref{Lyuk})
induce mass terms for the quarks,
\beq\label{mass}
{\cal L}_{\rm mass} = - (M_d)_{ij}\, \ov{d_{Li}^I}\, d_{Rj}^I
  - (M_u)_{ij}\, \ov{u_{Li}^I}\, u_{Rj}^I
  - (M_\ell)_{ij}\, \ov{\ell_{Li}^I}\, \ell_{Rj}^I + {\rm h.c.}\,,
\eeq
which is obtained by replacing $\phi$ with its VEV in Eq.~(\ref{Lyuk}), and
$M_f = (v/\sqrt2)\, Y^f$, where $f=u,d,\ell$ stand for up- and down-type
quarks and charged leptons, respectively.  To obtain the physical mass
eigenstates, we must diagonalize the matrices $M_f$.  As any complex matrix,
$M_f$ can be diagonalized by two unitary matrices, $V_{f\,L,R}$,
\beq
M_f^{\rm diag} \equiv V_{fL}\, M_f\, V_{fR}^\dagger \,.
\eeq
In this new basis the mass eigenstates are
\beq
f_{Li} \equiv (V_{fL})_{ij}\, f_{Lj}^I\,, \qquad
  f_{Ri} \equiv (V_{fR})_{ij}\, f_{Rj}^I\,.
\eeq
We see that the quark mass matrices are diagonalized by different
transformations for $u_{Li}$ and $d_{Li}$, which are part of the same $SU(2)_L$
doublet, $Q_L$,
\beq
\pmatrix{u_{Li}^I\cr d_{Li}^I} = (V_{uL}^\dagger)_{ij}
  \pmatrix{u_{Lj}\cr (V_{uL}V_{dL}^\dagger)_{jk}\, d_{Lk}} .
\eeq
The ``misalignment" between these two transformations,
\beq
V_{\rm CKM} \equiv V_{uL} V_{dL}^\dagger \,,
\eeq
is the Cabibbo-Kobayashi-Maskawa (CKM) matrix.\cite{KM,C}

This transformation makes the charged current weak interactions, that arise
from Eq.~(\ref{Lkin}), appear more complicated in the new basis
\beq
- {g\over 2}\, \ov{Q_{Li}^I}\, \gamma^\mu\, W_\mu^a\, \tau^a\, Q_{Li}^I 
  + {\rm h.c.} \ \Rightarrow \ 
- {g\over \sqrt2}\ \big(\ov{u_{L}},\ \ov{c_{L}},\, \ov{t_{L}}\big)\, 
  \gamma^\mu\, W_\mu^+ V_{\rm CKM}\pmatrix{d_L \cr s_L \cr b_L} + {\rm h.c.}\,,
\eeq
where $W_\mu^\pm = (W^1_\mu \mp W^2_\mu) / \sqrt2$.  As an exercise,
show that the neutral current interactions with the $Z^0$ remain flavor
conserving in the mass basis.  (This is actually true in all models with only
left handed doublet and right handed singlet quarks.)  Thus, in the SM all
flavor changing processes are mediated by charged current weak interactions,
whose couplings to the six quarks are given by a three-by-three unitary matrix,
the CKM matrix.  

As an aside, let us discuss briefly neutrino masses.  With the particle content
given in Eq.~(\ref{reps}), it is not possible to write down a renormalizable
mass term for neutrinos.  Such a term would require the existence of a
$\nu_R(1,1)_0$ field, a so-called sterile neutrino.  Omitting such a field from
Eq.~(\ref{reps}) is motivated by the prejudice that it would be unnatural for a
field that has no SM gauge interactions (is a singlet under all SM gauge
groups) to have mass of the order of the electroweak scale.  Viewing the SM as
an low energy effective theory, there is a single type of dimension-5 terms
made of SM fields that are gauge invariant and give rise to neutrino mass,
${1\over\Lambda_{\rm NP}} Y_{ij}^\nu L_i L_j\phi\phi$, where $\Lambda_{\rm NP}$
is a new physics scale.  This term violates lepton number by two units.  The
suppression of this term cannot be the electroweak scale, $1\over v$, instead
of $1\over\Lambda_{\rm NP}$, because such a term in the Lagrangian cannot be
generated from SM fields at arbitrary loop level, or even nonperturbatively. 
(The reason is that such a mass term violates $B-L$, baryon number minus lepton
number, which is an accidental symmetry of the SM that is not anomalous.)  The
above imply that neutrinos are Majorana fermions, since the mass term couples
the field $\nu_L$ to $\ov{(\nu_L)^c}$ and not to $\ov{\nu_R}$ [the latter
occurs for Dirac fermions, see Eq.~(\ref{mass})].  It can be shown that
$Y_{ij}^\nu$ has to be a real symmetric matrix.

\subsubsection{The CKM matrix and the unitarity triangle}

The nine complex entries of the CKM matrix depend on nine real parameters
because of unitarity.  However, five phases can be absorbed by redefining the
quark fields.  Thus we are left with four parameters, three mixing angles and a
phase.  This phase is the only source of $CP$ violation in flavor changing
transitions in the SM.  A cleaner way to count the number of physical
parameters is to note that the two Yukawa matrices, $Y_{i,j}^{u,d}$ in
Eq.~(\ref{Lyuk}), contain 18 real and 18 imaginary parameters.  They break
global $U(3)_Q \times U(3)_u \times U(3)_d$ symmetry to $U(1)_B$, so there is
freedom to remove $3 \times 3$ real and $3 \times 6 - 1$ imaginary parameters. 
This leaves us with $10$ physical quark flavor parameters: $9$ real ($6$ masses
and $3$ mixing angles) and a complex phase.  In the case on $N$ generations,
the CKM matrix depends on $N(N-1)/2$ mixing angles and $(N-1)(N-2)/2$ $CP$
violating phases.  (In the case of Majorana fermions, one can show following
either derivation that there are $N(N-1)/2$ $CP$ violating phases.)

\begin{figure}[t]
\centerline{\includegraphics*[width=.45\textwidth]{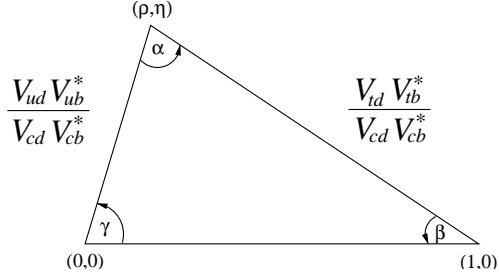}}
\caption{Sketch of the unitarity triangle.}
\label{fig:triangle}
\end{figure}

It has been observed experimentally that the CKM matrix has a hierarchical
structure, which is well exhibited in the Wolfenstein parameterization,
\begin{equation}\label{ckmdef}
V_{\rm CKM} = \pmatrix{ V_{ud} & V_{us} & V_{ub} \cr
  V_{cd} & V_{cs} & V_{cb} \cr
  V_{td} & V_{ts} & V_{tb} } 
= \pmatrix{ 1-\frac{1}{2}\lambda^2 & \lambda & A\lambda^3(\rho-i\eta) \cr
  -\lambda & 1-\frac{1}{2}\lambda^2 & A\lambda^2 \cr
  A\lambda^3(1-\rho-i\eta) & -A\lambda^2 & 1} + \ldots \,.
\end{equation}
This form is valid to order $\lambda^4$.  The small parameter is chosen as the
sine of the Cabibbo angle, $\lambda = \sin\theta_C \simeq 0.22$, while $A$,
$\rho$, and $\eta$ are order unity.  In the SM, the only source of $CP$
violation in flavor physics is the phase of the CKM matrix, parameterized by
$\eta$.  The  unitarity of $V_{\rm CKM}$ implies that the nine complex elements
of this matrix must satisfy $\sum_k V_{ik} V_{jk}^* = \sum_k V_{ki} V_{kj}^* =
\delta_{ij}$.  The vanishing of the product of the first and third columns
provides a simple and useful way to visualize these constraints,
\begin{equation}
V_{ud}\, V_{ub}^* + V_{cd}\, V_{cb}^* + V_{td}\, V_{tb}^* = 0 \,,
\end{equation}
which can be represented as a triangle (see Fig.~\ref{fig:triangle}). Making
overconstraining measurements of the sides and angles of this unitarity
triangle is one of the best ways to look for new physics.

It will be useful to define two angles in addition to those of the triangle in
Fig.~\ref{fig:triangle},
\beqa\label{angledef}
\ds \beta &\equiv& \phi_1 \equiv
  \arg\left(-{V_{cd}V_{cb}^*\over V_{td}V_{tb}^*}\right), \qquad
\beta_s \equiv \arg\left( -\frac{V_{ts}V_{tb}^*}{V_{cs}V_{cb}^*}\right), \qquad
\beta_K \equiv \arg\left( -\frac{V_{cs}V_{cd}^*}{V_{us}V_{ud}^*}\right),\nn\\*
\ds \alpha &\equiv& \phi_2 \equiv
  \arg\left(-{V_{td}V_{tb}^*\over V_{ud}V_{ub}^*}\right), \qquad
\gamma \equiv \phi_3 \equiv
  \arg\left(-{V_{ud}V_{ub}^*\over V_{cd}V_{cb}^*}\right).
\eeqa
Here $\beta_s$ ($\beta_K$) is the small angle of the ``squashed" unitarity
triangle obtained by multiplying the second column of the CKM matrix with the
third (first) column, and is of order $\lambda^2$ ($\lambda^4$).  $\beta_s$ is
the phase between $B_s$ mixing and the dominant $B_s$ decays, while $\beta_K$
is the phase between the charm contribution to $K$ mixing and the dominant $K$
decays.  Checking, for example, if $\beta_s$ is small is an equally important
test of the SM as comparing the sides and angles of the triangle in
Fig.~\ref{fig:triangle}.

To overconstrain the unitarity triangle, there are two very important clean
measurements which will reach precisions at the few, or maybe even one, percent
level.  One is $\sin2\beta$ from the $CP$ asymmetry in $B\to \psi K_S$, which
is becoming the most precisely known angle or side of the unitarity triangle. 
The other is $|V_{td}/V_{ts}|$ from the ratio of the neutral $B_d$ and $B_s$
meson mass differences, $\Delta m_d/\Delta m_s$.  These will be discussed in
detail in Sec.~\ref{sec:s2b} and Sec.~\ref{sec:mix}, respectively.

Compared to $\sin2\beta$ and $|V_{td}/V_{ts}|$, for which both the theory and
experiment are tractable, much harder is the determination of another side or
another angle, such as $|V_{ub}|$, or $\alpha$, or $\gamma$ ($|V_{cb}|$ is also
``easy" by these criteria).  However, our ability to test the CKM hypothesis in
$B$ decays will depend on a third best measurement besides $\sin2\beta$ and
$\Delta m_d/\Delta m_s$ (and on ``null observables", which are predicted to be
small in the SM). The accuracy of these measurements will determine the
sensitivity to new physics, and  the precision with which the SM is tested.  It
does not matter whether it is a side or an angle.  What is important is which 
measurements can be made that have theoretically clean interpretations for the
short distance physics we are after.

\subsection[$CP$ violation before Y2K]{\boldmath $CP$ violation before Y2K}

How do we know that $CP$ is violated in Nature?  Before the start of the $B$
factories, observations of $CP$ violation came from two sources.

\subsubsection[$CP$ violation in the universe]{\boldmath $CP$ violation in the
universe}

The visible Universe is dominated by matter, and antimatter appears to be much
more rare.  To quantify this asymmetry one usually compares the number of
baryons to the number of photons at the present time.  Following the evolution
of the universe back toward the big bang, this ratio is related to the
asymmetry between quarks and antiquarks at about $t \sim 10^{-6}$ seconds after
the big bang, when the temperature was $T \sim 1\,$GeV,
\beq\label{baryon}
{\#(\mbox{baryons}) \over \#(\mbox{photons})}\bigg|_{\mbox{\footnotesize now}} 
  \sim \ {n_q-n_{\ov q}\over n_q+n_{\ov q}}\bigg|_
  {t\sim10^{-6}\,\mbox{\footnotesize sec}} \sim 5\times 10^{-10}\,.
\eeq
It is usually assumed that at even earlier times the universe probably went
through an inflationary phase, which would have washed out any baryon asymmetry
that may have been present before inflation.  There are three conditions first
noted by Sakharov~\cite{sakharov} that any theory must satisfy in order to
allow for the possibility of dynamically generating the asymmetry in
Eq.~(\ref{baryon}).  The theory has to contain: (1) baryon number violating
interactions; (2) $C$ and $CP$ violation; and (3) deviation from thermal
equilibrium.

The first condition is obvious, and the second is required so that the
production rate of left (right) handed quarks and right (left) handed
antiquarks may differ.  The third condition is needed because in thermal
equilibrium the chemical potential for quarks and antiquarks is the same (the
$CPT$ theorem implies that the mass of any particle and its antiparticle
coincide), and so the production and annihilation rates of  quarks and
antiquarks would be the same even if the first two conditions are satisfied. 

The SM contains all three ingredients, but $CP$ violation is too small
(independent of the size of the CKM phase) and the deviation from thermal
equilibrium during electroweak phase transition is too small if there is only
one Higgs doublet.  Detailed analyses show that both of these problems can be
solved in the presence of new physics, that must contain new sources of $CP$
violation and have larger deviations from thermal equilibrium than that in the
SM.  However, for example, the allowed parameter space of the minimal
supersymmetric standard model is also getting very restricted to explain
electroweak baryogenesis (for details, see: Ref.~\citex{baryogen}).

While new physics may yield new $CP$ violating effects observable in $B$
decays, it is possible that the $CP$ violation responsible for baryogenesis
only affects flavor diagonal processes, such as electron or neutron electric
dipole moments.  Another caveat is that understanding the baryon asymmetry may
have nothing to do with the electroweak scale; in fact with the observation of
large mixing angles in the neutrino sector, leptogenesis~\cite{leptogen}
appears more and more plausible.  The idea is that at a very high scale a
lepton-antilepton asymmetry is generated, which is then converted to a baryon
asymmetry by $B+L$ violating but $B-L$ conserving processes present in the SM. 
The lepton asymmetry is due to $CP$ violating decays of heavy sterile
neutrinos, that live long enough to decay out of thermal equilibrium.  However,
the relevant $CP$ violating parameters may or may not be related to $CP$
violation in the light neutrino sector~\cite{Branco}.

\subsubsection[$CP$ violation in the kaon sector]{\boldmath $CP$ violation in
the kaon sector}

Prior to 1964, the explanation of the large lifetime ratio of the two neutral
kaons was $CP$ symmetry (before 1956, it was $C$ alone).  The argument is as
follows.  The flavor eigenstates,
\beq
|K^0\rangle = |\bar s d\rangle\,, \qquad  |\K0bar\rangle = |\bar d s\rangle\,,
\eeq
are clearly not $CP$ eigenstates.  If $CP$ was a good symmetry, then the
states with definite $CP$ would be the following linear combinations
\beq
|K_{1,2}\rangle = \frac1{\sqrt2} \left(|K^0\rangle \pm |\K0bar\rangle\right)\,,
  \qquad  CP|K_{1,2}\rangle = \pm |K_{1,2}\rangle\,.
\eeq
Then only the $CP$ even state could decay into two pions, $K_1 \to \pi\pi$,
whereas both states could decay to three pions, $K_{1,2} \to \pi\pi\pi$
(explain why!).  Therefore one would expect $\tau(K_1) \ll \tau(K_2)$, in
agreement with experimental data, since the phase space for the decay to two
pions is much larger than that to three pions.  The discovery of  $K_L \to
\pi\pi$ decay at the $10^{-3}$ level in 1964 was a big surprise~\cite{Kcpv}. 
The ``natural" explanation for the observed small $CP$ violation was a new
interaction, and, indeed, the superweak model\cite{superweak} was proposed less
than a year after the experimental discovery, whereas the Kobayashi-Maskawa
proposal\cite{KM} came nine years later (but still before even the charm quark
was discovered).

To analyze $CP$ violation in kaon decays, one usually defines the observables
\beq
\eta_{00} =  {\langle \pi^0\pi^0| {\cal H} |K_L\rangle \over
  \langle \pi^0\pi^0| {\cal H} |K_S\rangle}\,, \qquad
\eta_{+-} = {\langle \pi^+\pi^-| {\cal H} |K_L\rangle \over
  \langle \pi^+\pi^-| {\cal H} |K_S\rangle}\,,
\eeq
and the two $CP$ violating parameters,
\beq\label{epsdef}
\epsilon_K \equiv {\eta_{00}+2\eta_{+-}\over 3}\,, \qquad
\epsilon_K' \equiv {\eta_{+-}-\eta_{00}\over 3}\,.
\eeq
To understand these definitions, note that because of Bose statistics the
$|\pi\pi\rangle$ final state can only be in isospin $0$ (i.e., coming from the 
$\Delta I = \frac12$ part of the Hamiltonian, as the initial state is $I =
\frac12$) or isospin $2$ (i.e., $\Delta I = \frac32$) combination [see
discussion before Eq.~(\ref{ispinpipi1})].   Isospin is a symmetry of the
strong interactions, to a very good approximation.  The decomposition of
$|\pi\pi\rangle$ in terms of isospin is
\beqa\label{ispinpipi0}
|\pi^0 \pi^0\rangle &=& - \sqrt{\frac13} \, |(\pi \pi)_{I=0}\rangle
  + \sqrt{\frac23} \, |(\pi \pi)_{I=2}\rangle \,, \nn\\
|\pi^+ \pi^- \rangle &=& \sqrt{\frac23} \, |(\pi \pi)_{I=0}\rangle
  + \sqrt{\frac13} \, |(\pi \pi)_{I=2}\rangle \,.
\eeqa
(In kaon physics often an opposite sign convention is used for 
$|\pi^0 \pi^0\rangle$; Eq.~(\ref{ispinpipi0}) agrees with the Clebsch-Gordan
coefficients in the PDG, used in $B$ physics.)  Then the isospin amplitudes are
defined as
\beqa\label{ispinamp}
A_I &=& \langle (\pi \pi)_I |{\cal H}| K^0 \rangle 
  = |A_I|\, e^{i\delta_I}\, e^{i\phi_I}\,, \nn\\
\ov{A}_I &=& \langle (\pi \pi)_I |{\cal H}| \K0bar \rangle 
  = |A_I|\, e^{i\delta_I}\, e^{-i\phi_I}\,, 
\eeqa
where $I=0,2$, and $\delta_I$ and $\phi_I$ are the strong and weak phases,
respectively.  It is known experimentally that $|A_0| \gg |A_2|$, which is the
so-called $\Delta I = \frac12$ rule ($|A_0| \simeq 22\, |A_2|$).

The definition of $\epsilon_K$ in Eq.~(\ref{epsdef}) is chosen such that to
leading order in the $\Delta I = \frac12$ rule only the dominant strong
amplitude contributes, and therefore $CP$ violation in decay gives only
negligible contribution to $\epsilon_K$ (suppressed by $|A_2/A_0|^2$).  The
world average is $\epsilon_K = e^{i(0.97\pm0.02)\pi/4}\, (2.28 \pm 0.02)\times
10^{-3}$~\citex{pdg}.  Concerning $\epsilon'_K$, to first order in $|A_2/A_0|$,
\beqa\label{epsprime}
\epsilon'_K = \frac{\eta_{+-} - \eta_{00}}{3} 
&=& \frac{\epsilon_K}{\sqrt{2}} \left[ 
  \frac{\langle (\pi\pi)_{I=2} |{\cal H}| K_L \rangle}
  {\langle (\pi\pi)_{I=0} |{\cal H}| K_L \rangle} 
  - \frac{\langle (\pi\pi)_{I=2} |{\cal H}| K_S \rangle}
  {\langle (\pi\pi)_{I=0} |{\cal H}| K_S \rangle} \right] \nn\\
&=& \frac{i}{\sqrt{2}} \left|\frac{A_2}{A_0}\right| e^{i(\delta_2-\delta_0)}\,
  \sin(\phi_2-\phi_0) \,.
\eeqa
A non-vanishing value of $\epsilon'_K$ implies different $CP$ violating phases
in the two isospin amplitudes.  The quantity that is actually measured
experimentally is $|\eta_{00} / \eta_{+-}|^2 = 1- 6\, {\rm Re} (\epsilon'_K /
\epsilon_K)$.  The world average is ${\rm Re} (\epsilon'_K / \epsilon_K) = (1.8
\pm 0.4)\times 10^{-3}$~\citex{pdg}.

These two observed $CP$ violating parameters in the $K$ system are at the level
expected in the SM.  The value of $\epsilon_K$ can be described with an ${\cal
O}(1)$ value of the CKM phase and provides a useful constraint.  However,
precision tests are not yet possible, as $\epsilon_K'$ is notoriously hard to
calculate in the SM because of enhanced hadronic uncertainties due to
contributions that are comparable in magnitude and opposite in sign.  (The
measurement of $\epsilon_K'$ does provide useful constraints on new physics.)  

\begin{figure}[t]
\centerline{\includegraphics*[width=.55\textwidth]{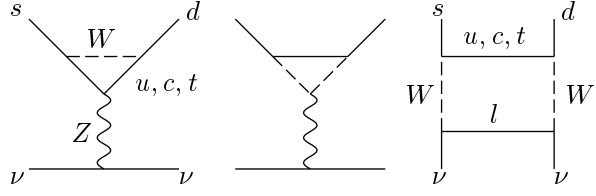}}
\caption{Diagrams contributing to $K\to \pi\nu\bar\nu$ decay (from
Ref.~\protect\citex{gino}).}
\label{fig:kpinunu}
\end{figure}

Precision tests of the SM flavor sector in $K$ decays will come from
measurements of $K \to \pi \nu\bar\nu$, planned in both the neutral and charged
modes.  These observables are theoretically clean, but the rates are very
small, $\sim 10^{-10} \ (10^{-11})$ in $K^\pm \ (K_L)$ decay.  They arise from
the diagrams in Fig.~\ref{fig:kpinunu}, with intermediate up-type quarks.  Due
to the GIM mechanism~\cite{GIM}, the rate would vanish in the limit where the
up, charm, and top quarks had the same mass.  Therefore each contribution to
the amplitude is proportional approximately to $m_q^2/m_W^2$, and we have
schematically
\beq\label{kpinunu}
A \propto \cases{ 
(\lambda^5\, m_t^2) + i (\lambda^5\, m_t^2)~~~~  &  $t$\,: CKM suppressed\,,\cr
(\lambda\, m_c^2) + i (\lambda^5\, m_c^2)  &  $c$\,: GIM suppressed\,,\cr
(\lambda\, \lqcd^2)  &  $u$\,: GIM suppressed\,, \cr}
\eeq
where we used the phase convention and parameterization in Eq.~(\ref{ckmdef}).
Each contribution is either GIM or CKM suppressed.  So far two  $K^+ \to \pi^+
\nu\bar\nu$ events have been observed~\cite{bnl}, corresponding to a branching
ratio ${\cal B} (K^+ \to \pi^+ \nu\bar\nu) = \Big(1.57^{+1.75}_{-0.82}\Big)
\times 10^{-10}$.  The decay $K_L \to \pi^0 \nu\bar\nu$ is even cleaner than
the charged mode because the final state is $CP$ even~\cite{Littenberg}, and
therefore only the imaginary parts in Eq.~(\ref{kpinunu}) contribute, where the
charm contribution is negligible and the top contribution is  a precisely
calculable short distance process.  (For a more detailed discussion, see
Ref.~\citex{rt}.)

\subsection[The $B$ physics program and the present status]{\boldmath The
$B$ physics program and the present status}

In comparison with kaons, the $B$ meson system has several features which makes
it well-suited to study flavor physics and $CP$ violation.  Because the top
quark in loop diagrams is neither GIM nor CKM suppressed, large $CP$ violating
effects and large mixing are possible in the neutral $B_d$ and $B_s$ systems,
some of which have clean interpretations.  For the same reason, a variety of
rare decays have large enough branching fractions to allow for detailed
studies.  Finally, some of the hadronic physics can be understood model
independently because $m_b \gg \lqcd$.

The goal of this program is to precisely test the flavor sector via redundant
measurements, which in the SM determine CKM elements, but can be sensitive to
different short distance physics.  New physics is most likely to modify $CP$
violating observables and decays that proceed in the SM via loop diagrams only,
such as mixing and rare decays.  Therefore, we want to measure $CP$ violating
asymmetries, mixing and rare decays, and compare the constraints on the CKM
matrix from tree and loop processes.

In the SM all $CP$ violation in flavor changing processes arises from the phase
in the CKM matrix.  The CKM elements with large (and related) phases in the
usual convention are $V_{td}$ and $V_{ub}$, and all large $CP$ violating
phenomena comes from these.  In the presence of new physics, many independent
$CP$ violating phases are possible; e.g., the phases in $B_d$ and $B_s$ mixing
may be unrelated.  Then using $\alpha$, $\beta$, $\gamma$ is only a language,
as two ``would-be" $\gamma$ measurements, for example, can be sensitive to
different new physics contributions.  Similarly, measurements of $|V_{td}|$ and
$|V_{ts}|$ from mixing may be unrelated to their values measured in rare
decays.  Thus, to search for new physics, all possible measurements which have
clean interpretations are important; their correlations and the pattern of
possible deviations from the SM predictions may be crucial to narrow down type
of new physics we are encountering.  The $B$ physics program is so broad
because independent measurements are the best way to search for new physics.

\begin{figure}[t]
\centerline{\includegraphics*[width=.6\textwidth]{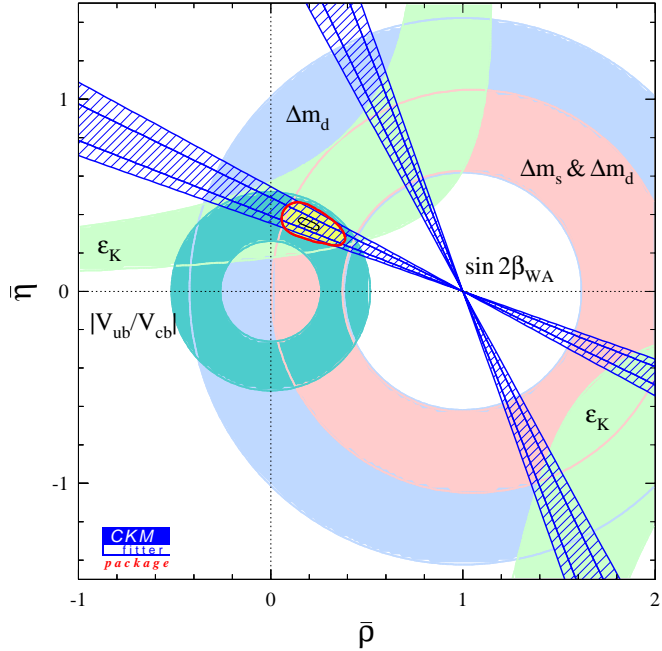}}
\caption{Present constraints on the CKM matrix (from
Ref.~\protect\citex{ckmfitter}).}
\label{fig:ckmfit}
\end{figure}

The allowed regions of $\rho$ and $\eta$, imposed by the constraints on
$\epsilon_K$, $B_{d,s}$ mixing, $|V_{ub}/V_{cb}|$, and $\sin2\beta$ are shown
in Fig.~\ref{fig:ckmfit}.  There is a four-fold discrete ambiguity in the
$\sin2\beta$ measurement.  Assuming the SM, this is resolved by $|V_{ub}|$:
there is only one allowed region using the $|V_{ub}|$ and $\sin2\beta$
constraints, whereas there would be four allowed regions if the $|V_{ub}|$
constraint is removed from the fit.  

Figure~\ref{fig:ckmfit} clearly shows that with the recent precise measurements
of $\sin2\beta$, the CKM picture passed its first real test, and the angle
$\beta$ has become the most precisely known ingredient in the unitarity
triangle.  Thus, it is very likely that the CKM matrix is the dominant source
of $CP$ violation in flavor changing processes at the electroweak scale.  This
implies a paradigm change in that we can no longer claim to be looking for new
physics alternatives of the CKM picture, but to seek corrections to it (a
possible exception is still the $B_s$ system).  The question is no longer
whether the CKM paradigm is right, but whether it is the only observable source
of $CP$ violation and flavor change near the electroweak scale.

In looking for modest deviations from the SM, the key measurements are those
that are theoretically clean and experimentally doable.  Measurements whose
interpretation depends on hadronic models cannot indicate unambiguously the
presence of new physics.  Our ability to test CKM in $B$ decays below the
$10\%$ level will depend on the 3rd, 4th, etc., most precise measurements
besides $\beta$ and $|V_{td}/V_{ts}|$ that are used to overconstrain it.  (The
error of $|V_{td}/V_{ts}|$ is expected to be below $10\%$ once the $B_s$ mass
difference is measured, as discussed in Sec.~\ref{sec:mix}.)  Prospects to
measure the $|V_{ub}/V_{cb}|$ side of the UT with small error are discussed in
the second lecture, while clean determinations of angles other than $\beta$ are
discussed in the third.  Certain observables that are (near) zero in the SM,
such as $a_{CP}(B_s\to\psi\phi)$, $a_{CP}(B\to\psi K_S) - a_{CP}(B\to\phi
K_S)$, $a_{dir}(B\to s\gamma)$, are also sensitive to new physics and some will
be discussed.

\subsection[$B_d$ and $B_s$ mixing]{\boldmath $B_d$ and $B_s$
mixing}\label{sec:mix}

Similar to the neutral kaon system, there are also two neutral $B^0$ flavor
eigenstates,
\beq
|B^0\rangle = |\bar b\, d\rangle \,, \qquad
  |\B0bar\rangle = |b\, \bar d\rangle \,.
\eeq
The time evolution of a state is described by the Schr\"odinger equation,
\beq
i\, {\d \over \d t} \pmatrix{|B^0(t)\rangle\cr |\B0bar(t)\rangle} 
= \Big(M - {i\over2}\,\Gamma\Big)\! 
  \pmatrix{|B^0(t)\rangle\cr |\B0bar(t)\rangle} \,,
\eeq
where the mass mixing matrix, $M$, and the decay mixing matrix, $\Gamma$, are
$2\times2$ Hermitian matrices.  $CPT$ invariance implies $M_{11} = M_{22}$ and
$\Gamma_{11} = \Gamma_{22}$.  The heavier and lighter mass eigenstates are the
eigenvectors of $M - i\Gamma/2$,
\beq\label{defpq}
|B_{H,L}\rangle = p\, |B^0\rangle \mp q\, |\B0bar\rangle\,,
\eeq
and their time dependence is 
\beq\label{timedep}
|B_{H,L}(t)\rangle = e^{-(iM_{H,L} + \Gamma_{H,L}/2)t}\, |B_{H,L}\rangle\,.
\eeq

\begin{figure}[t]
\centerline{\includegraphics*[height=.15\textwidth]{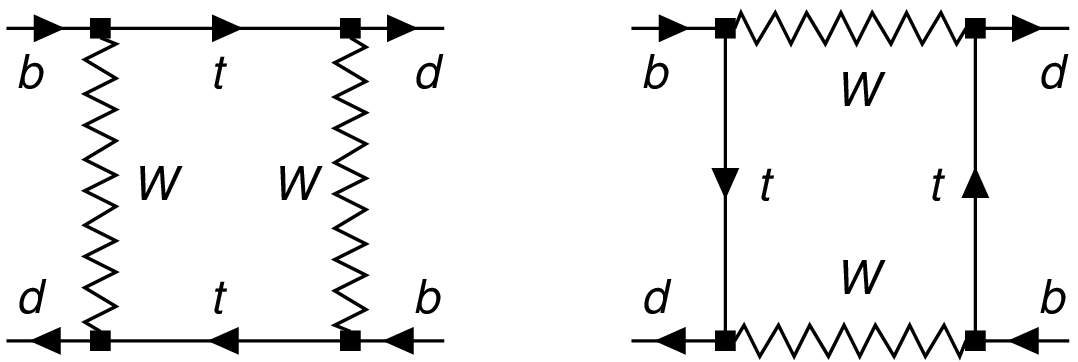}
\hspace{.8cm} \raisebox{1cm}{$\Longrightarrow$} \hspace{.6cm}
\includegraphics*[height=.15\textwidth]{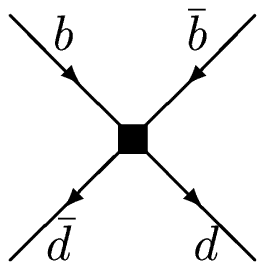}}
\caption{Left: box diagrams that give rise to the $B^0-\B0bar$ mass difference;
Right: operator in the effective theory below $m_W$ whose $B$ meson matrix
element determines $\Delta m_{B_d}$.}
\label{fig:boxes}
\end{figure}

The solution of the eigenvalue equation is
\beqa\label{eigenstates}
(\Delta m)^2 - \frac{1}{4}\, ( \Delta\Gamma)^2 
  &=& 4\, |M_{12}|^2 - |\Gamma_{12}|^2\,, \qquad
\Delta m\, \Delta\Gamma = -4\, {\rm Re} (M_{12} \Gamma_{12}^*)\,, \nn\\
{q\over p} &=& - \frac{\Delta m + i\, \Delta\Gamma/2}{2M_{12} -i\, \Gamma_{12}}
  = - \frac{2 M_{12}^* -i\, \Gamma_{12}^*}{\Delta m + i\, \Delta\Gamma/2}\,,
\eeqa
where $\Delta m = M_H - M_L$ and $\Delta\Gamma = \Gamma_L - \Gamma_H$.  This
defines $\Delta m$ to be positive, and the choice of $\Delta\Gamma$ is such
that it is expected to be positive in the SM (this sign convention for
$\Delta\Gamma$ agrees with Ref.~\citex{tevbook} and is opposite to
Ref.~\citex{babook}).  Note that $M_{H,L}$ ($\Gamma_{H,L}$) are not the
eigenvalues of $M$ ($\Gamma$).  The off-diagonal elements $M_{12}$ and
$\Gamma_{12}$ arise from virtual and on-shell intermediate states,
respectively.  The contributions to $M_{12}$ are dominated in the SM by box
diagrams with top quarks (see Fig.~\ref{fig:boxes}), while $\Gamma_{12}$ is
determined by physical states (containing $c$ and $u$ quarks) to which both
$B^0$ and $\B0bar$ can decay.

Simpler approximate solutions can be obtained expanding about the limit
$|\Gamma_{12}| \ll |M_{12}|$.  This is a good approximation in both $B_d$ and
$B_s$ systems.  $|\Gamma_{12}| < \Gamma$ always holds, because $\Gamma_{12}$
stems from decays to final states common to $B^0$ and $\B0bar$.  For the $B_s$
meson the experimental lower bound on $\Delta m_{B_s}$ implies $\Gamma
_{B_s}\ll \Delta m_{B_s}$, and hence $\Gamma_{12}^s \ll \Delta m_{B_s}$ [the
theoretical expectation is $\Delta\Gamma_s / \Gamma_s \sim 16\pi^2
(\lqcd/m_b)^3$].  For the $B_d$ meson, experiments give $\Delta m_{B_d} \approx
0.75\, \Gamma_{B_d}$.  However, $\Gamma_{12}^d$ arises  only due to
CKM-suppressed decay channels (giving common final states in $B_d^0$ and
$\B0bar_d$ decay), and so $|\Gamma_{12}^d| / \Gamma_{B_d}$ is expected to be at
or below the few percent level (and many experimental analyses assume that it
vanishes). In this approximation Eqs.~(\ref{eigenstates}) become
\beqa\label{limes}
\Delta m &=& 2\, |M_{12}|\,, \qquad
  \Delta\Gamma = -2\, {{\rm Re} (M_{12} \Gamma_{12}^*)\over |M_{12}|}\,, \nn\\
{q\over p} &=& - {M_{12}^*\over M_{12}} \left[1 - \frac12\, {\rm Im}\left(
  {\Gamma_{12}\over M_{12}} \right)\right]\,,
\eeqa
where we kept the second order term in $q/p$ because it will be needed later. 
Table~\ref{tab:Bdsmix} summarizes the expectations and data for the $B_{d,s}$
systems. 

\begin{table}[t]
\begin{center}
\begin{tabular}{ccccccc} \hline\hline
  &  \multicolumn{2}{c}{$x_q = \Delta m/\Gamma$}
  &  \multicolumn{2}{c}{$y_q = \Delta\Gamma/\Gamma$}
  &  \multicolumn{2}{c}{$A_q = 1 - |q/p|^2$}  \\[-6pt]
  &  theory  &  data  &  theory  &  data  &  theory  &  data \\ \hline\hline
$B_d$~  &  ${\cal O}(1)$  &  $\approx 0.75$
  &  $y_s\, |V_{td}/V_{ts}|^2$  &  $< 0.2$  &  $-0.001$  &  $|A_d| < 0.02$ \\
$B_s$~  &  $x_d\, |V_{ts}/V_{td}|^2$  &  $>20$  
  &  $0.1$  &  $< 0.4$  &  $-A_d\, |V_{td}/V_{ts}|^2$  &  --- \\  \hline\hline
\end{tabular}
\end{center}
\caption{Mixing and $CP$ violation in $B_{d,s}$ mesons.  The theory entries
indicates rough SM estimates.  Data are from the PDG~\cite{pdg} (bounds are
$90\%$ or $95\%$\,CL).}
\label{tab:Bdsmix}
\end{table}

A simple and important implication is that if $\Gamma_{12}$ is given by the SM,
then new physics cannot enhance the $B_{d,s}$ width differences.  To see this,
rewrite $\Delta\Gamma$ in Eq.~(\ref{limes}) as $\Delta\Gamma = 2\,
|\Gamma_{12}| \cos[{\rm arg} (-M_{12} / \Gamma_{12})]$.  In the SM, ${\rm arg}
(-M_{12} / \Gamma_{12})$ is suppressed by $m_c^2/m_b^2$ in both $B_{d,s}$
systems (in the $B_s$ system it is further suppressed by the small angle 
$\beta_s$).  Consequently, by modifying the phase of $M_{12}$, new physics
cannot enhance $\cos[{\rm arg} (-M_{12} / \Gamma_{12})]$, which is near unity
in the SM.  However, new physics can easily enhance $CP$ violation in mixing,
which is suppressed by the small quantity $\sin[{\rm arg} (-M_{12} /
\Gamma_{12})]$ in the SM, and is especially tiny in the $B_s$ system.

The $B_H - B_L$ mass difference dominated by the box diagrams with top quarks
(see Fig.~\ref{fig:boxes}) is a short distance process sensitive to physics at
high scales (similar to $\Delta m_K$). The calculation of $\Delta m_B$ is a
good example of the use of effective theories.  The first step is to ``match"
at the scale of order $m_W$ the box diagrams on the left in
Fig.~\ref{fig:boxes} onto the local four-fermion operator, $Q(\mu) = (\bar b_L
\gamma_\nu d_L) (\bar b_L \gamma^\nu d_L)$, represented by the diagram on the
right.  In this step one computes the Wilson coefficient of $Q(\mu=m_W)$.  In
the second step, one ``runs" the scale of the effective theory down from $m_W$
to a scale around $m_b$ using the renormalization group.  In the third step one
has to compute the matrix element of $Q(\mu)$ at a scale around $m_b$.  The
result is
\beq\label{M12}
M_{12} = \underbrace{{(V_{tb} V_{td}^*)^2}}_{\mbox{\footnotesize WANTED}}
\times \underbrace{\frac{G_F^2}{8 \pi^2}\, \frac{M_W^2}{m_B}}_
  {\mbox{\footnotesize known}}
\times \underbrace{S\bigg( \frac{m_t^2}{M_W^2} \bigg)\, \eta_B\, b_B(\mu)}_
  {\hspace{-1cm}\mbox{\footnotesize calculable perturbatively}\hspace{-1cm}}
\times \underbrace{\langle B^0 | Q (\mu) | \B0bar \rangle}_
{\mbox{\footnotesize nonperturbative}}\,,
\eeq
where the first term is the combination of CKM matrix elements we want to
measure, and the second term contains known factors.  The third term contains
the matching calculation at the high scale, $S(m_t^2/M_W^2)$ (an Inami-Lim
function~\cite{Inami-Lim}), and the calculable QCD corrections that occur in
running the effective Hamiltonian down to a low scale.  It is $\eta_B b_B(\mu)$
that contain the QCD corrections including resummation of the series of leading
logarithms, $\alpha_s^n \ln^n(m_W/\mu)$, $\mu \sim m_b$, which is often very
important.  The last term in Eq.~(\ref{M12}) is the matrix element,
\beq\label{fBBB}
\langle B^0 | Q (\mu) | \B0bar \rangle =
  \frac23\, m_B^2\, f_B^2\, \frac{\widehat{B}_B}{b_B (\mu)}\,,
\eeq
which is a nonperturbative quantity.  It is here that hadronic uncertainties
enter, and $f_B^2\, \widehat{B}_B$ has to be determined from lattice QCD. 
Eq.~(\ref{M12}) applies for $B_s$ mixing as well, replacing $V_{td} \to
V_{ts}$, $m_{B_d} \to m_{B_s}$, $f_{B_d} \to f_{B_s}$, and  $\widehat{B}_{B_d}
\to \widehat{B}_{B_s}$.

\begin{figure}[t]
\centerline{\includegraphics*[width=.46\textwidth]{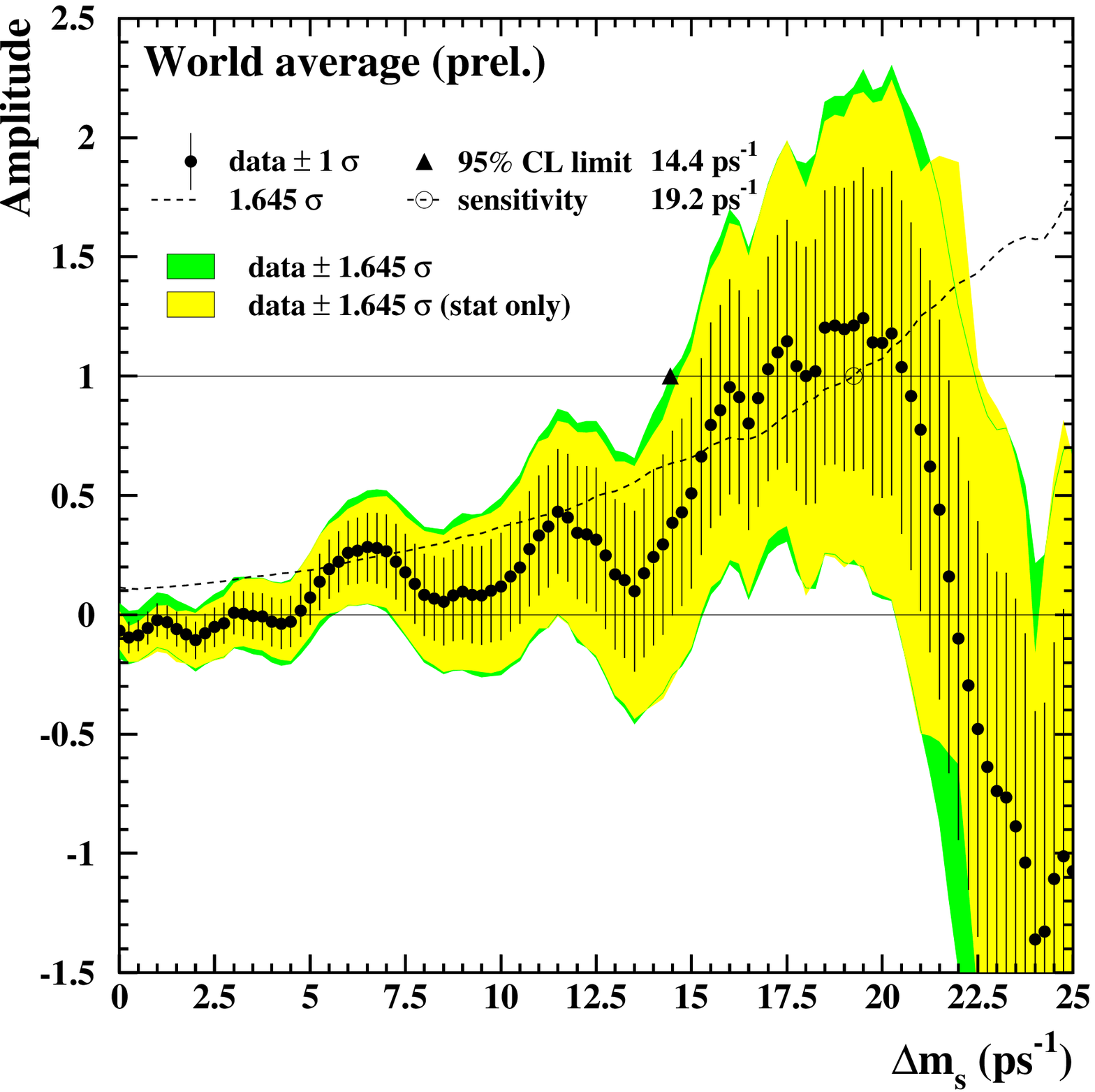}
\raisebox{-5pt}{\includegraphics*[width=.51\textwidth]{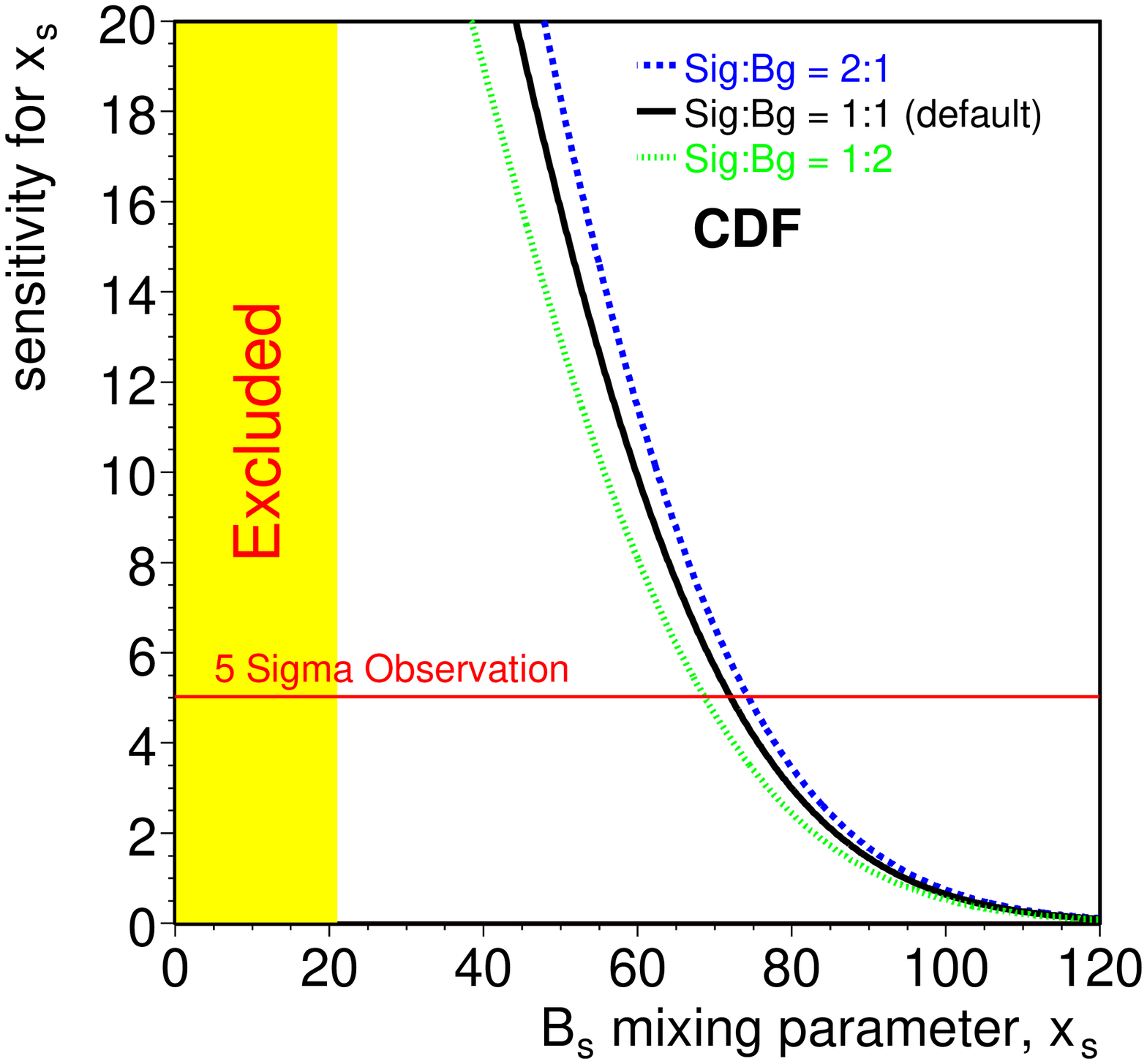}}}
\caption{Left: present $B_s$ oscillation amplitude analysis~\cite{lepbosc}.
Right: projected CDF sensitivity with $2\,{\rm fb}^{-1}$ data~\cite{tevbook}.
Note that $x_s \equiv \Delta m_s/\Gamma_{B_s} \approx \Delta m_s \times
1.46\,\mbox{ps}$.}
\label{fig:bdsmix}
\end{figure}

A clean determination of $|V_{td}/V_{ts}|$ will be possible from the ratio of
the $B_d$ and $B_s$ mass differences, $\Delta m_{B_d}/\Delta m_{B_s}$.  The
reason is that some of the hadronic uncertainties can be reduced by considering
the ratio $\xi^2 \equiv (f_{B_s}^2B_{B_s})/(f_{B_d}^2 B_{B_d})$ which is unity
in the flavor $SU(3)$ symmetry limit.  Figure~\ref{fig:bdsmix} shows the
preliminary LEP/SLD/CDF combined $B_s$ oscillation amplitude
analysis~\cite{lepbosc} that yields $\Delta m_s > 14.4\,{\rm ps}^{-1}$ at
$95\%$\,CL.  Probably $B_s$ mixing will be discovered at the Tevatron, and soon
thereafter the experimental error of $\Delta m_s$ is expected to be at the few
percent level~\cite{tevbook}.  The uncertainty of $|V_{td}/V_{ts}|$ will then
be dominated by the error of $\xi$ from lattice QCD.  For the last few years
the lattice QCD averages~\cite{latticerev} have been around $f_{B_s}/f_{B_d} =
1.18 \pm 0.04^{+0.12}_{-0}$ and $B_{B_s}/B_{B_d} = 1.00 \pm 0.03$, in agreement
with the chiral log calculation~\cite{Bslogs}.  The last error in the quoted
lattice result of $f_{B_s}/f_{B_d}$ reflects an increased appreciation of
uncertainties associated with the chiral extrapolation, that may reduce the
present results for $f_{B_d}$ but is unlikely to significantly affect
$f_{B_s}$.  It is very important to reliably control light quark effects, and
to do simulations with three light flavors.\footnote{Sorting this out reliably
may be challenging, since the leading chiral logarithms need not be a good
guide to the chiral behavior of quantities involving heavy hadrons.  Chiral
perturbation theory for processes with heavy hadrons may have a cutoff as low
as $500\,\MeV$ instead of $4\pi f_\pi \sim 1\,\GeV$, leading to large ``higher
order" effects~\cite{chiralsubl}.  Using chiral perturbation theory to
extrapolate lattice calculations with heavy ``light'' quarks to the chiral
limit may then be questionable~\cite{gbml}.}

\subsection[$CP$ violation in the $B$ meson system]{\boldmath $CP$ violation in
the $B$ meson system}

\subsubsection[The three types of $CP$ violation]{\boldmath The three types of
$CP$ violation}

\paragraph{\boldmath $CP$ violation in mixing}

If $CP$ were conserved, then the mass eigenstates would be proportional to
$|B^0\rangle \pm |\B0bar\rangle$, corresponding to $|q/p| = 1$ and  ${\rm
arg}(M_{12}/\Gamma_{12}) = 0$.  If $|q/p| \neq 1$, then $CP$ is violated.  This
is called $CP$ violation in mixing, because it results from the mass
eigenstates being different from the $CP$ eigenstates.  It follows from
Eq.~(\ref{defpq}) that $\langle B_H | B_L\rangle = |p|^2 - |q|^2$, and so if
there is $CP$ violation in mixing then the two physical states are not
orthogonal.  This is clearly a quantum mechanical effect, impossible in a
classical system.

\begin{figure}[t]
\centerline{\includegraphics*[width=.46\textwidth]{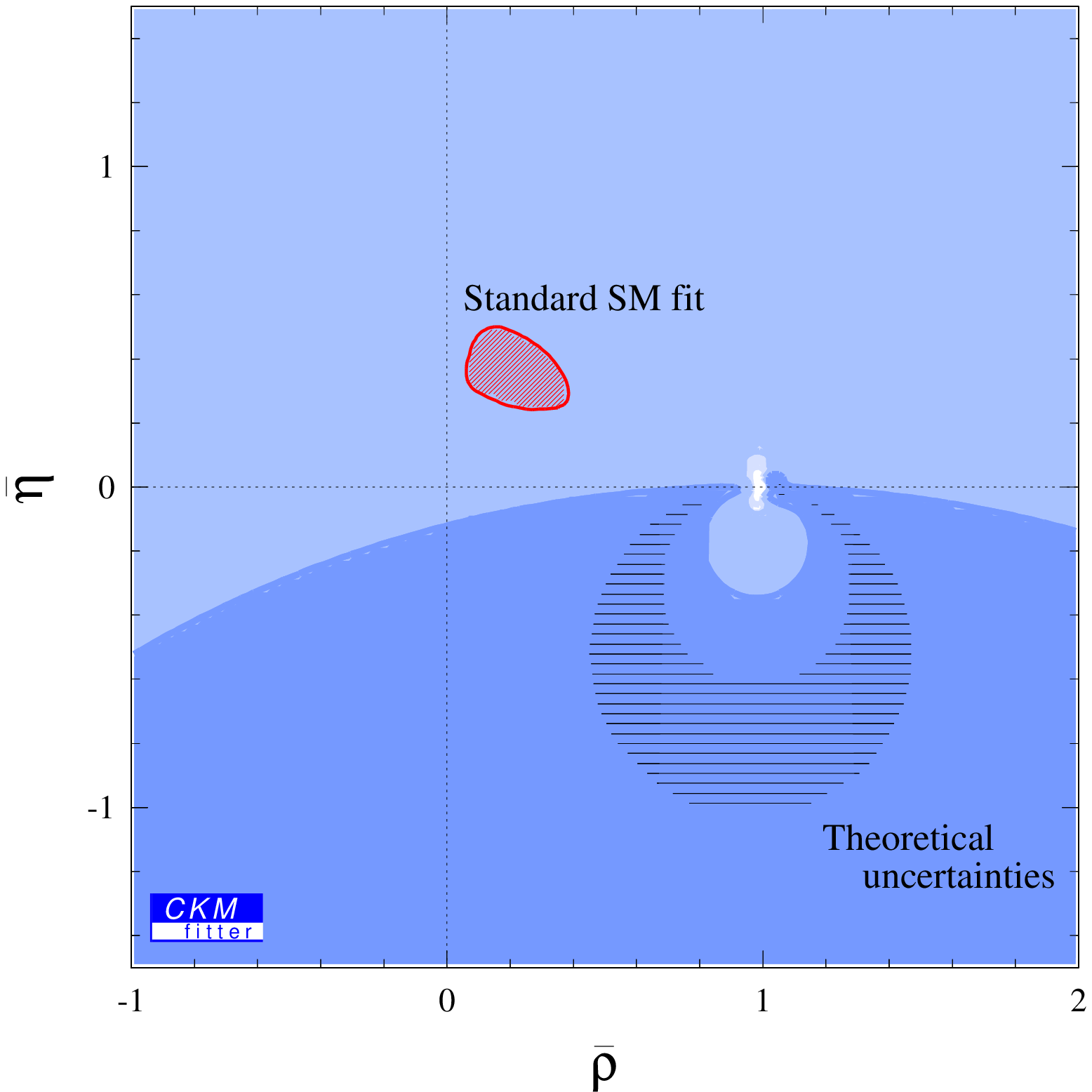}\hfil
\includegraphics*[width=.46\textwidth]{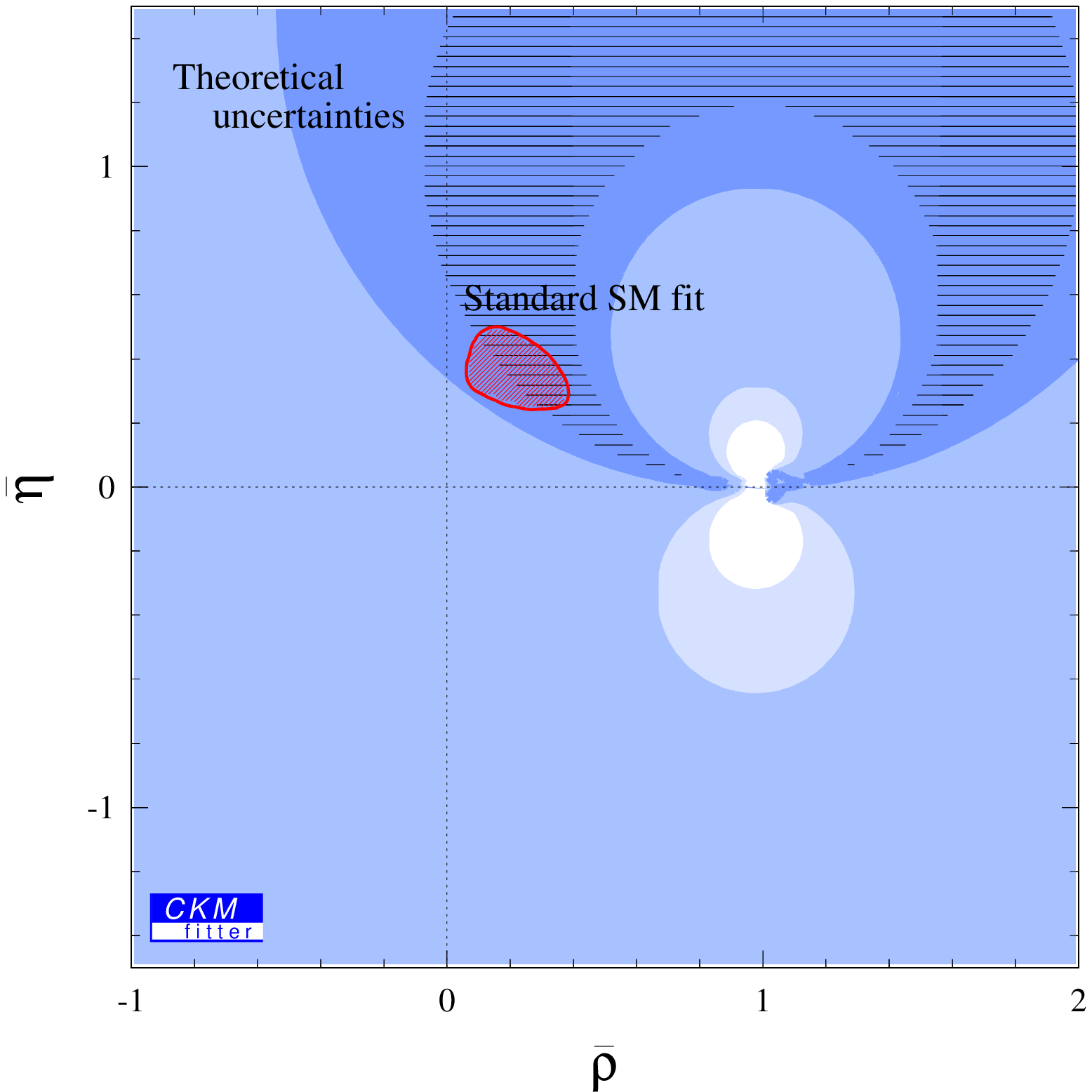}}
\caption{Left: present constraint from $A_{\rm SL} = (0.2 \pm 1.4) \times
10^{-2}$.  Right: constraint that would follow from $A_{\rm SL} = (-1 \pm
3)\times 10^{-3}$ (that may be achieved by 2005).  The dark-, medium-, and
light-shaded regions have CL $> 90\%$, $32\%$, and $10\%$.  (From
Ref.~\protect\citex{llnp}.)}
\label{fig:asl}
\end{figure}

The simplest example of this type of $CP$ violation is the  semileptonic decay
asymmetry of neutral mesons to ``wrong sign" leptons,
\beq
A_{\rm SL}(t) = {\Gamma(\B0bar(t) \to \ell^+ X) - \Gamma(B^0(t) \to \ell^- X)
  \over \Gamma(\B0bar(t) \to \ell^+ X) + \Gamma(B^0(t) \to \ell^- X) }
= {1 - |q/p|^4 \over 1 + |q/p|^4} 
  = {\rm Im}\, {\Gamma_{12}\over M_{12}} \,.
\eeq
To obtain the right-hand side, we used Eqs.~(\ref{defpq}) and (\ref{timedep})
for the time evolution, and Eq.~(\ref{limes}) for $|q/p|$.  In kaon decays this
asymmetry was recently measured~\cite{cplear}, in agreement with the
expectation that it should be equal to $4\,{\rm Re}\, \epsilon_K$.  In $B$
decays the asymmetry is expected to be~\cite{llnp} $-1.3\times10^{-3} < A_{\rm
SL} < -0.5\times10^{-3}$.  Figure~\ref{fig:asl} shows the (weak) constraints on
the $\rho-\eta$ plane from the present data on $A_{SL}$, and what may be
achieved by 2005.  One can only justify the calculation of ${\rm Im}
(\Gamma_{12} / M_{12})$ from first principles in the the $m_b \gg \lqcd$ limit,
since it depends on inclusive nonleptonic rates.  Such a calculation has
sizable hadronic uncertainties (by virtue of our limited understanding of $b$
hadron lifetimes), an estimate of which is shown by the horizontally stripped
regions.  However, the constraints on new physics are already
interesting~\cite{llnp}, as the $m_c^2/m_b^2$ suppression of $A_{SL}$ in the SM
can be avoided if new physics modifies the phase of $M_{12}$.

\paragraph{\boldmath $CP$ violation in decay}

For most final states $f$, the $B\to f$ and $\Bbar\to \ov f$ decay amplitudes
can, in general, receive several contributions
\beq
A_f = \langle f | {\cal H} |B\rangle = 
  \sum_k A_k\, e^{i\delta_k}\, e^{i\phi_k} , \qquad
\ov{A}_{\ov f} = \langle \ov f | {\cal H} |\Bbar\rangle = 
  \sum_k A_k\, e^{i\delta_k}\, e^{-i\phi_k} .
\eeq
There are two types of complex phases which can occur [a similar situation was
already encountered in Eq.~(\ref{ispinamp})].  Complex parameters in the
Lagrangian which enter a decay amplitude also enter the $CP$ conjugate
amplitude but in complex conjugate form.  In the SM such weak phases, $\phi_k$,
only occur in the CKM matrix.  Another type of phases are due to absorptive
parts of decay amplitudes, and give rise to $CP$ conserving strong phases,
$\delta_k$.  These arise from on-shell intermediate states rescattering into
the desired final state.  The individual phases $\delta_k$ and $\phi_k$ are
convention dependent, but the phase differences, $\delta_i - \delta_j$ and
$\phi_i - \phi_j$, and therefore $|\ov{A}_{\ov{f}}|$ and $|A_f|$, are physical.

Clearly, if $|\ov{A}_{\ov{f}}| \neq |A_f|$ then $CP$ is violated.  This is
called $CP$ violation in decay, or direct $CP$ violation.  Such $CP$ violation can also arise in charged meson and baryon decays, and in
$B^0$ decays in conjunction with the other types.  It occurs due to
interference between various terms in the decay amplitude, and requires that at
least two terms differ both in their strong and in their weak phases,
\beq
|A|^2 - |\ov A|^2 = -4 A_1 A_2 \sin(\delta_1-\delta_2) \sin(\phi_1-\phi_2)\,.
\eeq
The only unambiguous observation of direct $CP$ violation to date is ${\rm
Re}\, \epsilon'_K$ in kaon decay.  It can be seen from Eq.~(\ref{epsprime})
that ${\rm Im}\, \epsilon'_K$ is not a sign of $CP$ violation in decay, since
it may be nonzero even if there is no strong phase difference between the two
amplitudes.  Note that in $B^0$ decays different interference type $CP$
violation (see below) in two final states, ${\rm Im} \lambda_{f_1} \neq {\rm
Im} \lambda_{f_2}$, would also be a sign of direct $CP$ violation.

To extract the interesting weak phases from $CP$ violation in decay, one needs
to know the amplitudes $A_k$ and their strong phases $\delta_k$.  The problem
is that theoretical calculations of $A_k$ and $\delta_k$ usually have large
model dependences.  However, direct $CP$ violation can still be very
interesting for looking for new physics, especially when the SM prediction is
small, e.g., in $b\to s\gamma$.

\paragraph{\boldmath $CP$ violation in the interference between decays with and
without mixing}

Another type of $CP$ violation is possible when both $B^0$ and $\B0bar$ can
decay to the same final state.  The simplest example is when this is a $CP$
eigenstate, $f_{CP}$.  If $CP$ is conserved, then not only $|q/p| = 1$ and
$|\ov{A}_f/A_f| = 1$, but the relative phase between $q/p$ and $\ov{A}_f/A_f$
also vanishes.  It is convenient to define
\beq
\lambda_{f_{CP}} = \frac qp\, \frac{\ov{A}_{f_{CP}}}{A_{f_{CP}}} 
  = \eta_{f_{CP}}\, \frac qp\, \frac{\ov{A}_{\ov{f}_{CP}}}{A_{f_{CP}}} \,,
\eeq
where $\eta_{f_{CP}} = \pm 1$ is the $CP$ eigenvalue of $f_{CP}$ [$+1$ ($-1$)
for $CP$-even (-odd) states].  The second form is useful for calculations,
because $A_{f_{CP}}$ and $\ov{A}_{\ov{f}_{CP}}$ are related by $CP$
transformation.  If ${\rm Im}\lambda_{f_{CP}} \neq 0$ then it is a
manifestation of $CP$ violating interference between $B^0 \to f_{CP}$ decay and
$B^0 - \B0bar$ mixing followed by $\B0bar \to f_{CP}$ decay.

The time dependent asymmetry, neglecting $\Delta\Gamma$, is given by 
\beqa\label{SCdef}
a_{f_{CP}} &=& { \Gamma[\B0bar(t)\to f] - \Gamma[B^0(t)\to f]\over
  \Gamma[\B0bar(t)\to f] + \Gamma[B^0(t)\to f] } \nn\\
&=& - {(1-|\lambda_f|^2) \cos(\Delta m\, t) 
  - 2\,{\rm Im}\,\lambda_f \sin(\Delta m\, t) \over 1+|\lambda_f|^2} \nn\\[2pt]
&\equiv& S_f \sin(\Delta m\, t) - C_f \cos(\Delta m\, t) \,.
\eeqa
The last line defines the $S$ and $C$ terms that will be important later on
(note that the BELLE notation is $S \equiv {\cal S}$ and $C \equiv -{\cal
A}$).  This asymmetry can be nonzero if any type of $CP$ violation occurs.  In
particular, if $|q/p| \simeq 1$ and $|\ov{A}_f/A_f| \simeq 1$ then it is
possible that ${\rm Im} \lambda_f \neq 0$, but $|\lambda_f| = 1$ to a good
approximation.  In both the $B_d$ and $B_s$ systems $|q/p| - 1 < {\cal
O}(10^{-2})$, so the question is usually whether $|\ov A/A|$ is near unity.
Even if we cannot compute hadronic decay amplitudes model independently, $|\ov
A/A| = 1$  is guaranteed if amplitudes with a single weak phase dominate a
decay.  In such cases we can extract the weak phase difference between $B^0 \to
f_{CP}$ and $B^0 \to \B0bar \to f_{CP}$ in a theoretically clean way,
\beq
a_{f_{CP}} = {\rm Im} \lambda_f \sin(\Delta m\, t)\,.
\eeq

\subsubsection[$\sin2\beta$ from $B\to \psi K_{S,L}$]{\boldmath
$\sin2\beta$ from $B\to \psi K_{S,L}$}\label{sec:s2b}

This is the cleanest example of $CP$ violation in the interference between
decay with and without mixing, because $|\ov A/A| - 1 \lsim 10^{-2}$.  
Therefore, $\sin2\beta$ will be the theoretically cleanest measurement of a CKM
parameter other than $|V_{ud}|$ (and maybe $\eta$ from $K_L\to
\pi^0\nu\bar\nu$, which, however, is unlikely to be ever measured at the
percent level).

\begin{figure}[t]
\centerline{\includegraphics*[width=.3\textwidth]{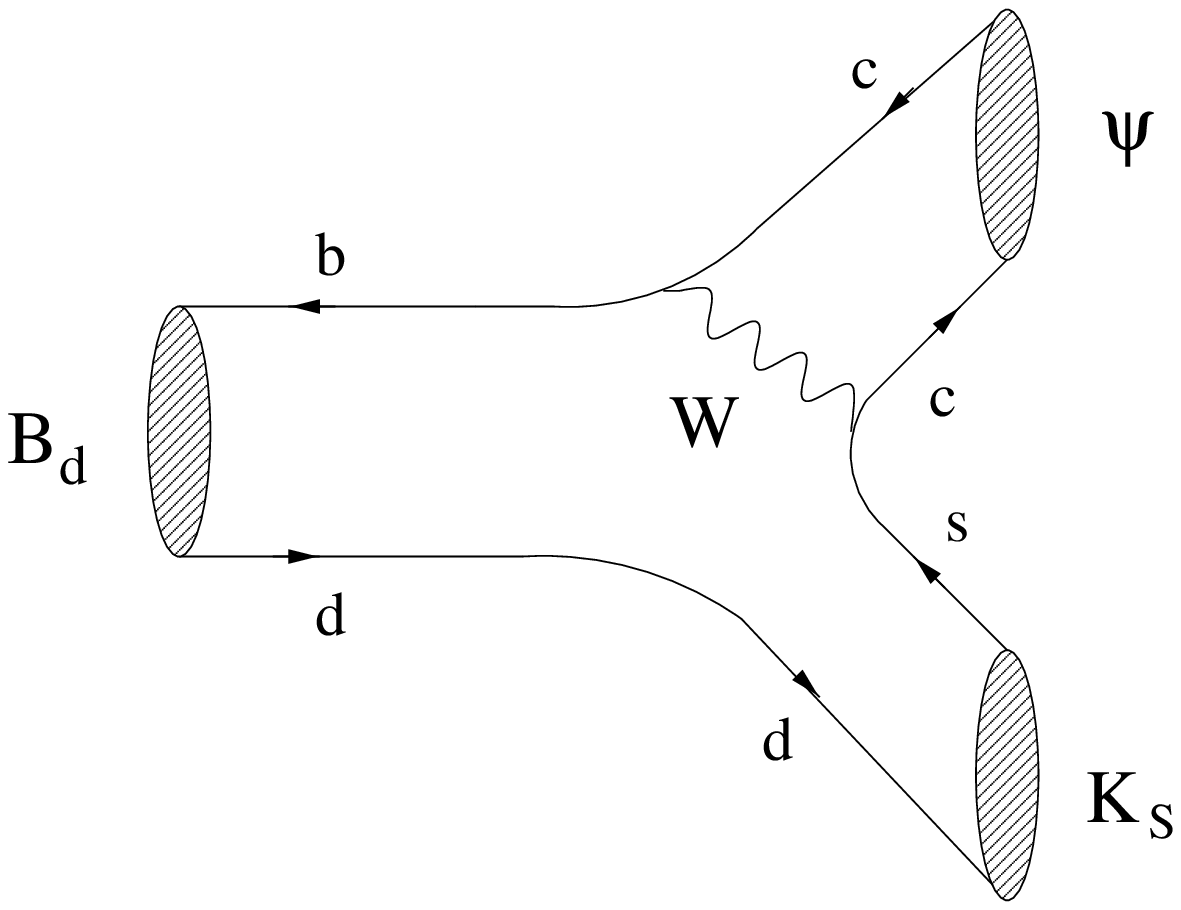}\hspace{1cm}
\includegraphics*[width=.3\textwidth]{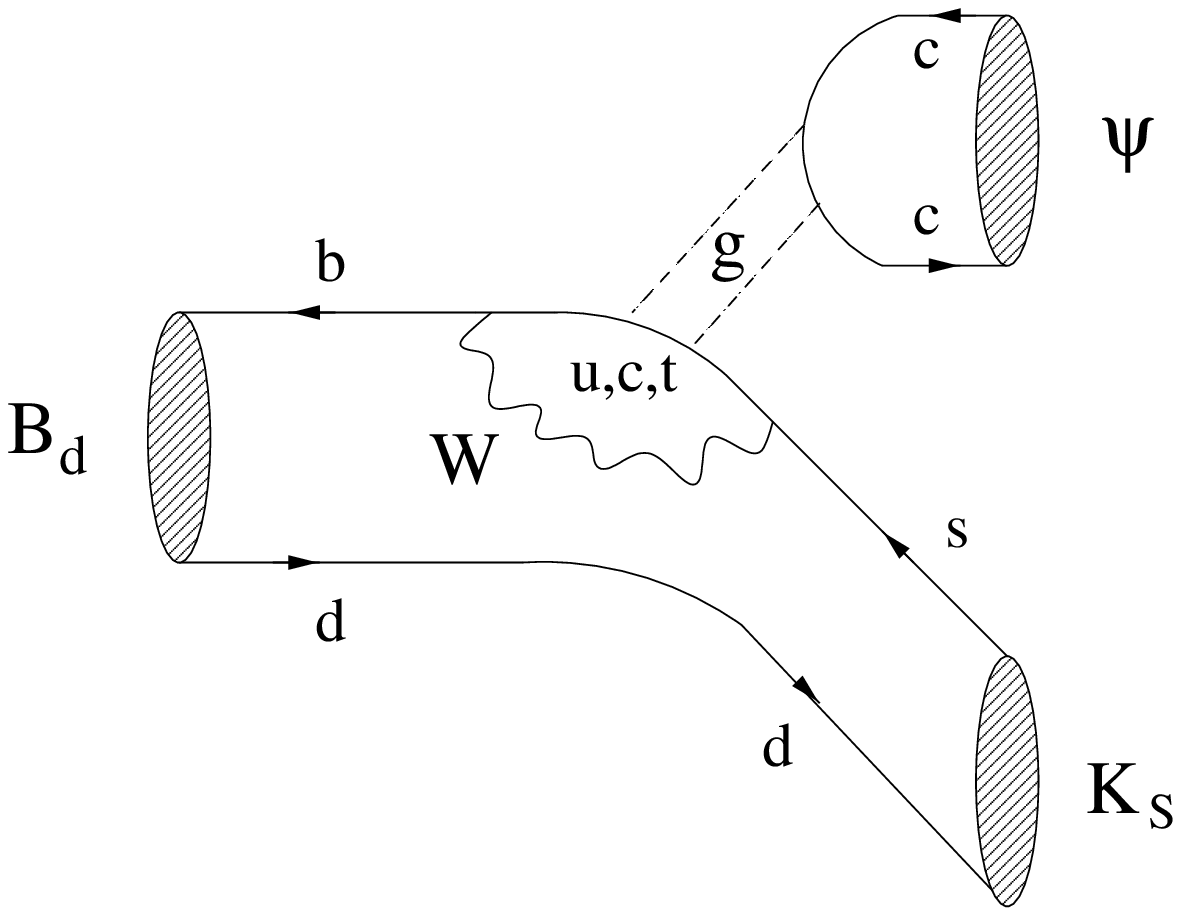}}
\caption{``Tree" (left) and ``Penguin" (right) contributions to $B\to \psi
K_S$ (from Ref.~\protect\citex{rf}).}
\label{fig:BpsiK}
\end{figure}

There are ``tree" and ``penguin" contributions to $B\to \psi K_{S,L}$ as shown
in Fig.~\ref{fig:BpsiK}.  The tree diagram arises from $b\to c\bar c s$
transition, while there are penguin contributions with three different
combinations of CKM elements,
\beq
\ov A_T = V_{cb} V_{cs}^*\, T_{c\bar cs}\,, \qquad
  \ov A_P = V_{tb} V_{ts}^*\, P_t + V_{cb} V_{cs}^*\, P_c 
  + V_{ub} V_{us}^*\, P_u\,.
\eeq
We can rewrite the penguin amplitude using $V_{tb} V_{ts}^* + V_{cb} V_{cs}^*
+ V_{ub} V_{us}^* =0$ to obtain
\beqa\label{BpsiKamp}
\ov A &=& V_{cb} V_{cs}^*\, (T_{c\bar cs} + P_c - P_t)
  + V_{ub} V_{us}^*\, (P_u - P_t) \nn\\
&\equiv& V_{cb} V_{cs}^*\, T + V_{ub} V_{us}^*\, P\,,
\eeqa
where the second line defines $T$ and $P$.  We expect  $|\ov{A}/A|-1 <
10^{-2}$, because $|(V_{ub} V_{us}^*) / (V_{cb} V_{cs}^*)| \simeq 1/50$ and
model dependent estimates of $|P/T|$ are well below unity.  So the amplitude
with weak phase $V_{cb} V_{cs}^*$ dominates.  The $CP$ asymmetry
measures
\beq\label{BpsiKlam}
\lambda_{\psi K_{S,L}} 
= \mp \bigg( { V_{tb}^* V_{td} \over V_{tb} V_{td}^*} \bigg)
  \bigg( {V_{cb} V_{cs}^* \over V_{cb}^* V_{cs}} \bigg)
  \bigg( {V_{cs} V_{cd}^* \over V_{cs}^* V_{cd}} \bigg) 
= \mp e^{-2i\beta} \,,
\eeq
and so ${\rm Im} \lambda_{\psi K_{S,L}} = \pm \sin2\beta$.  The first term is
the SM value of $q/p$ in $B_d$ mixing, the second is $\ov A/A$, and the last
one is $p/q$ in the $K^0$ system.  In the absence of $K^0 - \K0bar$ mixing
there could be no interference between $\B0bar\to \psi \K0bar$ and $B^0\to \psi
K^0$.

The first evidence for $CP$ violation outside the kaon sector was the recent
BABAR and BELLE measurements~\cite{babelle} of $a_{\psi K}$, whose average,
$\sin2\beta = 0.731 \pm 0.055$, completely dominates the world
average~\cite{yossi_ams} already, $\sin2\beta = 0.734 \pm 0.054$.

\subsubsection[$\sin2\beta$ from $B\to \phi K_S$]{\boldmath $\sin2\beta$ from
$B\to \phi K_{S,L}$} \label{sec:phiK}

\begin{figure}[t]
\centerline{\includegraphics*[height=.2\textwidth]{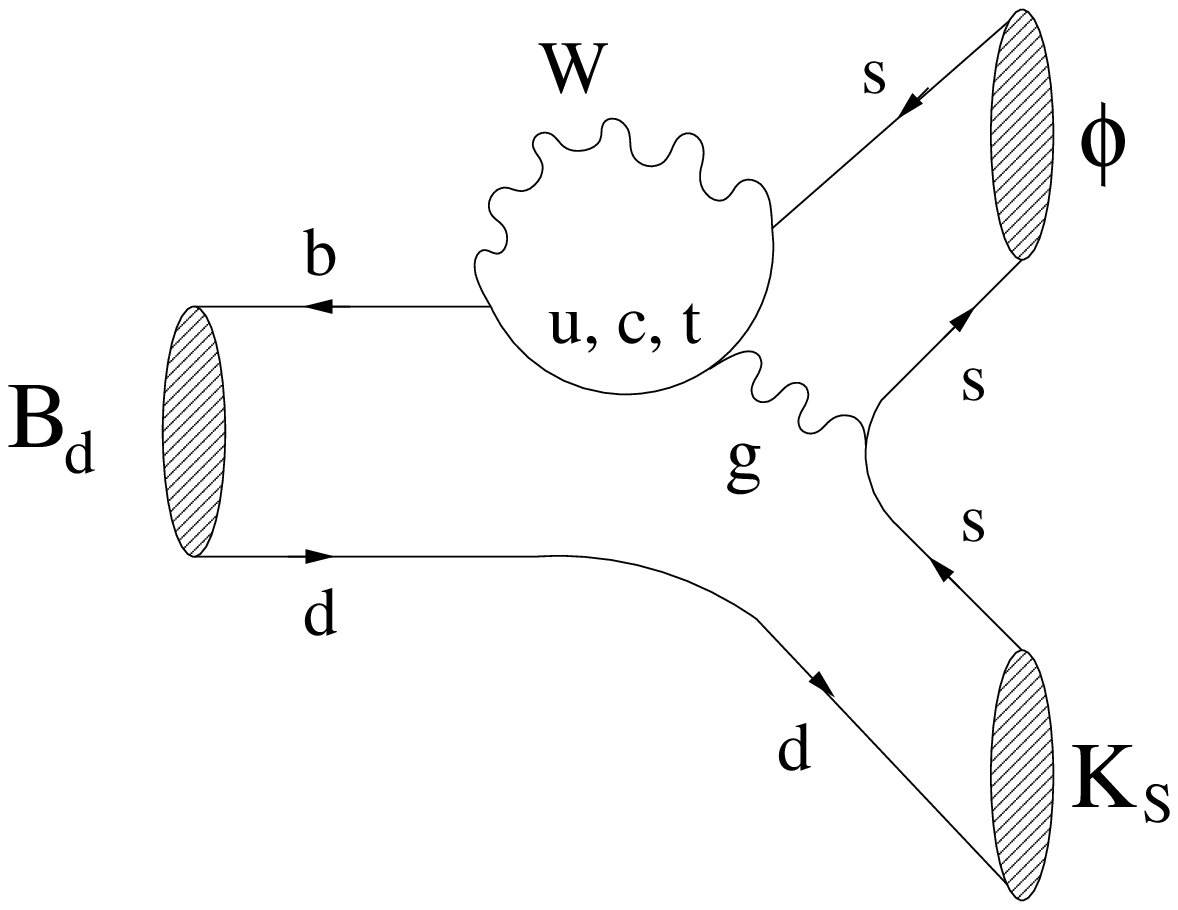}\hspace{1cm}
\includegraphics*[height=.2\textwidth]{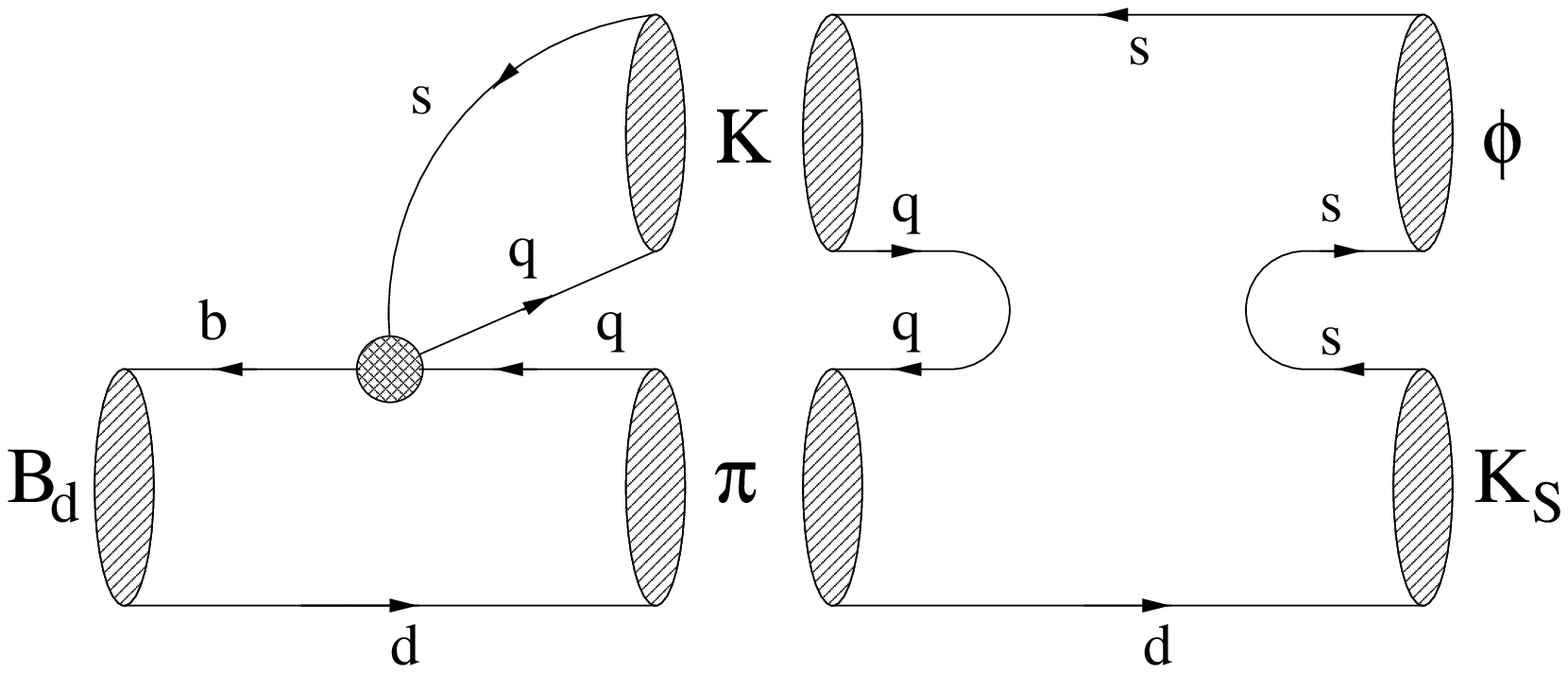}}
\caption{``Penguin" (left) and ``Tree" (right) contributions to $B\to \phi
K_S$ (from Ref.~\protect\citex{rf}).}
\label{fig:BphiK}
\end{figure}

The $CP$ violation in this channel is believed to be a very sensitive probe of
new physics.  Naively, tree contributions to $b\to s\bar s s$ transition are
absent, and the penguin contributions (see Fig.~\ref{fig:BphiK}) are
\beq
\ov A_P = V_{cb} V_{cs}^*\, (P_c - P_t) + V_{ub} V_{us}^*\, (P_u - P_t)\,.
\eeq
Due to $|(V_{ub} V_{us}^*)/(V_{cb} V_{cs}^*)| \sim {\cal O}(\lambda^2)$ and
because we expect $|P_c - P_t| \sim |P_u - P_t|$, the $B\to \phi K_S$ amplitude
is also dominated by a single weak phase, $V_{cb} V_{cs}^*$.  Therefore,
$|\ov{A}/A| - 1$ is small, although not as small as in $B\to \psi K_{S,L}$.  
There is also a ``tree" contribution to $B\to \phi K_S$, from $b\to u\bar u s$
decay followed by $u\bar u \to s\bar s$ rescattering, shown in
Fig.~\ref{fig:BphiK} on the right.  This amplitude is also proportional to the
suppressed CKM combination, $V_{ub} V_{us}^*$, and it is not even clear how to
separate it from ``penguin" terms.  Unless rescattering provides an
enhancement, this should not upset the proximity of ${\rm Im} \lambda_{\phi
K_S}$ from $\sin2\beta$.  Thus we expect ${\rm Im} \lambda_{\phi K_S} =
\sin2\beta + {\cal O}(\lambda^2)$ in the SM.

At present ${\rm Im} \lambda_{\phi K} = {\rm Im} \lambda_{\psi K}$ is violated
at the $2.7\sigma$ level~\cite{babarphik,bellephik}.  This is interesting
because new physics could enter $\lambda_{\psi K}$ mainly through $q/p$,
whereas it could modify $\lambda_{\phi K}$ through both $q/p$ and $\ov A/A$. 
Note, however, that in the $\eta' K_S$ and $K^+ K^- K_S$ channels there is no
similarly large deviation from $\sin2\beta$~\cite{bellephik}.  The $CP$
asymmetries in $b\to s\bar s s$ modes remain some of the best examples that
measuring the same angle in several decays sensitive to different short
distance physics is one of the most promising ways to look for new physics. 
This will be very interesting as the errors decrease.

\subsection{Summary}

\begin{itemize}

\item Want experimentally precise and theoretically reliable measurements that
in the SM relate to CKM elements, but can probe different short distance
physics.

\item The CKM picture passed its first real test; we can no longer claim to
look for alternatives, but to seek corrections due to new physics (except maybe
$B_s$ mixing).

\item Very broad program --- a lot more interesting as a whole than any single
measurement alone;~ redundancy/correlations may be the key to finding new
physics.

\item $B_{d,s}$ mixing $(|V_{td}/V_{ts}|)$ and $B\to \psi K$ $(\sin2\beta)$ are
``easy", i.e., both theory and experiment are under control; in the next
lectures start looking at harder things.

\end{itemize}

\section{Heavy Quark Limit: Spectroscopy, Exclusive and Inclusive Decays}

Over the last decade, most of the theoretical progress in understanding $B$
decays utilized the fact that $m_b$ is much larger than $\lqcd$.  Semileptonic
and rare decays allow measurements of CKM elements important for testing the
SM, and are sensitive to new physics.  Improving the accuracy of the
theoretical predictions increases the sensitivity to new physics.  For example,
as can be seen from Fig.~\ref{fig:ckmfit}, $|V_{ub}|$ is the dominant
uncertainty of the side of the unitarity triangle opposite to the angle
$\beta$.  The constraint from the $K^0 - \K0bar$ mixing parameter $\epsilon_K$
is proportional to $|V_{cb}|^4$, and so is the constraint from the $K^+\to
\pi^+ \nu \bar\nu$ rate.  (The ratio of the $K_L\to \pi^0 \nu \bar\nu$ and
$K^+\to \pi^+ \nu \bar\nu$ rates is much less sensitive to $|V_{cb}|$.)  Most
examples in this lecture are related to the determination of $|V_{cb}|$ and
$|V_{ub}|$ from exclusive and inclusive semileptonic decays.  The same
theoretical tools are directly applicable to reducing the hadronic
uncertainties in rare decays mediated by flavor changing neutral currents as
well.

To believe at some point in the future that a discrepancy is a sign of new
physics, model independent predictions are essential.  Results which depend on
modeling nonperturbative strong interaction effects cannot disprove the
Standard Model.  Most model independent predictions are of the form,
\beq
\mbox{Quantity of interest} = (\mbox{calculable factors}) 
  \times \bigg[ 1 + \sum_k\, (\mbox{small parameters})^k \bigg] ,
\eeq
where the small parameter can be $\lqcd/m_b$, $m_s/\Lambda_{\chi SB}$,
$\alpha_s(m_b)$, etc.  For the purposes of these lectures we mean by (strong
interaction) model independent that the theoretical uncertainty is suppressed
by small parameters [so that theorists argue about ${\cal O}(1)\times$(small
numbers) instead of ${\cal O}(1)$ effects].  Still, in most cases, there are
theoretical uncertainties suppressed by some $(\mbox{small parameter})^N$,
which cannot be estimated model independently.  If the goal is to test the
Standard Model, one must assign sizable uncertainties to such ``small" 
corrections not known from first principles.

Throughout the following it should be kept in mind that the behavior of
expansions that are formally in powers of $\lqcd/m_b$ can be rather different
in practice.  (By $\lqcd$ we mean hereafter a generic hadronic scale, and not
necessarily the parameter in the running of $\alpha_s$.)  Depending on the
process under consideration, the physical scale that determines the behavior of
expansions may or may not be much smaller than $m_b$ (and, especially, $m_c$). 
For example, $f_\pi$, $m_\rho$, and $m_K^2/m_s$ are all of order $\lqcd$
formally, but their numerical values span an order of magnitude.  As it will
become clear below, in most cases experimental guidance is needed to decide how
well the theory works.

\subsection{Heavy quark symmetry and HQET}

In hadrons composed of heavy quarks, the dynamics of QCD simplifies.  Mesons
containing a heavy quark -- heavy antiquark pair, $Q\ov Q$, form
positronium-type bound states, which become perturbative in $m_Q \gg \lqcd$
limit~\cite{AP}.  In heavy mesons composed of a heavy quark, $Q$, and a light
antiquark, $\bar q$ (and gluons and $q\bar q$ pairs), there are also
simplifications in the $m_Q \gg \lqcd$ limit.  The heavy quark acts as a static
color source with fixed four-velocity, $v^\mu$, and the wave function of the
light degrees of freedom (the so-called brown muck) become insensitive to the
spin and mass (flavor) of the heavy quark, resulting in heavy quark spin-flavor
symmetries~\cite{HQS}.

The physical picture to understand these symmetries is similar to those
well-known from atomic physics, where simplifications occur due to the fact
that the electron mass, $m_e$, is much smaller than the nucleon mass, $m_N$. 
The analog of flavor symmetry is that isotopes have similar chemistry, because
the electrons' wave functions become independent of $m_N$ in the $m_N \gg m_e$
limit.  The analog of spin symmetry  is that hyperfine levels are almost
degenerate, because the interaction of the electron and nucleon spin diminishes
in the $m_N \gg m_e$ limit.

The theoretical framework to analyze the consequences of heavy quark symmetry
and the corrections to the symmetry limit is the heavy quark effective theory
(HQET).  One can do a field redefinition to introduce a new field, $h_v(x)$,
which annihilates a heavy quark with four-velocity $v$, and has no dependence
on the large mass of the heavy quark~\cite{Geor},
\beq\label{hqfield}
  h_v^{(Q)}(x) = e^{i m_Q v\cdot x}\, \frac{1+ v\!\!\!\slash}{2}\, Q(x) \,,
\label{field}
\eeq
where $Q(x)$ denotes the quark field in full QCD.  It is convenient to label
heavy quark fields by $v$, because $v$ cannot be changed by soft interactions. 
The physical interpretation of the projection operator $(1+ v\!\!\!\slash)/2$
is that $h_v^{(Q)}$ represents just the heavy quark (rather than antiquark)
components of $Q$.  If $p$ is the total momentum of the heavy quark, then the
field $h_v^{(Q)}$ carries the residual momentum $k = p - m_Q v \sim {\cal
O}(\lqcd)$. In terms of these fields the QCD Lagrangian simplifies
tremendously,
\beq\label{Lag}
{\cal L} = \bar h_v^{(Q)}\, i v\!\cdot\!D\, h_v^{(Q)}
  + {\cal O} \bigg({1\over m_Q}\bigg) ,
\eeq
where $D^\mu = \partial^\mu - i g_s T_a A_a^\mu$ is the covariant derivative. 
The fact that there is no Dirac matrix in this Lagrangian implies that both
the heavy quark's propagator and its coupling to gluons become independent of
the heavy quark spin.  The effective theory provides a well-defined framework
to calculate perturbative ${\cal O}(\alpha_s)$ and parameterize nonperturbative
${\cal O}(\lqcd/m_Q)$ corrections.

\subsubsection{Spectroscopy}

The spectroscopy of heavy hadrons simplifies due to heavy quark symmetry
because the spin of the heavy quark becomes a good quantum number in
$m_Q\to\infty$ limit; i.e., it becomes a conserved quantity in the interactions
with the brown muck, $[\vec s_Q, {\cal H}] = 0$.  Of course, the total angular
momentum is conserved, $[\vec J, {\cal H}] = 0$, and therefore the spin of the
light degrees of freedom, $\vec s_l = \vec J - \vec s_Q$, also becomes
conserved in the heavy quark limit, $[\vec s_l, {\cal H}] = 0$.

This implies that hadrons containing a single heavy quark can be labeled with
$s_l$, and for any value of $s_l$ there are two (almost) degenerate states with
total angular momentum $J_\pm = s_l \pm \frac12$.  (An exception occurs for
baryons with $s_l = 0$, where there is only a single state with $J =
\frac12$.)  The ground state mesons with $Q\bar q$ flavor quantum numbers
contain light degrees of freedom with spin-parity $s_l^{\pi_l} = \frac12^-$,
giving a doublet containing a spin zero and spin one meson.  For $Q=c$ these
mesons are the $D$ and $D^*$, while $Q=b$ gives the $B$ and $B^*$ mesons.

\begin{figure}[t]
\centerline{\includegraphics*[width=.5\textwidth]{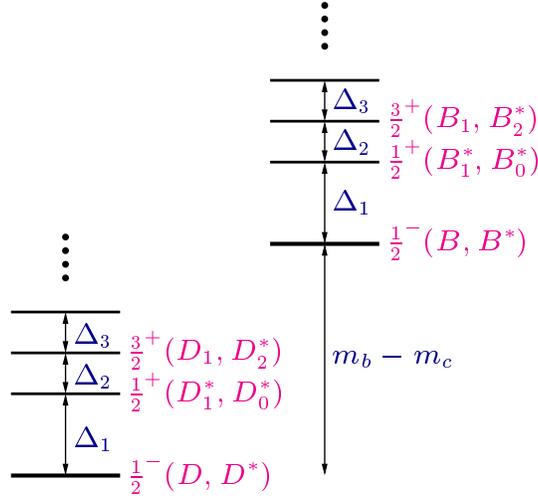}}
\caption{Spectroscopy of $B$ and $D$ mesons.  For each doublet level the
spin-parity of the brown muck, $s_l^{\pi_l}$, and the names of the physical
states are indicated.}
\label{fig:spectra}
\end{figure}

The mass splittings between the doublets, $\Delta_i$, are of order $\lqcd$, and
are the same in the $B$ and $D$ sectors at leading order in $\lqcd/m_Q$, as
shown in Fig.~\ref{fig:spectra}.  The mass splittings within each doublet are
of order $\lqcd^2/m_Q$.  This is supported by experimental data: for example,
for the $s_l^{\pi_l} = \frac12^-$ ground state doublets $m_{D^*}-m_D \approx
140\,$MeV while $m_{B^*}-m_B \approx 45\,$MeV, and their ratio, $0.32$, is
consistent with $m_c/m_b$.

As an aside, I cannot resist mentioning a well-known puzzle.  Since the ground
state vector-pseudoscalar mass splitting is proportional to $\lqcd^2/m_Q$, we
expect $m_V^2-m_P^2$ to be approximately constant.  This argument relies on
$m_Q \gg \lqcd$.  The data are
\beq
\begin{array}{rclrcl}
m_{B^*}^2 - m_B^2 &=& 0.49\,\GeV^2\,,\qquad  &
  m_{B_s^*}^2 - m_{B_s}^2 &=& 0.50\,\GeV^2\,, \\
m_{D^*}^2 - m_D^2 &=& 0.54\,\GeV^2\,,  &
  m_{D_s^*}^2 - m_{D_s}^2 &=& 0.58\,\GeV^2\,, \\
m_\rho^2 - m_\pi^2 &=& 0.57\,\GeV^2\,,  &
  m_{K^*}^2 - m_K^2 &=& 0.55\,\GeV^2\,.
\end{array}
\eeq
It is not understood why the light meson mass splittings satisfy the same
relation (although this would be expected in the nonrelativistic constituent
quark model).  There must be something more going on than just heavy quark
symmetry, and if this was the only prediction of heavy quark symmetry then we
could not say that there is strong evidence that it is a useful idea.

\subsubsection{\boldmath Strong decays of excited charmed mesons}

Heavy quark symmetry has implication for the strong decays of heavy mesons as
well, because the strong interaction Hamiltonian conserves the spin of the
heavy quark and the light degrees of freedom separately.  

\begin{figure}[t]
\centerline{\includegraphics*[width=0.5\textwidth]{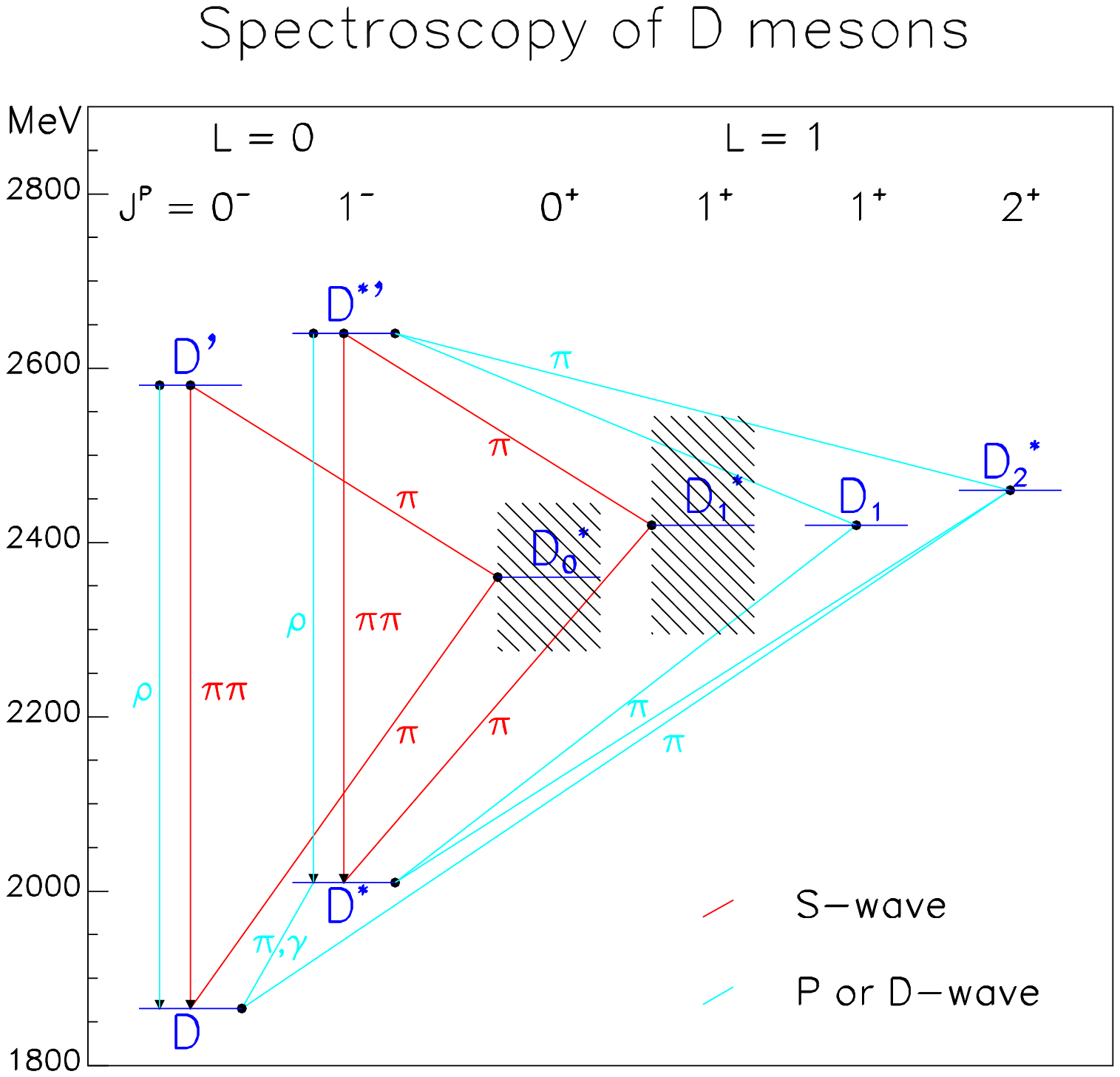}}
\caption{Spectroscopy and strong decays of $D$ mesons 
(from Ref.~\protect\citex{cleobroad}).}
\label{fig:Dspectra}
\end{figure}

Excited charmed mesons with $s_l^{\pi_l} = \frac32^+$ have been observed. 
These are the $D_1$ and $D_2^*$ mesons with spin one and two, respectively. 
They are quite narrow with widths around $20\,$MeV.  This is because their
decays to $D^{(*)}\pi$ are in $D$-waves.  An $S$-wave $D_1\to D^*\pi$ amplitude
is allowed by total angular momentum conservation, but forbidden in the
$m_Q\to\infty$ limit by heavy quark spin symmetry~\cite{IWprl}.  Members of the
$s_l^{\pi_l} = \frac12^+$ doublet, $D_0^*$ and $D_1^*$, can decay to $D\pi$ and
$D^*\pi$ in $S$-waves, and therefore these states are expected to be broad. 
The $D_1^*$ has been observed~\cite{cleobroad} with a width around $290 \pm
110\,$MeV.\footnote{In the nonrelativistic constituent quark model the
$s_l^{\pi_l}=\frac12^+$ and $\frac32^+$ doublets are $L=1$ orbital excitations
(sometimes collectively called $D^{**}$), and the two doublets arise from
combining the orbital angular momentum with the spin of the light antiquark. 
In the quark model the mass splittings of orbitally excited states vanish as
they come from $\langle \vec s_Q \cdot \vec s_{\bar q}\, \delta^3(\vec r\,)
\rangle$ interaction.  This is supported by the data: $m_{D_2^*} - m_{D_1} =
37\,{\rm MeV} \ll m_{D^*} - m_D$.}  The various allowed decays are shown in
Fig.~\ref{fig:Dspectra}.

It is possible to make more detailed predictions for the $(D_1, D_2^*) \to (D,
D^*) \pi$ decays, since the four amplitudes are related by spin symmetry.  The
ratios of rates are determined by Clebsch-Gordan coefficients, which are
convenient to write in terms of $6j$ symbols,
\beq\label{6j}
\Gamma(J\to J' \pi) \propto (2s_l+1)(2J'+1) \left| \left\{
  \matrix{L & s_l' & s_l \cr \frac12 & J & J'} \right\} \right|^2 ,
\eeq
given in the upper row in Table~\ref{tab:Dwidths}.  Since these decays are in
$L=2$ partial waves, the phase space depends on the pion momentum as
$|p_\pi|^5$ (one can check using Eq.~(\ref{6j}) that the $S$-wave $D_1\to
D^*\pi$ rate indeed vanishes).  This is a large but calculable heavy quark
symmetry breaking, which is included in the bottom line of
Table~\ref{tab:Dwidths}.  It changes the prediction for $\Gamma(D_2^*\to D\pi)
/ \Gamma(D_2^*\to D^*\pi)$ from $2/3$ to $2.5$; the latter agrees well with the
data, $2.3\pm0.6$.

\begin{table}[t] \begin{center}
\begin{tabular}{ccccccc} \hline\hline
$\Gamma(D_1\to D\pi)$ & : & $\Gamma(D_1\to D^*\pi)$ & : & 
  $\Gamma(D_2^*\to D\pi)$ & : & $\Gamma(D_2^*\to D^*\pi)$ \\ \hline\hline
$0$ & : & $1$ & : & $2/5$ & : & $3/5$ \\ \hline
$0$ & : & $1$ & : & $2.3$ & : & $0.92$ \\ \hline\hline
\end{tabular}
\end{center}
\caption{Ratio of $(D_1, D_2^*) \to (D,D^*) \pi$ decay rates without (upper
row) and with (lower row) corrections due to phase space differences (from
Ref.~\protect\citex{book1}).}
\label{tab:Dwidths}
\end{table}

The ratio of the $D_1$ and $D_2^*$ widths works less well: the prediction
$1/(2.3+0.9) \simeq 0.3$ is much smaller than the data, $\Gamma(D_1^0) /
\Gamma(D_2^{*0}) \simeq 0.7$.  The simplest explanation would be that $D_1$
mixes with the broad $D_1^*$, due to ${\cal O}(\lqcd/m_c)$ spin symmetry
violating effects; however, there is no indication of an $S$-wave component in
the $D_1 \to D^*\pi$ angular distribution.  The larger than expected $D_1$
width can be explained with other spin symmetry violating effects~\cite{FM}.
This is important because otherwise it would indicate that we cannot trust the
treatment of the charm quark as heavy in other contexts.

\subsection[Exclusive semileptonic $B$ decays]{\boldmath Exclusive semileptonic
$B$ decays}

Semileptonic and radiative rare decays can be used to determine CKM elements,
such as $|V_{cb}|$ and $|V_{ub}|$, and are sensitive probes of new physics. 
The difficulty is that the hadronic matrix elements that connect exclusive
decay rates to short distance weak interaction parameters are not accessible in
general theoretically.  Important exceptions occur in certain situations due to
enhanced symmetries, when some form factors are model independently related to
one another, and in the case of $B\to D^*$ decay even the rate is determined at
one point in phase space.

\subsubsection[$B\to D^{(*)}\ell\bar\nu$ decay and $|V_{cb}|$]{\boldmath $B\to
D^{(*)}\ell\bar\nu$ decay and $|V_{cb}|$}

Heavy quark symmetry is very predictive for $B\to D^{(*)}$ semileptonic form
factors.  In the $m_{b,c} \gg \lqcd$ limit, the configuration of the brown muck
only depends on the four-velocity of the heavy quark, but not on its mass and
spin.  In the decay of the $b$ quark, the weak current changes suddenly (on a
time scale $\ll \lqcd^{-1}$) the flavor $b\to c$, the momentum $\vec p_b \to
\vec p_c$, and possibly flips the spin, $\vec s_b\to \vec s_c$.  In the
$m_{b,c} \gg \lqcd$ limit, because of heavy quark symmetry, the brown muck only
feels that the four-velocity of the static color source in the center of the
heavy meson changed, $v_b \to v_c$.  Therefore, the form factors that describe
the wave function overlap between the initial and final mesons become
independent of Dirac structure of weak current, and can only depend on a scalar
quantity, $w \equiv v_b \cdot v_c$.  Thus all form factors are related to a
single universal function, $\xi(v_b\cdot v_c)$, the Isgur-Wise function, which
contains all the low energy nonperturbative hadronic physics relevant for these
decays.  Moreover, $\xi(1)=1$, because at the ``zero recoil" point, $w=1$,
where the $c$ quark is at rest in the $b$ rest frame, the configuration of the
brown muck does not change at all.

Using only Lorentz invariance, six form factors parameterize $B\to D^{(*)} \ell
\bar\nu$ decay,
\beqa\label{BDffdef}
\bra{D(v')}V_\nu\ket{B(v)} &=& \sqrt{m_Bm_D}\, \Big[ h_+\,(v+v')_\nu
  + h_-\, (v-v')_\nu \Big] , \nn\\
\bra{D^*(v')}V_\nu\ket{B(v)} &=& i \sqrt{m_Bm_{D^*}}\, h_V\,
  \epsilon_{\nu\alpha\beta\gamma}\epsilon^{*\alpha}v'^\beta v^\gamma , \nn\\
\bra{D(v')}A_\nu\ket{B(v)} &=& 0 , \\
\bra{D^*(v')}A_\nu\ket{B(v)} &=& \sqrt{m_Bm_{D^*}}\,
  \Big[h_{A_1}\,(w+1)\epsilon^*_\nu - h_{A_2}\,(\epsilon^*\cdot v)v_\nu
  - h_{A_3}\,(\epsilon^*\cdot v)v'_\nu\Big] , \nn
\eeqa
where the $h_i$ are functions of $w \equiv v\cdot v' = (m_B^2 + m_{D^{(*)}}^2 -
q^2)/(2m_B m_{D^{(*)}})$.  The currents relevant for semileptonic decay are
$V_\nu = \bar c\gamma_\nu b$ and $A_\nu = \bar c\gamma_\nu\gamma_5 b$.  In the
$m_Q\to\infty$ limit,
\beq
h_+(w) = h_V(w) = h_{A_1}(w) = h_{A_3}(w) = \xi(w)\,, \qquad
  h_-(w) = h_{A_2}(w) = 0\,.
\eeq
There are corrections to these relations for finite $m_{c,b}$, suppressed by
powers of $\alpha_s$ and $\lqcd/m_{c,b}$.  The former are calculable, while the
latter can only be parameterized, and that is where model dependence enters.

The determination of $|V_{cb}|$ from exclusive $B\to D^{(*)} \ell \bar\nu$
decay uses an extrapolation of the measured decay rate to zero recoil, $w=1$.
The rates can be schematically written as
\begin{equation}\label{rates}
{\d\Gamma(B\to D^{(*)} \ell\bar\nu)\over \d w} = (\mbox{known factors})\,
  |V_{cb}|^2 \cases{\! (w^2-1)^{1/2}\, \FDs^2(w)\,,\quad & for $B\to D^*$, \cr
  \!(w^2-1)^{3/2}\, \FD^2(w)\,, & for $B\to D$. \cr}
\end{equation}
Both $\FD(w)$ and $\FDs(w)$ are equal to the Isgur-Wise function in the $m_Q
\to\infty$ limit, and in particular $\FDt(1) = 1$, allowing for a model
independent determination of $|V_{cb}|$.  The corrections are again suppressed
by powers of $\alpha_s$ and $\lqcd/m_{c,b}$ and are of the form
\beqa\label{F1}
\FDs(1) &=& 1_{\mbox{\footnotesize (Isgur-Wise)}} + c_A(\alpha_s) 
  + {0_{\mbox{\footnotesize (Luke)}}\over m_{c,b}} 
  + {(\mbox{lattice or models})\over m_{c,b}^2} + \ldots \,, \nn\\
\FD(1) &=& 1_{\mbox{\footnotesize (Isgur-Wise)}} + c_V(\alpha_s) 
  + {(\mbox{lattice or models})\over m_{c,b}} + \ldots \,.
\eeqa
The perturbative corrections, $c_A = -0.04$ and $c_V = 0.02$, have been
computed to order $\alpha_s^2$~\citex{Czar}, and the yet higher order
corrections should be below the $1\%$ level.  The order $\lqcd/m_Q$ correction
to $\FDs(1)$ vanishes due to Luke's theorem~\cite{Luke}.  The terms indicated
by $(\mbox{lattice or models})$ in Eqs.~(\ref{F1}) are only known using
phenomenological models or quenched lattice QCD at present.  This is why the
determination of $|V_{cb}|$ from $B\to D^* \ell\bar\nu$ is theoretically more
reliable than that from $B\to D\ell\bar\nu$, although both QCD sum
rules~\cite{LNN} and quenched lattice QCD~\cite{latticeD} suggest that the
order $\lqcd/m_{c,b}$ correction to $\FD(1)$ is small (giving $\FD(1) = 1.02
\pm 0.08$ and $1.06 \pm 0.02$, respectively).  Due to the extra $w^2-1$
helicity suppression near zero recoil, $B\to D\ell\bar\nu$ is also harder
experimentally than $B\to D^*\ell\bar\nu$.  Reasonable estimates of $\FDs(1)$
are around
\beq
\FDs(1) = 0.91 \pm 0.04\,.
\eeq
This value is unchanged for over five years~\cite{babook}, and is supported by
a recent lattice result~\cite{latticeDs}.  The zero recoil limit of the
$B\to D^* \ell \bar\nu$ rate is measured to be~\cite{pdg}
\beq
|V_{cb}|\, \FDs(1) = (38.3 \pm 1.0) \times 10^{-3} \,,
\eeq
yielding $|V_{cb}| = (42.1 \pm 1.1_{\rm exp} \pm 1.9_{\rm th}) \times 10^{-3}$.

Another important theoretical input is the shape of $\FDt(w)$ used to fit the
data.  It is useful to expand about zero recoil and write $\FDt(w) = \FDt(1)\,
[1 - \rho_{(*)}^2 (w-1) + c_{(*)} (w-1)^2 + \ldots]$.  Analyticity imposes
stringent constraints between the slope, $\rho^2$, and curvature, $c$, at zero
recoil~\cite{BGL}, which is used in the experimental fits to the data. 
Measuring the $B\to D\ell\bar\nu$ rate is also important, because computing
$\FD(1)$ on the lattice is not harder than $\FDs(1)$.  Other cross-checks will
come from ratios of the form factors in $B\to D^*\ell\bar\nu$, and comparing
the shapes of the $B\to D^*$ and $B\to D$ spectra~\cite{BGZL}.  These can give
additional constraints on $\rho^2$, which is important because the correlation
between $\rho^2$ and the extracted value of $|V_{cb}|\, {\cal F}_*(1)$ is very
large.

\subsubsection[$B\to \mbox{light}$ form factors and SCET]{\boldmath $B\to
\mbox{light}$ form factors and SCET}
\label{sec:SCET}

In $B$ decays to light mesons, there is a much more  limited use of heavy quark
symmetry, since it does not apply for the final state.  One can still derive
relations between the $B\to \rho\ell\bar\nu$, $K^*\ell^+\ell^-$, and
$K^*\gamma$ form factors in the large $q^2$ region~\cite{IsWi}.  One can also
relate the form factors that occur in $B$ and $D$ decays to one another.  But
the symmetry neither reduces the number of form factors, nor does it determine
their normalization at any value of $q^2$.  For example, it is possible to
predict $B\to \rho\ell\bar\nu$ from the measured $D\to K^*\ell\bar\nu$ form
factors, using the symmetries:

\beq
\matrix{ &  \Bbar  &  \stackrel{ \bar u \Gamma b\, V_{ub}}
  {-\!\!\!-\!\!\!-\!\!\!\longrightarrow}  &  \rho\,\ell\bar\nu & \cr
\stackrel{\rm flavor}{\ds _{SU(2)}}\!\! &
  \updownarrow & & \updownarrow  & 
  \!\!\!\!\stackrel{\rm chiral}{\ds _{SU(3)}} \cr
& D  & \stackrel{\bar d\Gamma c\, V_{cs}}
  {-\!\!\!-\!\!\!-\!\!\!\longrightarrow}  &  K^*\ell\bar\nu & }
\eeq
The form factor relations hold at fixed value of $v\cdot v'$, that is, at the
same energy of the light mesons in the heavy meson rest frame.  The validity of
these relations is also limited to order one values of $v\cdot v'$.  (While
maximal recoil in $B\to D^*$ and $B\to D$ decays are $v\cdot v' \simeq 1.5$ and
$1.6$, respectively, it is $3.5$ in $B\to \rho$ and $18.9$ in $B\to \pi$.)  A
limitation of this approach is that corrections to both heavy quark symmetry
and chiral symmetry could be $\sim 20\%$ or more each.  It may ultimately be
possible to eliminate all first order symmetry breaking corrections~\cite{lw}
forming a ``Grinstein type double ratio"~\cite{Gtdr} of the form factors that
occur in the four decays $(B,D) \to (\rho,K^*)$, but this method will require
very large data sets.  The same region of phase space (large $q^2$ and modest
light meson energy) is also the most accessible to lattice QCD calculations.

\begin{figure}[t]
\centerline{\includegraphics*[width=.33\textwidth]{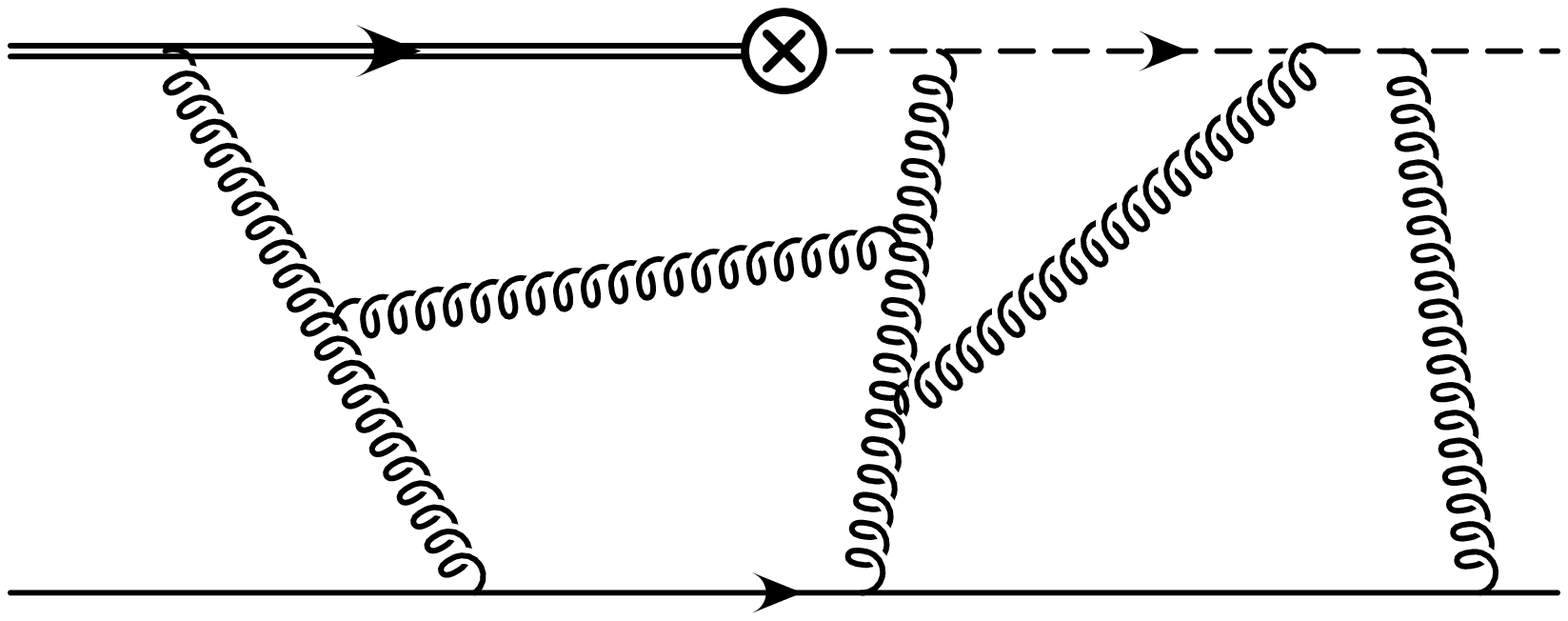}
\hspace*{1.5cm}
\includegraphics*[width=.33\textwidth]{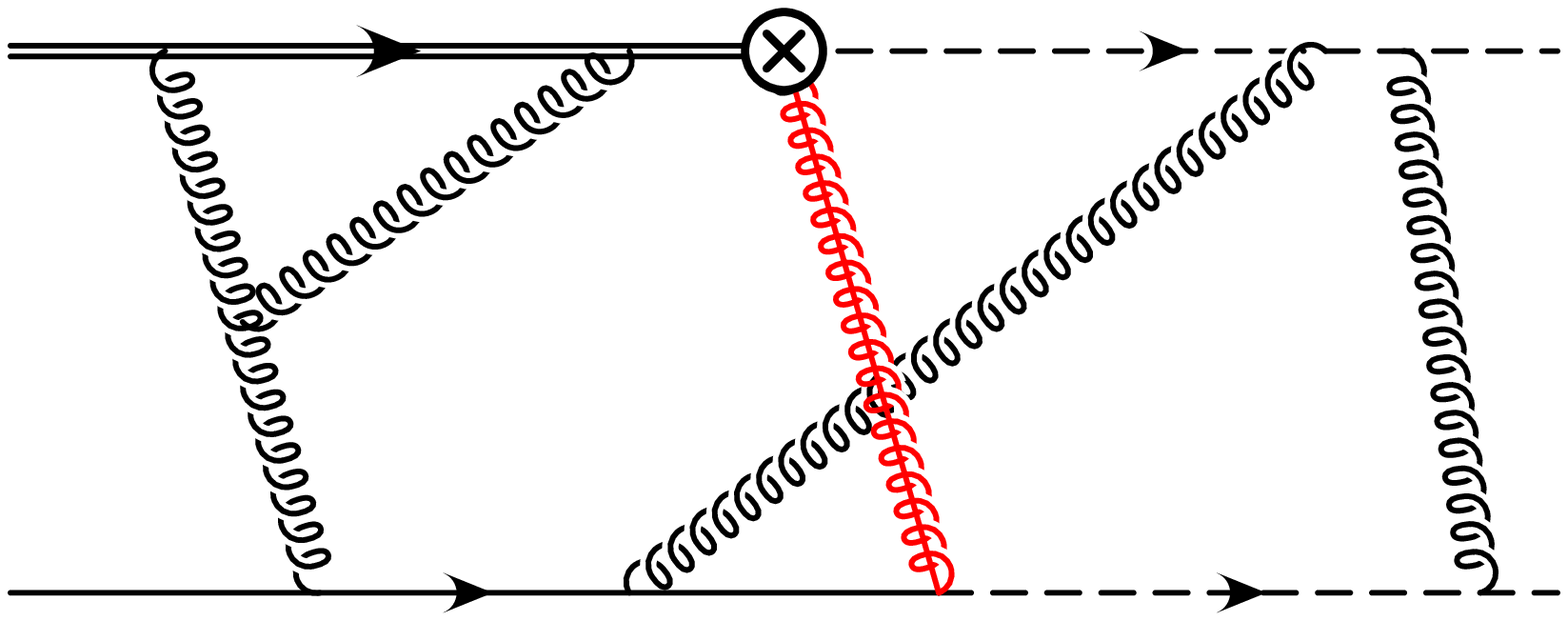}}
\caption{Contributions to heavy-to-light form factors: ``soft" nonfactorizable
part (left), and ``hard" factorizable part (right).  Note that these pictures
are somewhat misleading, as explained in the text.  (From
Ref.~\protect\citex{iain}.)}
\label{fig:scet}
\end{figure}

There have been important recent developments toward a better understanding of
these form factors in the $q^2 \ll m_B^2$ region.  It was proposed some time
ago that in the heavy mass limit heavy-to-light semileptonic form factors
become calculable in perturbative QCD~\cite{Szczepaniak}.  There were several
problems justifying such a proposal; for example, diagrams of the type in
Fig.~\ref{fig:scet} can give contributions proportional to $1/x^2$ leading to
singular integrals ($x$ is the momentum fraction of one of the quarks).  There
have been many attempts to separate ``soft" and ``hard" contributions and
understand how Sudakov effects might regulate the singularities~\cite{hmmm}.

It was recently proposed~\cite{charles} that the 7 form factors that
parameterize matrix elements of all possible currents ($V$, $A$, $S$, $P$, $T$)
in $B\to$ vector meson ($\rho$ or $K^*$) transitions have extra symmetries and
can be expressed in terms of two functions, $\xi_\perp(E)$ and $\xi_\|(E)$, in
the limit where $m_b\to\infty$ and $E_{\rho,K^*} = {\cal O}(m_b)$.  In the same
limit, the 3 form factors that parameterize decays to pseudoscalars ($\pi$ or
$K$) are related to one function, $\xi_P(E)$.  Loosely speaking, these
relations were expected to arise because soft gluons cannot flip the helicity
of the energetic light quark emerging from the weak decay.  

A new effective field theory, the soft-collinear effective theory
(SCET)~\cite{SCET0,SCET1,SCET2}, is being developed, that is a systematic
framework to describe from first principles the interactions of energetic but
low invariant mass particles with soft quanta.  The dynamics of a light quark
moving along the $z$ direction with large energy $Q$ is simplest to describe
decomposing its momentum in terms of the light-cone coordinates, $p = (p^-,
p_\perp, p^+)$,
\beq\label{scetdecomp}
p^\mu = \bar n\cdot p\, {n^\mu\over 2} + p_\perp^\mu 
  + n\cdot p\, {\bar n^\mu\over 2}
\equiv p^-\, {n^\mu\over 2} + p_\perp^\mu + p^+\, {\bar n^\mu\over 2}
\sim \Big[ {\cal O}(\lambda^0) + {\cal O}(\lambda^1) 
  + {\cal O}(\lambda^2)\Big]\, Q \,,
\eeq
where $n = (1,0,0,1)$ and $\bar n = (1,0,0,-1)$ are light-cone vectors
($n^2=0$), and $\lambda \sim {\cal O}(|p_\perp|/p^-)$ is a small parameter
(please do not confuse it with the Wolfenstein parameter!).  We have used that
the on-shell condition imposes $p^+ p^- \sim p_\perp^2 \sim \lambda^2\, Q^2$. 
In most applications $\lambda \sim \sqrt{\lqcd/m_b}$ or $\lqcd/m_b$. The goal
is to separate contributions from the scales $p^2 \sim Q^2$, $Q\lqcd$, and
$\lqcd^2$.

Similar to the field redefinition in HQET in Eq.~(\ref{hqfield}), one can
remove the  large component of the momentum of a collinear quark by a filed
redefinition~\cite{SCET1}
\beq\label{colfield}
\psi(x) = e^{-i \widetilde p\cdot x}\, \xi_n(x)\,,
\eeq
where $\widetilde p = p^-\, n/2 + p_\perp$ contain the parts of the light quark
momentum that can be parametrically larger than $\lqcd$.  An important
complication compared to HQET is that $\widetilde p$ is not a fixed label on
the collinear quark fields (in the sense that the four-velocity, $v$, is on
heavy quarks), since emission of collinear gluons by a massless quark is not
suppressed and changes $\widetilde p$.  Therefore, one has to introduce
separate collinear gluon fields in addition to collinear quarks and
antiquarks.  SCET gives an operator formulation of this complicated dynamics
with well-defined power counting that simplifies all order proofs of
factorization theorems, while previously such processes were analyzed only in
terms of Feynman diagrams.

As far as heavy-to-light form factors are concerned, the relevant region of
phase space is the small $q^2$ region, when $m_M/E_M$ is small.  The goal is to
have a clean separation of contributions from momentum regions $p^2 \sim
E_M^2$, $E_M\lqcd$, and $\lqcd^2$.  There are two crucial questions when
setting up such a framework.  First, it has to be proven that such a separation
is possible to all orders in the strong interaction.  It was first shown at
leading order in $\alpha_s$ that the infrared divergences can be absorbed into
the soft form factors~\cite{BF}.  However, the relative size of the soft and
hard contributions depend on assumptions about the tail of the pion wave
function~\cite{BF} or on the suppression of the soft part due to Sudakov
effects~\cite{Lirecent}.  SCET allows to prove factorization without such
assumptions, to all orders in $\alpha_s$ and to leading order in $1/Q (\equiv
1/E_M)$~\cite{BPSsl}.  A generic form factor can be split to two contributions
$F(Q) = f^{\rm F}(Q) + f^{\rm NF}(Q)$, where the two terms arise from matrix
elements of distinct operators between the same states.  One can
write~\cite{BPSsl}
\beqa\label{htlff}
f^{\rm F}(Q) &=& \frac{f_B f_M}{Q^2} \int_0^1\! \d z \int_0^1\! \d x 
  \int_0^\infty\! \d r_+\, T(z,Q,\mu_{\rm 0}) \nn\\
&& \qquad \times J(z,x,r_+,Q,\mu_{\rm 0},\mu)\, \phi_M(x,\mu)\,
  \phi^{+}_B(r_+,\mu) \,, \nn\\
f^{\rm NF}(Q) &=& C_k(Q,\mu)\, \zeta_k^M(Q,\mu) \,.
\eeqa
The hard coefficients, $T$, $C_k$, and $J$ are process dependent; $C_k$ and $T$
can be calculated in an expansion in $\alpha_s(Q)$, while the so-called jet
function, $J$, is dominated by momenta $p^2 \sim Q\lqcd$ and starts at order
$\alpha_s(\sqrt{Q\lqcd})$.  In Eq.~(\ref{htlff}) $\phi_M$ and $\phi_B^{+}$ are
nonperturbative distribution amplitudes for the final meson $M$ and the initial
$B$, on which both contributions depend.  The nonfactorizable part depends
on three soft form factors, $\zeta_k^M$, which are universal nonperturbative
functions.  Only one occurs for decays to pseudoscalars, and two for decays to
vector mesons, thus reproducing the heavy-to-light form factor
relations~\cite{charles}.  The second question is to understand the power
counting of the two contributions, including possible suppressions by
$\alpha_s$.  Both terms in Eq.~(\ref{htlff}) scale as $(\lqcd/Q)^{3/2}$. It is
yet unknown whether the $f^{\rm NF}$ term might also have an
$\alpha_s(\sqrt{Q\lqcd})$ suppression~\cite{BPSsl}, similar to that present in
$J$.  Progress in theory is expected to answer this in the formal $m_b \gg
\lqcd$ limit, and testing the one relation between the three experimentally
measurable $B\to \rho\ell\bar\nu$ form factors could tell us about the relative
size of the two contributions for the physical $b$ quark mass.

There are many possible applications.  For example, one could use the $B\to
K^*\gamma$ rate to constrain the $B\to \rho\ell\bar\nu$ and $B\to
K^*\ell^+\ell^-$ form factors relevant for the determination of $|V_{ub}|$ and
searches for new physics~\cite{BH}.  Some others are discussed in
Sec.~\ref{sec:addrare}.

\subsection[Inclusive semileptonic $B$ decays]{\boldmath Inclusive semileptonic
$B$ decays}

Sometimes, instead of identifying all particles in a decay, it is convenient to
be ignorant about some details.  For example, we might want to specify the
energy of a charged lepton or a photon in the final state, or restrict the
flavor of the final hadrons.  These~decays are inclusive in the sense that we
sum over final states which can be produced by strong interactions, subject to
a limited set of constraints determined by short distance perturbative
physics.  Typically we are interested in a quark-level transition, such as
$b\to c\ell\bar\nu$, $b\to s\gamma$, etc., and we would like to extract the
corresponding short distance parameters, $|V_{cb}|$, $C_7(m_b)$, etc., from the
data.  To do this, we need to be able to model independently relate the
quark-level operators to the experimentally accessible observables.

\subsubsection[The OPE, total rates, and $|V_{cb}|$]{\boldmath The OPE, total
rates, and $|V_{cb}|$}

In the large $m_b$ limit, when the energy released in the decay is large, there
is a simple heuristic argument that the inclusive rate may be modeled simply by
the decay of a free $b$ quark.  The argument is again based on a separation of
time (or distance) scales.  The $b$ quark decay mediated by weak interaction
takes place on a time scale that is much shorter than the time it takes the
quarks in the final state to form physical hadronic states.  Once the $b$ quark
has decayed on a time scale $t \ll \lqcd^{-1}$, the probability that the final
states will hadronize somehow is unity, and we need not know the (uncalculable)
probabilities of hadronization into specific final states.

Let us consider inclusive semileptonic $b\to c$ decay, mediated by the operator
\beq
O_{\rm sl} = -{4G_F\over \sqrt2}\, V_{cb}\, 
  (J_{bc})^\alpha\, (J_{\ell\nu})_\alpha \,,
\eeq
where $J^\alpha_{bc} = (\ov c\, \gamma^\alpha P_L\, b)$ and
$J^\beta_{\ell\nu} = (\ov\ell\, \gamma^\beta P_L\, \nu)$.
The decay rate is given by the square of the matrix element, integrated over
phase space, and summed over final states,
\beq
\ds \Gamma(B\to X_c\ell\bar\nu) \sim \sum_{X_c} \int\! \d [{\rm PS}]\,
  \big| \bra{X_c\ell\bar\nu} O_{\rm sl} \ket{B} \big|^2 \,.
\eeq
Since the leptons have no strong interaction, it is convenient to factorize the
phase space into $B\to X_c W^*$ and a perturbatively calculable leptonic part,
$W^* \to \ell\bar\nu$.  The nontrivial part is the hadronic tensor,
\beqa
W^{\alpha\beta} &\sim& \sum_{X_c} \delta^4(p_B-q-p_X)\,
  \big| \bra{B} J^{\alpha\dagger}_{bc} \ket{X_c}\, 
  \bra{X_c} J^\beta_{bc} \ket{B} \big|^2 \nn\\*
&\sim& {\rm Im} \int\! \d x\, e^{-iq\cdot x}\,
  \bra{B}\, T \big\{ J^{\alpha\dagger}_{bc}(x)\, 
  J^\beta_{bc}(0) \big\}\, \ket{B} \,.
\eeqa
where the second line is obtained using the optical theorem, and $T$ denotes the
time ordered product of the two operators.  This is convenient, because it is
this time ordered product that can be expanded in local operators in the $m_b
\gg \lqcd$ limit~\cite{OPE}.  In this limit the time ordered product is
dominated by short distances, $x \ll \lqcd^{-1}$, and one can express the
nonlocal hadronic tensor $W^{\alpha\beta}$ as a sum of local operators. 
Schematically,
\beq\label{opesketch}
\raisebox{-36pt}{\includegraphics*[width=.34\textwidth]{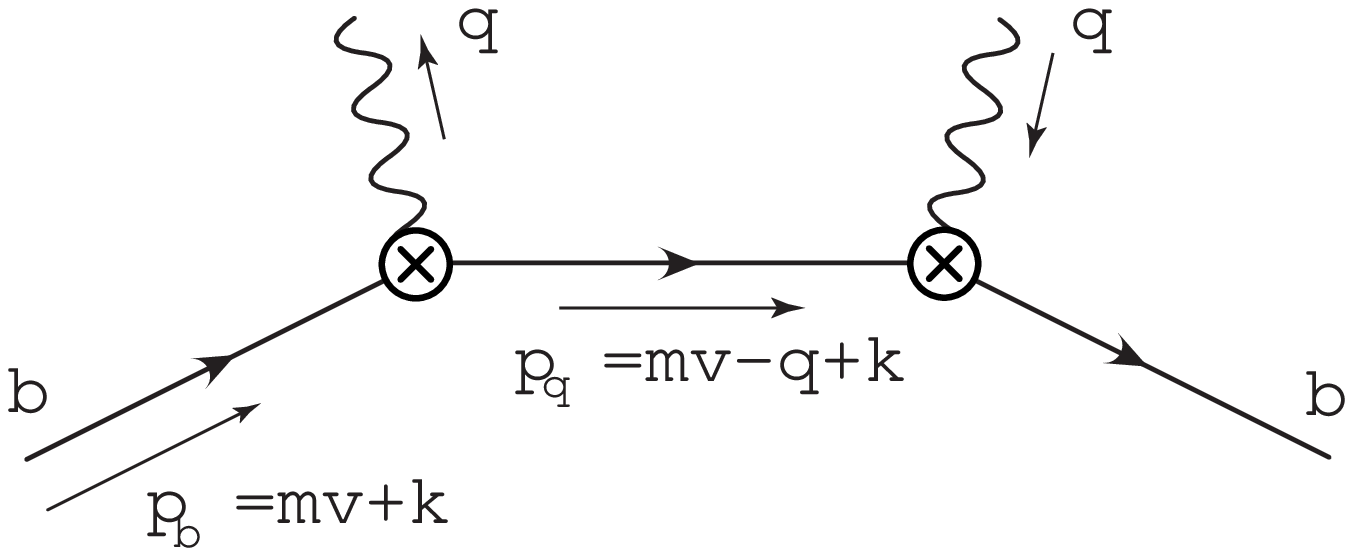}} = 
  \raisebox{-22pt}{\includegraphics*[width=.12\textwidth]{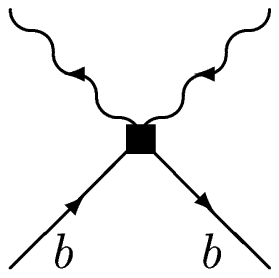}}
 + {0\over m_b} \raisebox{-22pt}{\includegraphics*[width=.12\textwidth]{ope2}}
 + {1\over m_b^2} \raisebox{-22pt}{\includegraphics*[width=.12\textwidth]{ope2}}
 + \ldots \,.
\eeq
This is analogous to a multipole expansion.  At leading order the decay rate is
determined by the $b$ quark content of the initial state, while subleading
effects are parameterized by matrix elements of operators with increasing
number of derivatives that are sensitive to the distribution of chromomagnetic
and chromoelectric fields.

At lowest order in $\lqcd/m_b$ this operator product expansion (OPE) leads to
operators of the form $\bar b\, \Gamma\, b$, where $\Gamma$ is some
(process-dependent) Dirac matrix.  For $\Gamma = \gamma^\mu$ or
$\gamma^\mu\gamma_5$ their matrix elements are known to all orders in
$\lqcd/m_b$
\beqa\label{1:eq:hqe3}
\langle B(p_B) |\, \bar b\, \gamma^\mu b \, |B(p_B)\rangle &=&
  2p_B^\mu = 2m_B\, v^\mu \,, \nn\\*
\langle B(p_B) |\, \bar b\, \gamma^\mu\gamma_5\, b \, |B(p_B)\rangle &=& 0 \,,
\eeqa
because of conservation of the $b$ quark number and parity invariance of strong
interactions.  The matrix elements for other $\Gamma$'s can be related by heavy
quark symmetry to these plus ${\cal O}(\lqcd^2/m_b^2)$ terms.  Thus the OPE
justifies that inclusive $B$ decay rates in the $m_b\to \infty$ limit are given
by free $b$ quark decay.

To compute subleading corrections, it is convenient to use HQET.  There are no
${\cal O}(\lqcd/m_b)$ corrections, because the $B$ meson matrix element of any
dimension-4 operator vanishes, $\langle B(v) |\, \bar h_v^{(b)} iD_\alpha
\Gamma\, h_v^{(b)} |B(v) \rangle = 0$.  The leading nonperturbative effects
suppressed by $\lqcd^2/m_b^2$ are parameterized by two HQET matrix elements,
\beq
\lambda_1 = {1\over 2m_B}\, \langle B|\, \bar h_v^{(b)} (iD)^2\, h_v^{(b)}\, 
  |B\rangle \,, \qquad 
  \lambda_2 = {1\over 6m_B}\, \langle B |\, \bar h_v^{(b)}\, {g_s\over2}\,
  \sigma_{\mu\nu}\, G^{\mu\nu}\, h_v^{(b)}\,| B\rangle \,.
\eeq
The $B^*-B$ mass splitting determines $\lambda_2 = (m_{B^*}^2-m_B^2)/4 \simeq
0.12\, {\rm GeV}^2$, whereas the most promising way to determine $\lambda_1$
is from experimental data on inclusive decay distributions, as explained
below.  The result of the OPE can then be written schematically as
\beq\label{incl}
\d\Gamma = \pmatrix{b{\rm ~quark} \cr {\rm decay}\cr} \times 
\bigg\{ 1 + \frac0{m_b} + \frac{f(\lambda_1,\lambda_2)}{m_B^2} + \ldots
  + \alpha_s(\ldots) + \alpha_s^2(\ldots) + \ldots \bigg\} \,.
\eeq
At order $\lqcd^3/m_b^3$, six new and largely unknown hadronic matrix elements
enter, and usually naive dimensional analysis is used to estimate the
uncertainties related to them.  For most quantities of interest, the
perturbation series are known including the $\alpha_s$ and $\alpha_s^2\beta_0$
terms, where $\beta_0 = 11 - 2n_f/3$ is the first coefficient of the QCD
$\beta$-function (in many cases this term is expected to dominate the order
$\alpha_s^2$ corrections). 

In which regions of phase space can the OPE be expected to converge?  Near
boundaries of the Dalitz plot the assumption that the energy release to the
final hadronic state is large can be violated. It is useful to think of the OPE
as an expansion in the residual momentum of the $b$ quark, $k$, in the diagram
on the left-hand side of Eq.~(\ref{opesketch}).  Expanding the propagator,
\beq
{1\over (m_b v + k -q)^2-m_q^2} 
  = {1\over [(m_b v -q)^2-m_q^2] + [2k\cdot (m_b v-q)] + k^2} \,,
\eeq
we see that for the expansion in powers of $k$ to converge, the final state
phase space can only be restricted in a manner to still allow hadronic final
states $X$ to contribute with
\beq\label{converge}
m_X^2-m_q^2 \gg E_X \lqcd \gg \lqcd^2 \,.
\eeq

Before discussing the implications of this inequality, it has to be mentioned
that the OPE implicitly relies on quark-hadron duality~\cite{PQW}.  This is
simply the notion that averaged over sufficiently many exclusive final states,
hadronic quantities can be computed at the parton level.  Its violations are
believed to be small for fully inclusive semileptonic $B$ decay rates (although
this is not undisputed~\cite{NIdual}), however, exactly how small is very
hard to quantify.  Comparing differential distributions discussed below appears
to be the most promising way to constrain it experimentally.

The good news from Eq.~(\ref{converge}) is that the OPE calculation of total
rates should be under good  control.  The theoretical uncertainty is dominated
by the uncertainty in a short distance $b$ quark mass (whatever way it is
defined) and in the perturbation series.  Using the ``upsilon expansion", the
relation between the inclusive semileptonic rate and $|V_{cb}|$
is~\cite{upsexp}
\begin{equation}\label{Vcb}
|V_{cb}| = (41.9 \pm 0.8_{(\rm pert)} \pm 0.5_{(m_b)} \pm 0.7_{(\lambda_1)}) 
  \times 10^{-3}\, \bigg( {{\cal B}(B\to X_c \ell\bar\nu)\over0.105}\,
  {1.6\,{\rm ps}\over\tau_B} \bigg)^{1/2} .
\end{equation}
The first error is from the uncertainty in the perturbation series, the second
one from the $b$ quark mass, $m_b^{1S} = 4.73 \pm 0.05\,$GeV (a very
conservative range of $m_b$ may be larger~\cite{bmasses}), and the third one
from $\lambda_1 = -0.25 \pm 0.25\,{\rm GeV}^2$.  This result is in agreement
with Ref.~\citex{BSUreview}, where the central value is $40.8 \times 10^{-3}$
(including a small, 1.007, electromagnetic radiative correction).

Progress in the determinations of $m_b$ and $\lambda_1$ is likely to come from
measurements of shape variables in inclusive $B$ decays~\cite{shape}.  The idea
is to look at decay distributions independent of CKM elements to learn about
the hadronic parameters, that can in turn reduce the errors of the CKM
measurements.  Such observables are ratios of differently weighted integrals of
decay distributions (sometimes called ``moments"); specifically the charged
lepton energy~\cite{volo,gremmetal,GK,GS} and hadronic invariant
mass~\cite{FLSmass,GK} spectra in $B\to X_c\ell\bar\nu$ and the photon energy
spectrum in $B\to X_s\gamma$~\cite{kl,llmw,bauer}.  Comparing these shape
variables is also the most promising approach to constrain experimentally the
accuracy of OPE, including the possible size of quark-hadron duality violation.
The presently available measurements~\cite{momentsdata,CLEObsgmom} do not seem
to fit well together.  It appears crucial to determine the $B\to
D^{(*)}\ell\bar\nu$ branching ratios with higher precision, to model
independently map out the hadronic invariant mass distribution in $B\to
X_c\ell\bar\nu$ decay, and to try to measure the $B\to X_s\gamma$ spectrum to
as low photon energies as possible.  If the overall agreement improves, then
this program may lead to an error in $|V_{cb}|$ at the $\sim 2\%$ level.

The bad news from Eq.~(\ref{converge}) is that in certain restricted regions of
phase space the OPE breaks down.  This is a problem, for example, for the
determination of $|V_{ub}|$ from $B\to X_u\ell\bar\nu$, because severe cuts are
required to eliminate $\sim\!100$ times larger $b\to c$ background.  Similarly,
in $B\to X_s\gamma$, the rate can only be measured for energetic photons that
populate a modest region of phase space, $E_\gamma^{\rm max} - E_\gamma^{\rm
min}< 1\,$GeV.  Some of the new theoretical problems that enter in such
situations are discussed next.

\subsubsection[$B\to X_u\ell\bar\nu$ spectra and $|V_{ub}|$]{\boldmath
$B\to X_u\ell\bar\nu$ spectra and $|V_{ub}|$}

If it were not for the huge $B\to X_c\ell\bar\nu$ background, measuring
$|V_{ub}|$ would be as ``easy" as $|V_{cb}|$.  The total $B\to X_u \ell\bar\nu$
rate can be predicted in the OPE with small uncertainty~\cite{upsexp},
\begin{equation}\label{Vub}
|V_{ub}| = (3.04 \pm 0.06_{(\rm pert)} \pm 0.08_{(m_b)}) \times 10^{-3}\,
  \bigg( {{\cal B}(B\to X_u \ell\bar\nu)\over 0.001}
  {1.6\,{\rm ps}\over\tau_B} \bigg)^{1/2} ,
\end{equation}
where the errors are as discussed after Eq.~(\ref{Vcb}).  If this fully
inclusive rate is measured without significant cuts on the phase space, then
$|V_{ub}|$ may be determined with less than $5\%$ theoretical error.

When kinematic cuts are used to distinguish the $b\to u$ signal from the $b\to
c$ background, the behavior of the OPE can become significantly worse.  As
indicated by Eq.~(\ref{converge}), there are three qualitatively different
regions of phase space, depending on how the invariant mass and energy of the
hadronic final state (in the $B$ rest frame) is restricted:
\begin{itemize} \vspace*{-4pt}\itemsep -2pt
\item[(i)] $m_X^2 \gg E_X \lqcd \gg \lqcd^2$: the OPE converges, and the first
few terms are expected to give reliable result.  This is the case for the 
$B\to X_c\ell\bar\nu$ width relevant for measuring $|V_{cb}|$.
\item[(ii)] $m_X^2 \sim E_X \lqcd \gg \lqcd^2$: an infinite set of equally
important terms in the OPE must be resummed.  The OPE becomes a twist expansion
and nonperturbative input is needed.
\item[(iii)] $m_X \sim \lqcd$: the final state is dominated by resonances, and
it is not known how to compute any inclusive quantity reliably.
\end{itemize}\vspace*{-4pt}
The charm background can be removed by several different kinematic cuts:
\begin{enumerate} \vspace*{-4pt}\itemsep -2pt
\item $E_\ell > (m_B^2-m_D^2) / (2m_B)$: the lepton endpoint region that was
used to first observe $b\to u$ decay;
\item $m_X < m_D$: the small hadronic invariant mass
region;~\cite{mass,FLW,BDU,llrhadron}
\item $E_X < m_D$: the small hadronic energy region;~\cite{energy}
\item $q^2 \equiv (p_\ell + p_\nu)^2 > (m_B - m_D)^2$: the large dilepton
invariant mass region~\cite{BLL1}.
\end{enumerate}\vspace*{-4pt}
These contain roughly $10\%$, $80\%$, $30\%$, and $20\%$ of the rate,
respectively.  Measuring any other variable than $E_\ell$ requires the
reconstruction of the neutrino, which is challenging experimentally. 
Combinations of cuts have also been proposed, $q^2$ with $m_X$~\citex{BLL2},
$q^2$ with $E_\ell$~\citex{KoMe}, or $m_X$ with $E_X$~\citex{ugo}.

\begin{figure}[t]
\centerline{\includegraphics*[width=.95\textwidth]{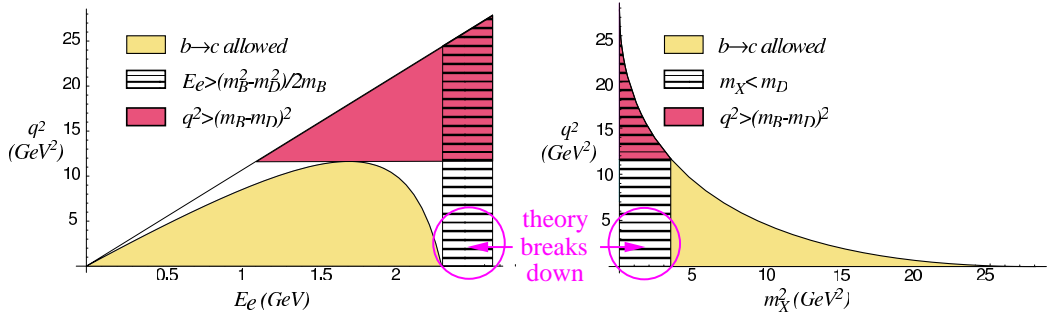}}
\caption{Dalitz plots for $B\to X\ell\bar\nu$ in terms of $E_\ell$ and
$q^2$ (left), and $m_X^2$ and $q^2$ (right).}
\label{fig:dalitz}
\end{figure}

The problem is that both phase space regions 1.\ and 2.\ belong to the regime
(ii), because these cuts impose $m_X \lsim m_D$ and $E_X \lsim m_B$, and
numerically $\lqcd\, m_B \sim m_D^2$.  The region $m_X < m_D$ is better than
$E_\ell > (m_B^2-m_D^2) / (2m_B)$ inasmuch as the expected rate is larger, and
the inclusive description is expected to hold better.  But nonperturbative
input is needed in both cases, formally at the ${\cal O}(1)$ level, which is
why the model dependence increases rapidly if the $m_X$ cut is lowered below
$m_D$~\cite{FLW}.  These regions of the Dalitz plot are shown in
Fig.~\ref{fig:dalitz}.

\begin{figure}[t]
\centerline{\includegraphics*[height=3cm]{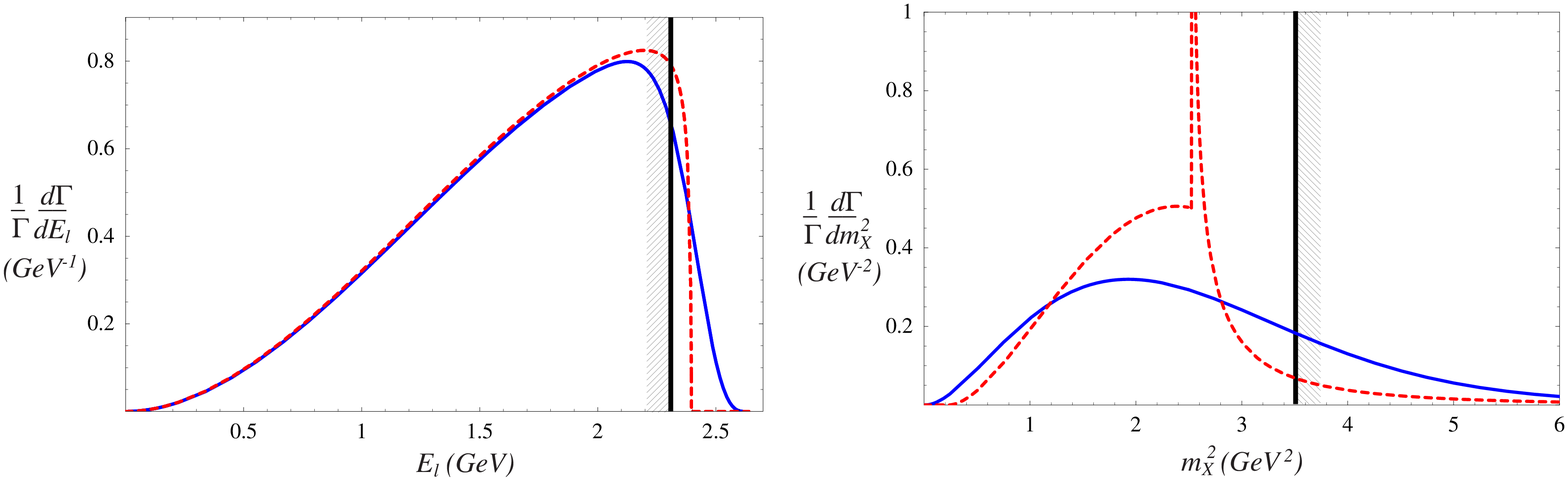}~
\includegraphics*[height=2.97cm]{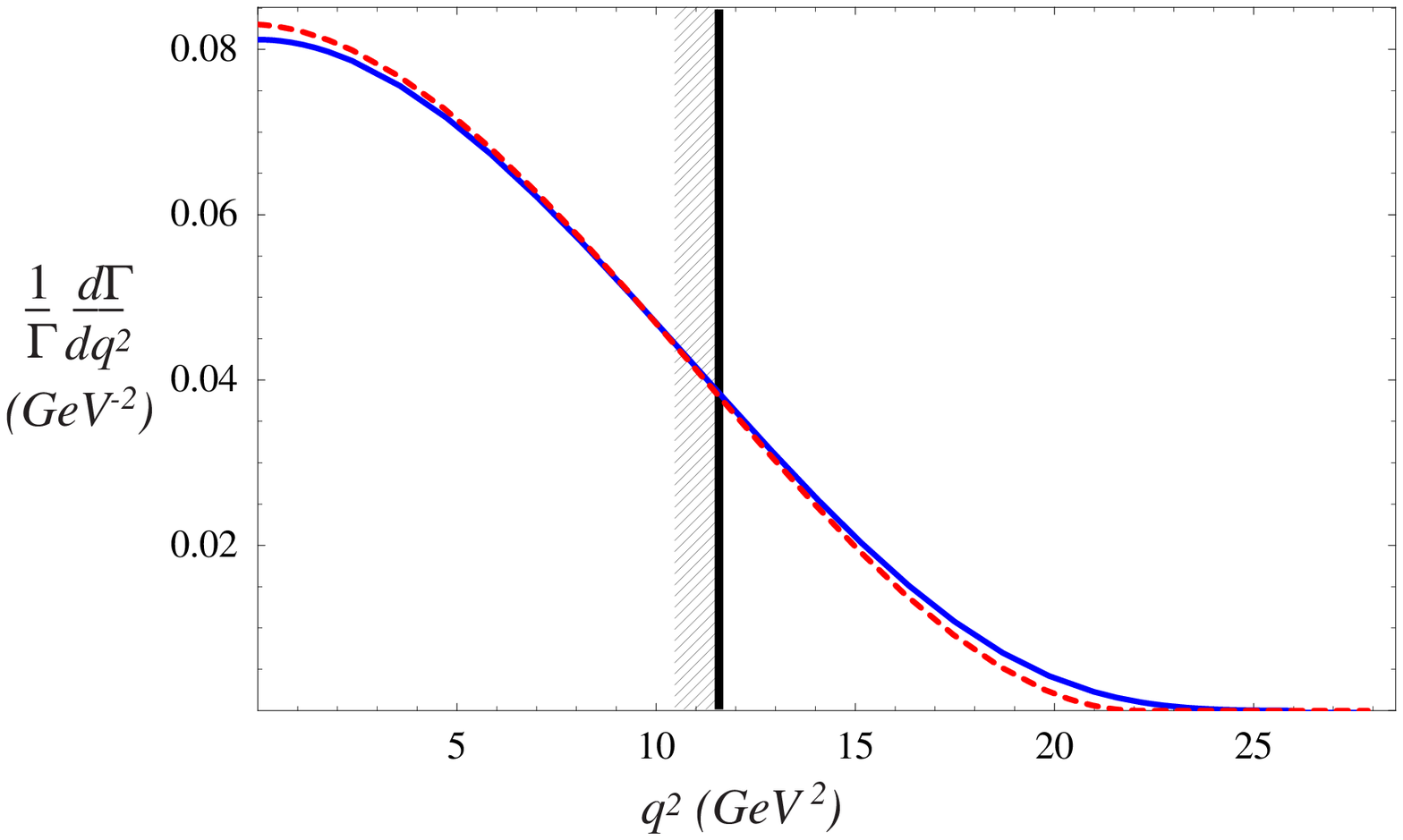}}
\caption{$B\to X_u\ell\bar\nu$ spectra --- $E_\ell$  (left), $m_X^2$ (middle),
and $q^2$ (right) --- as given by $b$ quark decay including ${\cal
O}(\alpha_s)$ terms (dashed curves), and including the Fermi motion model
(solid curves).}
\label{fig:smear}
\end{figure}

The large $E_\ell$ and small $m_X$ regions are determined by the $b$ quark
light-cone distribution function that describes the Fermi motion inside the $B$
meson (sometimes called the shape function).  Its effect on the spectra are
illustrated in Fig.~\ref{fig:smear}, where we also show the $q^2$ spectrum
unaffected by it.  This nonperturbative function is universal at leading order
in $\lqcd/m_b$, and is related to the $B\to X_s\gamma$ photon
spectrum~\cite{structure}.  These relations have been extended to the resummed
next-to-leading order corrections~\cite{extractshape}, and to include effects
of operators other than $O_7$ contributing to $B\to X_s\gamma$~\cite{notO7}. 
Weighted integrals of the $B\to X_s\gamma$ photon spectrum are equal to the
$B\to X_u\ell\bar\nu$ rate in the large $E_\ell$ or small $m_X$ regions. 
Recently CLEO~\cite{cleoVub} used the $B\to X_s\gamma$ photon spectrum as an
input to determine $|V_{ub}| = (4.08 \pm 0.63)\times 10^{-3}$ from the lepton
endpoint region.

The dominant theoretical uncertainty in this determinations of $|V_{ub}|$ are
from subleading twist contributions, which are not related to $B\to
X_s\gamma$~\cite{subltwistus}.  The $B\to X_u \ell\bar\nu$ lepton spectrum,
including dimension-5 operators and neglecting perturbative corrections, is
given by~\cite{OPE}
\beqa\label{slspec}
{\d \Gamma\over \d y} &=& {G_F^2\, m_b^5\, |V_{ub}|^2\, \over 192\,\pi^3}\,
\bigg\{ \bigg[ y^2(3-2y) + {5\lambda_1\over 3m_b^2}\, y^3
  + {\lambda_2\over m_b^2}\, y^2(6+5y) \bigg]\, 2\theta(1-y) \nn\\*
&&\qquad\qquad\quad\ \,\, - \bigg[ {\lambda_1\over 6m_b^2} 
  + {11\lambda_2\over 2m_b^2} \bigg]\, 2\delta(1-y)
  - {\lambda_1\over 6m_b^2}\, 2\delta'(1-y) + \ldots \bigg\} . 
\eeqa
The behavior near $y=1$ is determined by the leading order structure function,
which contains the terms $2 [\theta(1-y) - \lambda_1/(6m_b^2)\, \delta'(1-y) +
\ldots]$.  The derivative of the same combination occurs in the $B\to
X_s\gamma$ photon spectrum~\cite{FLS}, given by
\beqa\label{bsgspec} 
{\d \Gamma\over \d x} &=& 
  {G_F^2\, m_b^5 \,|V_{tb}V_{ts}^*|^2\, \alpha\, C_7^2 \over32\,\pi^4}\, 
  \bigg[ \bigg( 1 + {\lambda_1-9\lambda_2\over2m_b^2} \bigg)\, \delta(1-x) 
  - {\lambda_1 + 3\lambda_2 \over 2m_b^2}\, \delta'(1-x) \nn\\
&&\qquad\qquad\qquad\qquad\quad 
  - {\lambda_1 \over 6m_b^2}\, \delta''(1-x) + \ldots \bigg] ,
\eeqa
At subleading order, proportional to $\delta(1-y)$ in Eq.~(\ref{slspec}) and to
$\delta'(1-x)$ in Eq.~(\ref{bsgspec}), the terms involving $\lambda_2$ differ
significantly, with a coefficient $11/2$ in Eq.~(\ref{slspec}) and $3/2$ in
Eq.~(\ref{bsgspec}).  Because of the 11/2 factor, the $\lambda_2\, \delta(1-y)$
term is important in the lepton endpoint
region~\cite{subltwistus,subltwist,subltwistmn}.  There is also a significant
uncertainty at order $\lqcd^2/m_b^2$ from weak
annihilation~\cite{Voloshin,subltwistus}.  Moreover, if the lepton endpoint
region is found to be dominated by the $\pi$ and $\rho$ exclusive channels,
then the applicability of the inclusive description may be questioned.

In contrast to the above, in the $q^2 > (m_B-m_D)^2$ region the first few terms
in the OPE determine the rate~\cite{BLL1}.  This cut implies $E_X \lsim m_D$
and $m_X \lsim m_D$, and so the $m_X^2 \gg E_X \lqcd \gg \lqcd^2$ criterion of
regime (i) is satisfied.  This relies, however, on $m_c \gg \lqcd$, and so the
OPE is effectively an expansion in $\lqcd/m_c$~\cite{neubertq2}.  The largest
uncertainties come from order $\lqcd^3/m_{c,b}^3$ nonperturbative corrections,
the $b$ quark mass, and the perturbation series.  Weak annihilation (WA)
suppressed by $\lqcd^3/m_b^3$ is important, because it enters the rate as
$\delta(q^2-m_b^2)$~\cite{Voloshin}.  Its magnitude is hard to estimate,
because it is proportional to the difference of two matrix elements, which are
equal in the factorization limit.  Assuming a 10\% violation of factorization,
WA could be $\sim 2\%$ of the $B\to X_u\ell\bar\nu$ rate, and, in turn, $\sim
10\%$ of the rate in the $q^2 > (m_B-m_D)^2$ region.  The uncertainty of this
estimate is large.  Since this contribution is also proportional to
$\delta(E_\ell - m_b/2)$, it is even more important for the lepton endpoint
region.  Experimentally, WA can be constrained by comparing $|V_{ub}|$ measured
from $B^0$ and $B^\pm$ decays, and by comparing the $D^0$ and $D_s$
semileptonic widths~\cite{Voloshin}.

Combining the $q^2$ and $m_X$ cuts can significantly reduce the theoretical
uncertainties~\cite{BLL2}.  The right-hand side of Fig.~\ref{fig:dalitz} shows
that the $q^2$ cut can be lowered below $(m_B-m_D)^2$ by imposing an additional
cut on $m_X$.  This changes the expansion parameter from $\lqcd/m_c$ to
$m_b\lqcd/(m_b^2-q_{\rm cut}^2)$, resulting in a significant decrease of the
uncertainties from both the perturbation series and from the nonperturbative
corrections.  At the same time the uncertainty from the $b$ quark light-cone
distribution function only turns on slowly.  Some representative results are
give in Table~\ref{tab:dblcut}, showing that it may be possible to determine
$|V_{ub}|$ with a theoretical error at the $\sim 5\%$ level using up to $\sim
45\%$ of the semileptonic decays.

\begin{table}
\begin{center}
\begin{tabular}{ccc} \hline\hline
Cuts on  &  Fraction  &  Error of $|V_{ub}|$ \\[-6pt]
$q^2$ and $m_X$  &  of events  &  $\delta m_b=80/30\,\MeV$ \\ \hline\hline
$6\,\GeV^2,\, m_D$		& $46\%$ & $8\%/5\%$  \\
$8\,\GeV^2,\, 1.7\,\GeV$ 	& $33\%$ & $9\%/6\%$  \\ \hline
$(m_B-m_D)^2,\, m_D$ 		& $17\%$ & $15\%/12\%$ \\ \hline\hline
\end{tabular}
\end{center}
\caption{$|V_{ub}|$ from combined cuts on $q^2$ and $m_X$ 
(from Ref.~\protect\citex{BLL2}).}
\label{tab:dblcut}
\end{table}

\subsection{Some additional topics}

This section contains short discussions of three topics that there was no time
to cover during the lectures, but were included in the printed slides. 
Skipping this section will not affect the understanding of the rest of this
writeup.

\subsubsection[$B$ decays to excited $D$ mesons]{\boldmath $B$ decays to
excited $D$ mesons}

Heavy quark symmetry implies that in the $m_Q \to \infty$ limit, matrix
elements of the weak currents between a $B$ meson and an excited charmed meson
vanish at zero recoil.  However, in some cases at order $\Lambda_{\rm QCD}/m_Q$
these matrix elements are nonzero and calculable~\cite{dss}.  Since most of the
phase space is near zero recoil, $\Lambda_{\rm QCD}/m_Q$ corrections can be
very important.  

In the heavy quark limit, for each doublet of excited $D$ mesons, all
semileptonic decay form factors are related to a single Isgur-Wise
function~\cite{IWsr}.  At ${\cal O}(\lqcd/m_Q)$~many new functions occur. In
$B\to (D_1,D_2^*)\ell\bar\nu$ there are $8$ subleading Isgur-Wise functions
(neglecting time ordered products with subleading terms in the Lagrangian,
which are expected to be small or can be absorbed), but only $2$ of them are
independent~\cite{dss}.  Moreover, in $B\to$ orbitally excited $D$ decays, the
zero recoil matrix element at ${\cal O}(\lqcd/m_Q)$ is given by mass splittings
and the $m_Q\to \infty$ Isgur-Wise function.  For example, in $B\to
D_1\ell\bar\nu$ decay~\cite{dss},
\beq\label{BDssff}
f_{V_1}(1) = -{4\over \sqrt6\, m_c}\, (\bar\Lambda' - \bar\Lambda)\, \tau(1)\,.
\eeq
Here $f_{V_1}$ is the form factor defined by 
\beq
\langle D_1(v',\epsilon)| V^\mu
|B(v)\rangle = \sqrt{m_{D_1} m_B}\, \Big[ f_{V_1} \epsilon^{*\mu}  
  + (f_{V_2} v^\mu + f_{V_3} v'^\mu) (\epsilon^*\cdot v) \Big]\,, 
\eeq
which determines the rate at zero recoil, similar to $h_{A_1}$ in $B\to D^*$
decay defined in Eq.~(\ref{BDffdef}).  Here $\tau$ denotes the leading order
Isgur-Wise function, and $\bar\Lambda'$ is the $m_{D_1} - m_c$ mass splitting
in the heavy quark limit ($\bar\Lambda' - \bar\Lambda \equiv \Delta_1$ in
Fig.~\ref{fig:spectra}).  Using Eq.~(\ref{BDssff}), the decay rate can be
expanded simultaneously in powers of $\lqcd/m_Q$ and $w-1$ schematically as
\beqa
{\d \Gamma(B\to D_1\ell\bar\nu) \over \d w}
&\propto& \sqrt{w^2-1}\, [\tau(1)]^2 
  \bigg\{ 0 + 0\,(w-1) + (\ldots)(w-1)^2 + \ldots \nn\\*
&&\qquad\quad\ \, + {\lqcd\over m_Q}\, \Big[ 0
  + (\mbox{almost~calculable}) (w-1) + \ldots \Big] \nn\\*
&&\qquad\quad\ \, + {\lqcd^2\over m_Q^2}\, 
  \Big[ (\mbox{calculable}) + \ldots \Big] + \ldots \bigg\}
\eeqa
The zeros and the calculable terms are model independent  predictions of HQET,
while the ``almost calculable" term has a  calculable part that is expected to
be dominant.

There are many experimentally testable implications.  One of the least model
dependent is the prediction for
\beq
R \equiv 
  {{\cal B}(B\to D_2^*\ell\bar\nu) \over {\cal B}(B\to D_1\ell\bar\nu)} \,,
\eeq
because the leading order Isgur-Wise function drops out to a good
approximation.  This ratio is around $1.6$ in the infinite mass limit, and it
was predicted to be reduced to about $0.4 - 0.7$~\cite{dss}, because
$\lqcd/m_c$ corrections enhance the $B\to D_1$ rate significantly but hardly
affect $B\to D_2^*$.  The present world average is about $0.4\pm 0.15$.

To compare the $B\to (D_1,D_2^*)$ rates with $(D_0^*,D_1^*)$, we need to know
the leading Isgur-Wise functions.  Quark models and QCD sum rules predict that
the Isgur-Wise function for the broad $(D_0^*,D_1^*)$ doublet is not larger
than for the  narrow $(D_1,D_2^*)$ doublet~\cite{orsay}.  These arguments make
the large $B\to (D_0^*,D_1^*) \ell\bar\nu$ rates puzzling.

\begin{figure}[t]
\centerline{\includegraphics*[width=.5\textwidth,height=.45\textwidth]{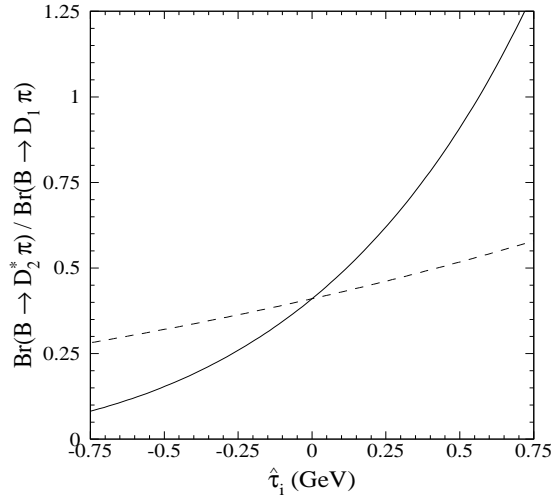}}
\caption{Factorization prediction for $R_\pi$ defined in Eq.~(\ref{belleRpi})
as a function of  $\hat\tau_1$ for $\hat\tau_2=0$ (solid curve), and as a 
function of $\hat\tau_2$ for $\hat\tau_1=0$ (dashed curve).  
(From Ref.~\protect\citex{dss}.)}
\label{fig:dssfact}
\end{figure}

Another way the theory of these decays can be tested is via nonleptonic
decays.  Factorization in $B\to D^{**}\pi$ is expected to work as well as in
$B\to D^{(*)}\pi$ (see Sec.~\ref{sec:bcfact}),
\beq\label{BDfact}
\Gamma_\pi = {3\pi^2\, |V_{ud}|^2\, C^2 f_\pi^2 \over m_B^2\, r} \times
  \left( {\d \Gamma_{\rm sl}\over \d w} \right)_{w_{\rm max}} \,,
\eeq
where $r=m_{D^{**}}/m_B$, $f_\pi\simeq131\,$MeV, $w_{\rm max} =
(1+r^2)/(2r)\simeq 1.3$ in these decays, and $C\,|V_{ud}| \simeq 1$.
(As we will see in Sec.~\ref{sec:factest}, this test would be more reliable in
$B^0$ decay, however that is harder to measure experimentally.) An interesting
ratio with little sensitivity to the leading order Isgur-Wise function was
recently measured with good precision~\cite{belleDss}
\beq\label{belleRpi}
R_\pi \equiv 
  {{\cal B}(B^-\to D_2^{*0}\pi^-) \over {\cal B}(B^-\to D_1^0\pi^-)} 
  = 0.89 \pm 0.14 \,,
\eeq
whereas the CLEO result was $1.8 \pm 0.9$~\cite{cleoDss}. 
Figure~\ref{fig:dssfact} shows that $R_\pi$ is very sensitive to the subleading
${\cal O}(\lqcd/m_Q)$ Isgur-Wise functions, $\hat\tau_1$ and $\hat\tau_2$. 
Assuming that they are below $500\,$MeV (which is not an unusually large value
by any means), the theory predicts $R_\pi < 1$.  Neglecting $\lqcd/m_Q$
corrections~\cite{mndss}, the prediction is $R_\pi \sim 0.35$, as also seen
from Fig.~\ref{fig:dssfact}.  We learn that the BELLE result in
Eq.~(\ref{belleRpi}) agrees well with theory, which is a success of HQET in a
regime with large sensitivity to $\lqcd/m_Q$ effects.  It constrains the
subleading Isgur-Wise functions, which has useful implications for the analysis
of $B\to D_1\ell\bar\nu$ and $D_2^*\ell\bar\nu$ decays.

Sorting out these semileptonic and nonleptonic decays to excited $D$'s will
provide important tests of HQET, factorization, and will also impact the
determinations of $|V_{cb}|$.

\subsubsection{\boldmath Exclusive rare decays}
\label{sec:addrare}

Exclusive rare decays are interesting for a large variety of reasons.  As any
flavor-changing neutral current process, they are sensitive probes of new
physics, and within the SM they are sensitive to $|V_{td}|$ and $|V_{ts}|$. For
example, $B\to K^{(*)}\ell^+\ell^-$ or $B\to X\,\ell^+\ell^-$ are sensitive to
SUSY, enhanced $bsZ$ penguins, right handed couplings, etc.

\begin{figure}[t]
\centerline{\includegraphics*[width=.45\textwidth]{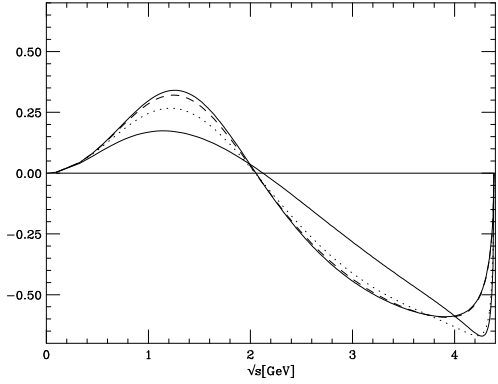}}
\caption{Forward-backward asymmetry in $B\to K^*\ell^+\ell^-$ decay in
different form factor models  ($s \equiv q^2$). (From
Ref.~\protect\citex{burdmanAfb}.)}
\label{fig:afb}
\end{figure}

Exclusive rare decays are experimentally easier to measure than inclusive
decays, but a clean theoretical interpretation requires model independent
knowledge of the corresponding form factors.  (However, certain $CP$
asymmetries are independent of them.)  It was originally observed that there is
an observable, the forward-backward asymmetry in $B\to K^*\ell^+\ell^-$,
$A_{FB}$, that vanishes at a value of the dilepton invariant mass, $q^2$,
independent of form factor models~\cite{burdmanAfb} (near $q_0^2 =
4\,\mbox{GeV}^2$ in the SM, see Fig.~\ref{fig:afb}).  This was shown to follow
from the large energy limit~\cite{charles,SCET1}, as far as the soft
contributions to the form factors are concerned.  One finds the following
implicit equation for $q_0^2$
\begin{equation}
C_9(q_0^2) = - C_7\, {2m_B\, m_b\over q_0^2}\,  
  \bigg[ 1 + {\cal O}\bigg(``\alpha_s", \frac\lqcd{m_b}\bigg) \bigg] \,.
\end{equation}
The quotation marks around the $\alpha_s$ corrections indicate that it is
actually not known yet whether these are formally suppressed compared to the
``leading" terms. The order $\alpha_s$ terms have been
calculated~\cite{BF,BFS}, but reliable estimates of the $\lqcd/E_{K^*}$ terms
are not available yet.  It is hoped that with future theoretical developments
the vanishing of $A_{FB}$ will allow to search for new physics ; $C_7$ is known
from $B\to X_s\gamma$, so the zero of $A_{FB}$ determines $C_9$, which is
sensitive to new physics ($C_{7,9}$ are the effective Wilson coefficients often
denoted by $C_{7,9}^{\rm eff}$, and $C_9$ has a mild $q^2$-dependence).

There has also been considerable progress refining predictions for $B\to
K^*\gamma$ and $\rho\gamma$.  The calculations of ${\cal O}(\alpha_s)$
corrections show a strong enhancement ($\sim 80\%$) of the $B\to K^*\gamma$
rate~\cite{BFS,BoBu}.  The counting of $\alpha_s$ factors is again not firmly
established yet.

The form factors also enter the prediction for the isospin splitting.  These
are power suppressed corrections, but were claimed to be calculable with some
assumptions~\cite{KN}.  The prediction,
\beq
\Delta_{0-} = 
{\Gamma(\B0bar\to\Kbar{}^{*0}\gamma) - \Gamma(B^-\to K^{*-}\gamma) \over
  \mbox{$\Gamma(\B0bar\to\Kbar{}^{*0}\gamma) + \Gamma(B^-\to K^{*-}\gamma)$}}
= {0.3\over T_1^{B\to K^*}} \times \Big(0.08^{+0.02}_{-0.03}\Big) ,
\eeq
is to be compared with the present world average, $0.02 \pm 0.07$.

Testing these predictions is important in their own rights, and may also help
to understand some assumptions entering factorization in charmless nonleptonic
$B$ decay.

\subsubsection{\boldmath Inclusive rare decays}

Rare $B$ decays are sensitive probes of new physics.  There are many
interesting modes sensitive to different extensions of the Standard Model.  For
example, $B\to X_s\gamma$ provides the best bound on the  charged Higgs mass in
type-II two Higgs doublet model, and also constrains the parameter space of
SUSY models.  Other rare decays such as $B\to X\ell^+\ell^-$ are sensitive
through the $bsZ$ effective coupling to SUSY and left-right symmetric models.
$B\to X\nu\bar\nu$ can probe models containing unconstrained couplings between
three 3rd generation fermions~\cite{GLN}.

We learned in the last year that the CKM contributions to rare decays are
probably the dominant ones, as they are for $CP$ violation in $B\to \psi K_S$. 
This is supported by the measurement of ${\cal B}(B\to X_s\gamma)$ which agrees
with the SM at the 15\% level~\cite{CLEObsgmom,babellebsg}; the measurements of
$B\to X_s\ell^+\ell^-$ and $B\to K\ell^+\ell^-$, which are in the ballpark of
the SM expectation~\cite{sll,Kll}; and the non-observation of direct $CP$
violation in $b\to s\gamma$, $A_{CP}(B\to X_s\gamma) = -0.08\pm
0.11$~\cite{acp_bsg} and $A_{CP}(B\to K^*\gamma) = -0.02 \pm
0.05$~\cite{acp_bKg}, which are expected to be tiny in the SM.  These results
make it unlikely that new physics yields order-of-magnitude enhancement of any
rare decay.  It is more likely that only a broad set of precision measurements
will be able to find signals of new physics.

\begin{table}
\begin{center}
\begin{tabular}{ccc} \hline\hline
Decay mode  &  Approximate SM rate  &   Present status \\ \hline\hline
$B\to X_s\gamma$  &  $3.6\times 10^{-4}$ & $(3.4 \pm 0.4)\!\times\! 10^{-4}$ \\
$B\to X_s\nu\bar\nu$  &  $4\times 10^{-5}$  &  $<7.7\times10^{-4}$ \\
$B\to \tau\nu$  &  $4\times 10^{-5}$  &  $<5.7\times10^{-4}$ \\
$B\to X_s \ell^+ \ell^-$  &  $5\times 10^{-6}$  &  $(6 \pm 2)\times10^{-6}$ \\
$B_s\to \tau^+\tau^-$  &  $1\times 10^{-6}$  & \\
$B\to X_s\tau^+\tau^-$  &  $5\times 10^{-7}$  &  \\
$B\to \mu\nu$  &  $2\times 10^{-7}$  &  $<6.5\times10^{-6}$ \\
$B_s\to \mu^+\mu^-$  &  $4\times 10^{-9}$  &  $<2\times10^{-6}$ \\
$B\to \mu^+\mu^-$  &  $1\times 10^{-10}$  &  $<2.8\times10^{-7}$\\ \hline\hline
\end{tabular}
\end{center}
\caption{Some interesting rare decays, their SM rates, and present status.}
\label{tab:rare}
\end{table}

At present, inclusive rare decays are theoretically cleaner than the exclusive
ones, since they are calculable in an OPE and precise multi-loop results exist
(see Ref.~\citex{Hurth} for a recent review).  Table~\ref{tab:rare} summarizes
some of the most interesting modes.  The $b\to d$ rates are expected to be
about a factor of $|V_{td}/V_{ts}|^2 \sim \lambda^2$ smaller than the
corresponding $b\to s$ modes shown.  As a guesstimate, in $b\to q\, l_1 l_2$
decays one expects $10-20\%$ $K^*/\rho$ and $5-10\%$ $K/\pi$.

A source of worry (at least, to me) is the long distance contribution, $B\to
\psi X_s$ followed by $\psi\to \ell^+\ell^-$, which gives a combined branching
ratio ${\cal B}(B\to X_s\ell^+\ell^-) \approx (4 \times 10^{-3}) \times (6
\times 10^{-2}) \approx 2\times 10^{-4}$.  This is about $30$ times the short
distance contribution.  Averaged over a large region of invariant masses (and
$0 < q^2 < m_B^2$ should be large enough), the $c\bar c$ loop is expected to be
dual to $\psi + \psi' + \ldots$.  This is what happens in $e^+e^- \to
\mbox{hadrons}$, in $\tau$ decay, etc., but apparently not here.  Is it then
consistent to ``cut out" the $\psi$ and $\psi'$ regions and then compare the
data with the short distance calculation?  Maybe yes, but our present
understanding is not satisfactory.

\subsection{Summary}

\begin{itemize}

\item $|V_{cb}|$ is known at the $\sim5\%$ level; error may become half of
this in the next few years using both inclusive and exclusive determinations
(latter will rely on lattice).

\item Situation for $|V_{ub}|$ may become similar to present $|V_{cb}|$; for
precise inclusive determination the neutrino reconstruction seems crucial (the
exclusive will use lattice).

\item For $|V_{cb}|$ and $|V_{ub}|$  important to pursue both inclusive
and exclusive measurements.

\item Progress in understanding heavy-to-light form factors in $q^2 \ll m_B^2$
region: $B\to \rho\ell\bar\nu$, $K^{(*)}\gamma$, and $K^{(*)}\ell^+\ell^-$
below the $\psi$ $\Rightarrow$ increase sensitivity to new physics.  Related to
certain questions in factorization in charmless decays.

\end{itemize}

\section{\boldmath Future Clean $CP$ Measurements, Nonleptonic $B$ Decays,
Conclusions}

This last lecture discusses several topics which will play important roles in
the future of $B$ physics.  First, the complications of a clean determination
of the CKM angle $\alpha$ from $B\to \pi\pi$ decays, and how those might be
circumvented.  Then we discuss some future clean $CP$ measurements, such as
$B_s\to D_s K$ and $B\to D K$.  Although some of these measurements are only
doable at a super-$B$-factory and/or LHCb/BTeV, their theoretical cleanliness
makes them important.  The second half of the lecture deals with factorization
in $B\to D^{(*)}X$ type decays and its tests, followed by the different
approaches to factorization in charmless decays and some possible applications.

\subparagraph{ Effective Hamiltonians}

Nonleptonic $B$ decays mediated by $\Delta B = -\Delta C = \pm 1$ transitions
are the simplest hadronic decays, described by the effective Hamiltonian
\beq\label{H1}
H = {4G_F \over\sqrt2}\, V_{cb} V_{uq}^*\, 
  \sum_{i=1}^2 C_i(\mu)\, O_i(\mu) + {\rm h.c.}\,,
\eeq
where $q = s$ or $d$, and
\beq\label{O12}
O_1(\mu) = (\bar q_L^\alpha \gamma_\mu  u_L^\beta) \, 
  (\bar c_L^\beta  \gamma^\mu  b_L^\alpha) \,, \qquad 
O_2(\mu) = (\bar q_L^\alpha \gamma_\mu  u_L^\alpha) \, 
  (\bar c_L^\beta \gamma^\mu  b_L^\beta) \,.
\eeq
Here $\alpha$ and $\beta$ are color indices.  The $\Delta B = \Delta C = \pm
1$ Hamiltonian is related to Eqs.~(\ref{H1})--(\ref{O12}) by the trivial $c
\leftrightarrow u$ interchange.

Decays with  $\Delta B = \pm 1$ and $\Delta C = 0$ are more complicated,
\beq\label{H2}
H = {4G_F \over\sqrt2} \sum_{j=u,c}\! V_{jb} V_{jq}^*\, 
  \sum_i C_i(\mu)\, O_i(\mu) + {\rm h.c.} \,,
\eeq
The $C_i$ are calculable Wilson coefficients, known to high precision.  To
write Eq.~(\ref{H2}), the unitarity relation $V_{tb}V_{tq}^* = - V_{cb}V_{cq}^*
- V_{ub}V_{uq}^*$ is used to rewrite the CKM elements that occur in penguin
diagrams with intermediate top quark in terms of the CKM elements that occur in
tree diagrams.  The operator basis is conventionally chosen as
\begin{equation}\label{basis}
\begin{array}{rclrcl}
O_1^j &=& (\bar q_L^\alpha \gamma_\mu  j_L^\beta) \, 
  (\bar j_L^\beta  \gamma^\mu  b_L^\alpha) \,, \qquad &
O_2^j &=& (\bar q_L^\alpha \gamma_\mu  j_L^\alpha) \, 
  (\bar j_L^\beta \gamma^\mu  b_L^\beta) \,, \\[4pt]
O_3 &=& \ds \bar q_L^\alpha \gamma_\mu  b_L^\alpha \, 
  \sum_{q'} \bar q_L^{\prime\beta} \gamma^\mu  q_L^{\prime\beta} \,, \qquad & 
O_4 &=& \ds \bar q_L^\alpha \gamma_\mu  b_L^\beta \, 
  \sum_{q'} \bar q_L^{\prime\beta} \gamma^\mu  q_L^{\prime\alpha} \,, \\
O_5 &=& \ds \bar q_L^\alpha \gamma_\mu  b_L^\alpha \, 
  \sum_{q'} \bar q_R^{\prime\beta} \gamma^\mu  q_R^{\prime\beta} \,, \qquad &
O_6 &=& \ds \bar q_L^\alpha \gamma_\mu  b_L^\beta \, 
  \sum_{q'} \bar q_R^{\prime\beta} \gamma^\mu  q_R^{\prime\alpha} \,, \\
O_8 &=& \ds - \frac{g}{16 \pi^2} \, m_b \, 
    \bar q_L \sigma^{\mu \nu} G_{\mu \nu} b_R \,,
\end{array}
\end{equation}
where $j = c$ or $u$, and the sums run over $q' = \{u,d,s,c,b\}$.  In $O_8$,
$G_{\mu\nu}$ is the chromomagnetic field strength tensor.  Usually $O_1$ and
$O_2$ are called current-current operators, $O_3-O_6$ are four-quark penguin
operators, and $O_8$ is the chromomagnetic penguin operator.  These operators
arise at lowest order in the electroweak interaction, i.e., diagrams involving
a single $W$ boson and QCD corrections to it.  In some cases, especially when
isospin breaking plays a role, one also needs to consider penguin diagrams
which are second order in $\alpha_{\rm ew}$.  They give rise to the electroweak
penguin operators,
\begin{equation}\label{basis2}
\begin{array}{rclrcl}
O_7 &=& \ds - {e \over 16\pi^2}\, m_b\, 
  \bar q_L^\alpha \, \sigma^{\mu\nu} F_{\mu\nu}\, b_R^\alpha \,, \\[4pt]
O_7^{\rm ew} &=& \ds \frac32\, \bar q_L^\alpha \gamma_\mu b_L^\alpha \, 
  \sum_{q'} e_{q'}\, \bar q_R^{\prime\beta} \gamma^\mu q_R^{\prime\beta} \,, 
  \qquad&
O_8^{\rm ew} &=& \ds \frac32\, \bar q_L^\alpha \gamma_\mu b_L^\beta \, 
  \sum_{q'} e_{q'}\, \bar q_R^{\prime\beta} \gamma^\mu q_R^{\prime\alpha} \,,\\
O_9^{\rm ew} &=& \ds \frac32\, \bar q_L^\alpha \gamma_\mu b_L^\alpha \, 
  \sum_{q'} e_{q'}\, \bar q_L^{\prime\beta} \gamma^\mu q_L^{\prime\beta} \,, 
  \qquad&
O_{10}^{\rm ew} &=& \ds \frac32\, \bar q_L^\alpha \gamma_\mu b_L^\beta \,
  \sum_{q'} e_{q'}\, \bar q_L^{\prime\beta} \gamma^\mu  q_L^{\prime\alpha} \,.
\end{array}
\end{equation}
Here $F^{\mu\nu}$ is the electromagnetic field strength tensor, and
$e_{q'}$ denotes the electric charge of the quark $q'$.  

Sometimes the contributions to decay amplitudes are classified by the
appearance of Feynman diagrams with propagating top quarks, $W$ and $Z$ bosons,
and people talk about tree (T), color-suppressed tree (C), penguin (P), and
weak annihilation or $W$-exchange (W) contributions.  While this may be
convenient in some cases, the resulting arguments can be misleading.  The
separation between these contributions is usually ambiguous, as the ``tree" and
``penguin" operators mix under the renormalization group.  At the scale $m_b$
the physics relevant for weak decays is described by the operators in
Eqs.~(\ref{O12}), (\ref{basis}), and (\ref{basis2}), and their Wilson
coefficients, and there are no propagating heavy particles.  Usually one calls
the $O_1$ and $O_2$ contributions (plus possibly a part of $O_3 - O_6$ and
$O_8$) ``tree", while $O_3 - O_6$ and $O_8$ (plus possibly a part of $O_1 -
O_2$) ``penguin".  Below we will try to state clearly what is meant in each
case.

\subsection[$B\to\pi\pi$ --- beware of penguins]{\boldmath $B\to\pi\pi$ ---
beware of penguins}

\begin{figure}[t]
\centerline{\includegraphics*[width=.3\textwidth]{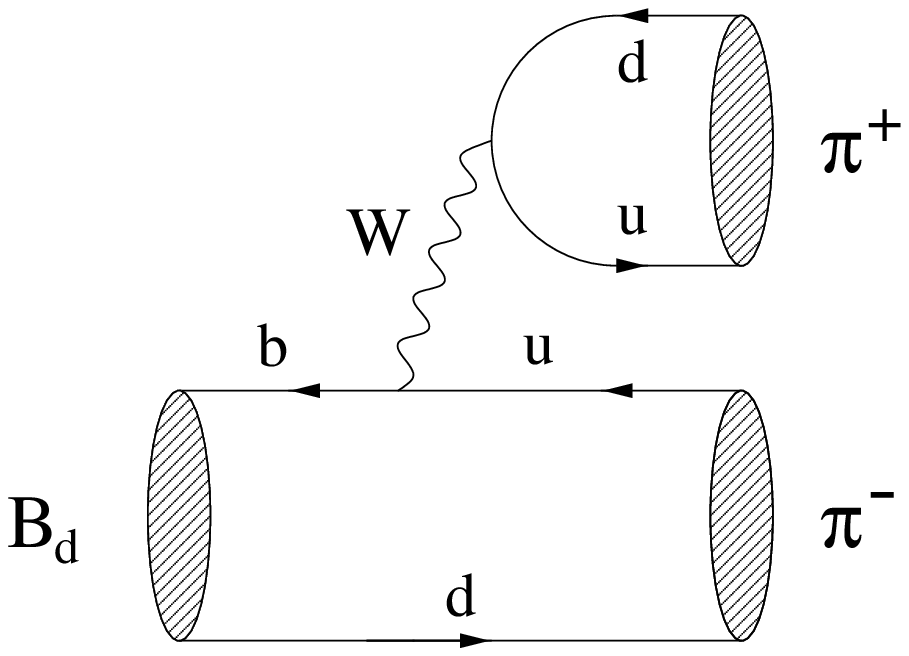}\hspace{1cm}
\includegraphics*[width=.3\textwidth]{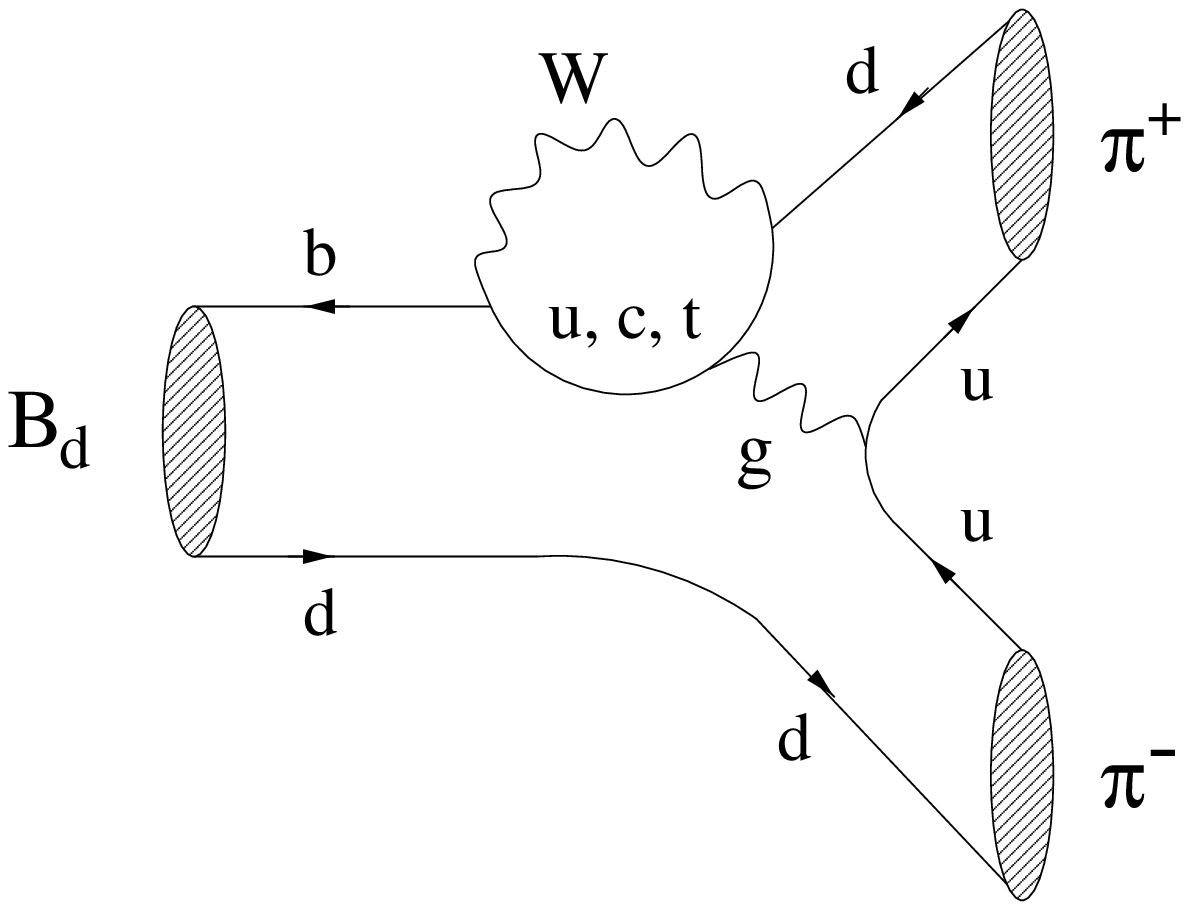}}
\caption{``Tree" (left) and ``Penguin" (right) contributions to $B\to \pi\pi$
(from Ref.~\protect\citex{rf}).}
\label{fig:Bpipi}
\end{figure}

We saw in Sec.~\ref{sec:s2b} that the $CP$ asymmetry in $B\to \psi K_S$ gives a
theoretically very clean determination of $\sin2\beta$, because the amplitude
is dominated by contributions with a single weak phase.  Similar to that case,
there are tree and penguin contributions to the $B\to \pi^+\pi^-$ amplitude as
well, as shown in Fig.~\ref{fig:Bpipi}.  The tree contribution comes from $b\to
u\bar u d$ transition, while there are penguin contributions with three
different CKM combinations
\beq
\ov A_T = V_{ub} V_{ud}^*\, T_{u\bar ud}\,, \qquad
  \ov A_P = V_{tb} V_{td}^*\, P_t + V_{cb} V_{cd}^*\, P_c 
  + V_{ub} V_{ud}^*\, P_u\,.
\eeq
The convention is to rewrite the penguin contributions in terms of $V_{ub}
V_{ud}^*$ and $V_{tb} V_{td}^*$ [instead of $V_{cb} V_{cd}^*$, as in
Eq.~(\ref{H2})] using CKM unitarity as
\beqa\label{Bpipiamp}
\ov A &=& V_{ub} V_{ud}^*\, (T_{u\bar ud} + P_u - P_c)
  + V_{tb} V_{td}^*\, (P_t - P_c) \nn\\
&\equiv& V_{ub} V_{ud}^*\, T + V_{tb} V_{td}^*\, P \,.
\eeqa
where the second line defines $T$ and $P$.  If the penguin contribution was
small, then the $CP$ asymmetry in $B\to \pi^+ \pi^-$ would measure ${\rm Im}
\lambda_{\pi\pi}^{\rm (tree)} = \sin2\alpha$, since
\beq\label{Bpipilam}
\lambda_{\pi\pi}^{\rm (tree)}
= \bigg( { V_{tb}^* V_{td} \over V_{tb} V_{td}^*} \bigg)
  \bigg( {V_{ub} V_{ud}^* \over V_{ub}^* V_{ud}} \bigg)
= e^{2i\alpha} \,.
\eeq
The first term is the SM value of $q/p$ in $B_d$ mixing and the second one is
$\ov A_T/A_T$.

The crucial new complication compared to $B\to \psi K_S$ is that the CKM
elements multiplying both contributions in Eq.~(\ref{Bpipiamp}) are of order
$\lambda^3$, and so $|(V_{tb} V_{td}^*) / (V_{ub} V_{ud}^*)| = {\cal O}(1)$,
whereas the analogous ratio in $B\to \psi K_S$ in Eq.~(\ref{BpsiKamp}) was
$|(V_{ub} V_{us}^*)/(V_{cb} V_{cs}^*)| \simeq 1/50$.  Therefore, we do not know
whether amplitudes with one weak phase dominate, and our inability to model
independently compute $P/T$ results in a sizable uncertainty in the relation
between ${\rm Im} \lambda_{\pi\pi}$ and $\sin2\alpha$.  If there are two
comparable amplitudes with different weak and strong phases, then sizable $CP$
violation in the $B\to \pi^+ \pi^-$ decay is possible in addition to that in
the interference between mixing and decay.

Present estimates of $|P/T|$ are around $0.2-0.4$.  The large $B\to K\pi$ decay
rate, which is probably dominated by the $b\to s$ penguin amplitudes, implies
the crude estimate $|P/T| \sim \lambda\, \sqrt{ {\cal B} (B\to K\pi) / {\cal B}
(B\to \pi\pi)} \sim 0.3$, i.e., $|P/T| \not\ll 1$.  The BABAR and BELLE
measurements~\cite{babelleaeff} do not yet show a consistent picture.  BELLE
measured a large value for $C_{\pi\pi}$ [see the definition in
Eq.~(\ref{SCdef})], while the BABAR result is consistent with zero.  If
$C_{\pi\pi}$ is sizable, that implies model independently that $|P/T|$ cannot
be small.  However, if $C_{\pi\pi}$ is small, that may be due to a small strong
phase between the $P$ and $T$ amplitudes and does not imply model independently
that $|P/T|$ is small, nor that $S_{\pi\pi}$ is close to $\sin2\alpha$.  The
central value of the BELLE measurement indicates that both the magnitude and
phase of $P/T$ has to be large, whereas the BABAR central value is consistent
with a modest $|P/T|$.

There are two possible ways to deal with a non-negligible penguin contribution:
(i)~eliminate $P$ (see the next section); or (ii) attempt to calculate $P$ (see
Sec.~\ref{sec:factnoc}).

\subsubsection{\boldmath Isospin analysis}
\label{sec:pipiispin}

Isospin is an approximate, global $SU(2)$ symmetry of the strong interactions,
violated by effects of order $(m_d-m_u)/(4\pi f_\pi) \sim 1\%$.  It allows the
separation of tree and penguin contributions~\cite{pipi}.  Let's see how this
works.  The $(u,d)$ quarks and  the $(\bar d, \bar u)$ antiquarks each form an
isospin doublet, while all other (anti)quarks are singlets under $SU(2)$
isospin.  Gluons couple equally to all quarks so they are also singlets.  The
$\gamma$ and the $Z$ are mixtures of $I = 0$ and $1$, as they have unequal
couplings to $u\bar u$ and $d\bar d$.

The transformation of $B$ mesons are determined by their flavor quantum
numbers, i.e., $(\B0bar,B^-)$ form an $I = \frac12$ doublet.  The pions form
an  $I=1$ triplet.  Since the $B$ meson and the pions are spinless particles,
the pions in $B\to \pi\pi$ decay must be in a state with zero angular
momentum.  Because of Bose statistics, the pions have to be in an even isospin
state.  While  $|\pi^0\pi^0\rangle$ is manifestly symmetric, when writing
$|\pi^+\pi^-\rangle$ and $|\pi^0\pi^-\rangle$ what is actually meant is the
symmetrized combinations $(|\pi^+ \pi^- \rangle  + |\pi^- \pi^+ \rangle) /
\sqrt2$ and $(|\pi^0 \pi^- \rangle  + |\pi^- \pi^0 \rangle) / \sqrt2$,
respectively. The isospin decompositions of $|\pi^0\pi^0\rangle$ and
$|\pi^+\pi^-\rangle$ were given in Eq.~(\ref{ispinpipi0}), so we only need in
addition
\beq\label{ispinpipi1}
|\pi^0 \pi^- \rangle = |(\pi \pi)_{I=2}\rangle \,.
\eeq

The $b\to u \bar u d$ Hamiltonian is a mixture of $I = \frac12$ and $\frac32$.
More precisely, it has $|I,I_z\rangle = |\frac12, -\frac12\rangle$ and 
$|\frac32, -\frac12\rangle$ pieces, which can only contribute to the $I=0$ and
$I=2$ final states, respectively.  The crucial point is that the penguin
operators ($O_3 - O_6$ and $O_8$) only contribute to the $|I,I_z\rangle =
|\frac12, -\frac12\rangle$ part of the Hamiltonian, because the gluon is
isosinglet (these operators involve a flavor sum, $\sum\! q'\bar q'$).  If we
can (effectively) isolate $CP$ violation in the $I = 2$ final state then the
resulting asymmetry would determine $\sin 2\alpha$.  However, electroweak
penguin operators [$O_7$ and $O_7^{\rm ew} - O_{10}^{\rm ew}$ in
Eq.~(\ref{basis2})] contribute to both $I = \frac12$ and $\frac32$ pieces of
the Hamiltonian, and their effects cannot be separated from the tree
contributions via the isospin analysis.

Besides the decomposition of the $\pi\pi$ final state in
Eqs.~(\ref{ispinpipi0}) and (\ref{ispinpipi1}), we also have to consider the
combination of the $\B0bar$ and $B^-$ with the Hamiltonian, where another
Clebsch-Gordan coefficient enters.  The $I = \frac12$ part of the Hamiltonian
only contributes to $\B0bar$ decay.  However, the $I = \frac32$ part has
different matrix elements in $\B0bar$ and $B^-$ decay: $\langle\B0bar | H_{I =
3/2} | (\pi\pi)_{I=2}\rangle = (1/\sqrt2) {\cal A}_2$, while $\langle B^- |
H_{I = 3/2} | (\pi\pi)_{I=2}\rangle = (\sqrt3/2) {\cal A}_2$.  Thus the $A_2
\equiv  {\cal A}_2/\sqrt2$ amplitude in $\B0bar$ decay has to be multiplied by
$\sqrt{3/2}$ to get the relative normalization of the $B^-$ amplitude right. 
We thus obtain
\beqa\label{Apipi}
\ov A{}^{00} \equiv A(\B0bar\to \pi^0 \pi^0) &=& - \sqrt{\frac13} \, A_0
  + \sqrt{\frac23} \, A_2 \,, \nn\\*
\ov A{}^{+-} \equiv A(\B0bar\to \pi^+ \pi^-) &=& \sqrt{\frac23} \, A_0
  + \sqrt{\frac13} \, A_2 \,, \nn\\*
\ov A{}^{0-} \equiv A(B^-\to \pi^0 \pi^-) &=& \sqrt{\frac32}\, A_2 \,.
\eeqa
This implies the triangle relation:
\beq\label{tri1}
{1\over \sqrt2}\, \ov A{}^{+-} + \ov A{}^{00} = \ov A{}^{0-}\,.
\eeq
Similar isospin decompositions hold for $B^0$ and $B^+$ decays, yielding
another triangle relation
\beq\label{tri2}
{1\over \sqrt2}\, A^{+-} + A^{00} = A^{0+}\,.
\eeq
Since only a single isospin amplitude contributes to $\ov A{}^{0-}$ and
$A^{0+}$, we have $|\ov A{}^{0-}| = |A^{0+}|$ (however, in general, $|\ov
A{}^{+-}| \neq |A^{+-}|$ and $|\ov A{}^{00}| \neq |A^{00}|$).  So one can
superimpose the two triangles by introducing $\widetilde A^{ij} \equiv
e^{-2i\phi_{\rm T}} \ov A{}^{ij}$, where $\phi_T = {\rm arg}(V_{ub} V_{ud}^*)$.

\begin{figure}[t]
\centerline{\includegraphics*[width=.6\textwidth]{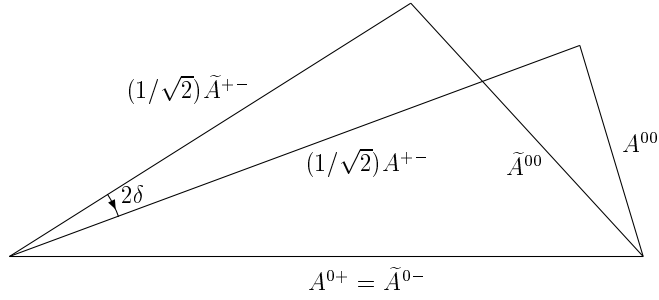}}
\caption{The isospin triangles of Eqs.~(\ref{tri1}) and (\ref{tri2}).}
\label{fig:pipitriangle}
\end{figure}

Measuring the six decay rates entering Eqs.~(\ref{tri1}) and (\ref{tri2})
allows the construction of the two triangles shown in
Fig.~\ref{fig:pipitriangle}.  Measuring in addition the time dependent $CP$
asymmetry in $B\to \pi^+\pi^-$ determines
\beq
{\rm Im}\, \lambda_{\pi^+\pi^-} = {\rm Im} \bigg(e^{2i\alpha}\,
  {\widetilde A^{+-}\over A^{+-}} \bigg)
  = {\rm Im}\, e^{2i(\alpha+\delta)} \,.
\eeq
Since $\delta$, the strong phase difference between $\ov A{}^{+-}$ and
$A^{+-}$, is known from the construction in Fig.~\ref{fig:pipitriangle}, this
provides a theoretically clean determination of the CKM angle $\alpha$. 
Probably the dominant remaining theoretical uncertainty is due to electroweak
penguins mentioned above, that cannot be eliminated with the isospin analysis.
This has been estimated to give a $\,\lsim 5\%$ uncertainty~\cite{babook}.
There is also a four-fold discrete ambiguity in $\delta$ corresponding to
reflections of each of the two triangles along the $A^{0+}$ side.

A similar analysis is also possible in $B\to \rho\pi$ decays.  A complication
is that the final state contains non-identical particles, so it can have $I=0$,
$1$, and $2$ pieces.  Then there are four amplitudes, and one obtains pentagon
relations~\cite{rhopi} instead of the $B\to \pi\pi$ triangle relations.  It may
be experimentally more feasible to do a Dalitz plot analysis that allows in
principle to eliminate the hadronic uncertainties due to the QCD penguin
contributions by considering only the $\pi^+\pi^-\pi^0$ final
state~\cite{rhopidalitz}.

\subsection{Some future clean measurements}

We discuss below a few theoretically clean measurements that may play important
roles in overconstraining the CKM picture (in addition to $B\to \phi K_S$
discussed in Sec.~\ref{sec:phiK}, and $B\to \pi\pi$ [$\rho\pi$] with isospin
[Dalitz plot] analysis discussed above).  These also indicate the
complementarity between high statistics $e^+ e^-$ and hadronic $B$ factories.

\subsubsection[$B_s\to \psi\phi$ and $B_s\to \psi\eta^{(\prime)}$]{\boldmath
$B_s\to \psi\phi$ and $B_s\to \psi\eta^{(\prime)}$}

\begin{figure}[t]
\centerline{\includegraphics*[width=0.55\textwidth]{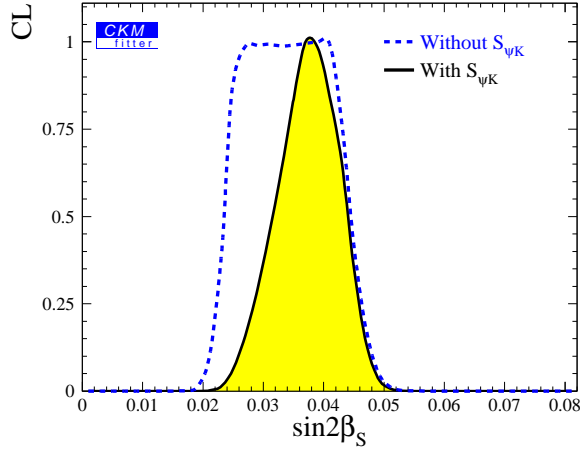}}
\caption{Confidence levels of $\sin 2\beta_s$ in the SM with and without
including the constraint from the $CP$ asymmetry in $B\to\psi K_S$ (from
Ref.~\protect\citex{llnp}).}
\label{fig:sin2bs}
\end{figure}

Similar to $B\to \psi K_{S,L}$, the $CP$ asymmetry in $B_s\to \psi\phi$
measures the phase difference between $B_s$ mixing and $b\to c\bar cs$ decay,
$\beta_s$, in a theoretically clean way.  The greater than 10\% CL range of
$\sin2\beta_s$ in the SM is~\cite{llnp} $0.026 < \sin2\beta_s < 0.048$ (see
Fig.~\ref{fig:sin2bs}).

The $\psi\phi$ final state is not a pure $CP$ eigenstates, but it has $CP$ self
conjugate particle content and can be decomposed into $CP$-even and odd partial
waves.  An angular analysis can separate the various components, and may
provide theoretically clean information on $\beta_s$.  Even before this can be
done, one can search for new physics, since the asymmetry measured without the
angular analysis can only be smaller in magnitude than $\sin2\beta_s$.  If
$\alpha^2$ is the $CP$-even fraction of the $\psi\phi$ final state (i.e.,
$|\psi\phi\rangle = \alpha\, | CP=+ \rangle + \sqrt{1-\alpha^2}\, | CP=-
\rangle$), then $S_{\psi\phi} = (2\alpha^2 - 1) \sin2\beta_s$.  Thus, the
observation of a large asymmetry would be a clear signature of new physics. 

The advantage of $B_s\to \psi\eta^{(\prime)}$ compared to $B_s\to \psi\phi$ is
that the final states are pure $CP$-even.  BTeV will be well-suited to measure
the $CP$ asymmetries in such modes.

\subsubsection[$B_s\to D_s^\pm K^\mp$ and $B_d\to D^{(*)\pm}
\pi^\mp$]{\boldmath $B_s\to D_s^\pm K^\mp$ and $B_d\to D^{(*)\pm} \pi^\mp$}
\label{sec:DsK}

In certain decays to final states which are not $CP$ eigenstates, it is still
possible to extract weak phases model independently from the interference
between mixing and decay.  This occurs if both $B^0$ and $\B0bar$ can decay
into a final state and its $CP$ conjugate, but there is only one contribution
to each decay amplitude.  In such a case no assumption about hadronic physics
is needed, even though $|\ov{A}_f/A_f| \neq 1$ and
$|\ov{A}_{\ov{f}}/A_{\ov{f}}| \neq 1$.

An important decay of this type is $B_s\to D_s^\pm K^\mp$, which allows a model
independent determination of the angle $\gamma$~\cite{ADK}.  Both $\B0bar_s$
and $B_s^0$ can decay to $D_s^+ K^-$ and $D_s^- K^+$, but there is  only one
amplitude in each decay corresponding to the tree level $b\to c\bar u s$ and
$b\to u \bar c s$ transitions, and their $CP$ conjugates.  There are no penguin
contributions to these decays.  One can easily see that
\beq\label{1:A12ratio}
{\ov{A}_{D_s^+ K^-} \over A_{D_s^+ K^-} } = \frac{A_1}{A_2}
  \bigg( {V_{cb} V_{us}^* \over V_{ub}^* V_{cs}} \bigg) \,, \qquad
{\ov{A}_{D_s^- K^+} \over A_{D_s^- K^+} } = \frac{A_2}{A_1}
  \bigg( {V_{ub} V_{cs}^* \over V_{cb}^* V_{us}} \bigg) \,,
\eeq
where the ratio of hadronic amplitudes, $A_1/A_2$, includes the strong (but not
the weak) phases, and is an unknown complex number that is expected to be of
order unity.  It is important for the feasibility of this method that $|V_{cb}
V_{us}|$ and $|V_{ub} V_{cs}|$ are both of order $\lambda^3$, and so are
comparable in magnitude.  Measuring the four time dependent decay rates
determine $\lambda_{D_s^+K^-}$ and $\lambda_{D_s^-K^+}$.  The ratio of unknown
hadronic amplitudes, $A_1/A_2$, drops out from their product,
\beq
\lambda_{D_s^+ K^-}\, \lambda_{D_s^- K^+} = 
  \bigg( {V_{tb}^* V_{ts} \over V_{tb} V_{ts}^*} \bigg)^{\!2}
  \bigg( {V_{cb} V_{us}^* \over V_{ub}^* V_{cs}} \bigg)
  \bigg( {V_{ub} V_{cs}^* \over V_{cb}^* V_{us}} \bigg)
= e^{-2i(\gamma-2\beta_s-\beta_K)}\,.
\eeq
The first factor is the Standard Model value of $q/p$ in $B_s$ mixing.  The
angles $\beta_s$ and $\beta_K$ defined in Eq.~(\ref{angledef}) occur in
``squashed" unitarity triangles; $\beta_s$ is of order $\lambda^2$ and
$\beta_K$ is of order $\lambda^4$.  Thus we can get a theoretically clean
measurement of $\gamma-2\beta_s$.

In analogy with the above, the time dependent $B_d\to D^{(*)\pm} \pi^\mp$ rates
may be used to measure $\gamma + 2\beta$, since $\lambda_{D^+\pi^-}\,
\lambda_{D^-\pi^+} = \exp\, [-2i(\gamma + 2\beta)]$.  In this case, however,
the ratio of the two decay amplitudes is of order $\lambda^2$, and therefore
the $CP$ asymmetries are expected to be much smaller, at the percent level,
making this measurement in $B_d$ decays rather challenging.

\subsubsection[$B^\pm \to (D^0, \D0bar) K^\pm$ and $\gamma$]{\boldmath $B^\pm
\to (D^0, \D0bar) K^\pm$ and $\gamma$}

\begin{figure}[t]
\centerline{\includegraphics*[width=0.8\textwidth]{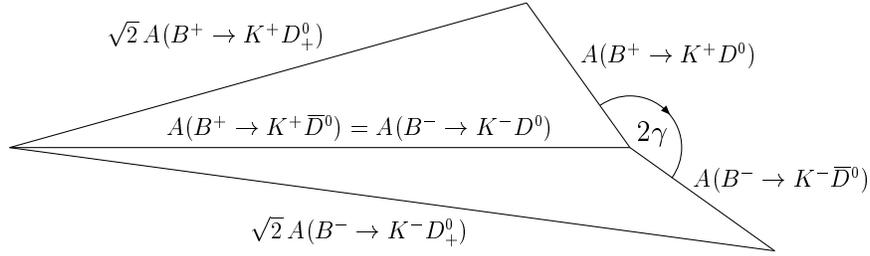}}
\caption{Relations between $B^\pm \to D K^\pm$ amplitudes that allow
determination of $\gamma$.}
\label{fig:gw}
\end{figure}

Some of the theoretically cleanest determinations of the weak phase $\gamma$
rely on $B\to D K$ and related decays.  The original idea of Gronau and Wyler
was to measure two rates arising from $b \to c \bar u s$ and $b \to u \bar c s$
amplitudes, and a third one that involves~their interference~\cite{GW}.  Thus
one can gain sensitivity to the weak phase between the two amplitudes, which is
$\gamma$ in the usual phase convention.  Assuming that there is no $CP$
violation in the $D$ sector (which is a very good approximation in the SM), and
defining the $CP$-even and odd states as
\beq
|D_\pm^0\rangle = \frac1{\sqrt2}\, \Big( |D^0 \rangle \pm |\D0bar\rangle \Big),
\eeq
imply the following amplitude relations,
\beqa\label{gwtri}
\sqrt2\, A(B^+\to K^+ D^0_+) &=& A(B^+\to K^+ D^0) + A(B^+\to K^+ \D0bar)\,,
  \nn\\
\sqrt2\, A(B^-\to K^- D^0_+) &=& A(B^-\to K^- D^0) + A(B^-\to K^- \D0bar) \,.
\eeqa
In the first relation, for example, $B^+\to K^+ \D0bar$ is a $b\to c$
transition, $B^+\to K^+ D^0$ is a $b\to u$ transition, and $B^+\to K^+ D^0_+$
receives contributions from both.  Then the triangle construction in
Fig.~\ref{fig:gw} determines the weak phase between the $\bar b\to \bar u$ and
$b\to u$ transitions, which is $2\gamma$ (in the usual phase convention). 
There is again a four-fold discrete ambiguity corresponding to the reflections
of the triangles.  Since all the quarks which appear in $B\to D K$ decays have
distinct flavors, the theoretical uncertainty arises only from higher order
weak interaction effects (including, possibly, $D-\Dbar$ mixing).  There are
again no penguin contributions, as in Sec.~\ref{sec:DsK}.

In practice there are significant problems in the application of this method.  
Although the amplitudes in Eq.~(\ref{gwtri}) are the same order in the
Wolfenstein parameter, the triangles in Fig.~\ref{fig:gw} are expected to be
squashed because $|V_{ub} / V_{cb}| < \lambda$ and the $B^+\to K^+ D^0$ decay
is color suppressed.  The ``long" sides of the triangles have been measured,
including reconstruction of the $D$ in $CP$ eigenstates~\cite{DCP-K}.  The 
amplitude ratio is estimated based on naive factorization as
\beq\label{naive}
{|A(B^+\to K^+D^0)|\over |A(B^+\to K^+ \D0bar| }
  \sim \left|{V_{ub} V_{cs}^*\over V_{cb} V_{us}^*}\right| \frac1{N_c}
  \sim 0.15 \,,
\eeq
where $N_C = 3$ is the number of colors.  As a result, the measurement of
$|A(B^+\to K^+D^0)|$ using hadronic $D$ decays is hampered by a significant
contribution from the decay $B^+\to K^+ \D0bar$, followed by a doubly
Cabibbo-suppressed decay of the $\D0bar$.  

This problem can be avoided by making use of large final state interactions in
$D$ decays.  One can consider common final states of $D^0$ and $\D0bar$ decay,
such that~\cite{ADS}
\beqa
B^+ &\to& K^+ \D0bar \to K^+ f_i \,, \qquad \D0bar \to K^+ f_i\
  \mbox{ doubly Cabibbo-suppressed}\,, \nn\\
B^+ &\to& K^+ D^0 \to K^+ f_i \,, \qquad D^0 \to K^+ f_i\
  \mbox{ Cabibbo-allowed}\,,
\eeqa
which reduces the difference of the magnitudes of the two interfering
amplitudes.  By using at least two final states (e.g., $f_1 = K^-\pi^+$ and
$f_2 = K^-\rho^+$) one can determine all strong phases directly from the
analysis~\cite{ADS}.

It may be advantageous, especially if the amplitude ratio in Eq.~(\ref{naive})
is not smaller than its naive estimate, to consider only singly
Cabibbo-suppressed $D$ decays~\cite{GLS}.  In this case the two final states
can be $K^\pm K^{*\mp}$, corresponding to simply flipping the charge
assignments, because the $D^0 \to K^+ K^{*-}$ and $D^0 \to K^- K^{*+}$ rates
differ significantly.  This measurement is less sensitive to $D^0-\D0bar$
mixing than considering doubly Cabibbo-suppressed $D$ decays~\cite{GLS}. 
Moreover, all the modes that need to be measured for this method are accessible
in the present data sets.

\subsection[Factorization in $b\to c$ decay]{\boldmath Factorization in $b\to
c$ decay}\label{sec:bcfact}

Until recently little was known model independently about exclusive nonleptonic
$B$ decays.  Crudely speaking, factorization is the hypothesis that, starting
from the effective nonleptonic Hamiltonian, one can estimate matrix elements of
four-quark operators by grouping the quark fields into a pair that can mediate
$B\to M_1$ decay ($M_1$ inherits the spectator quark from the $B$), and another
pair that can describe $\mbox{vacuum} \to M_2$ transition.  For $M_1 = D^{(*)}$
and $M_2 = \pi$, this amounts to the assumption that the contributions of
gluons between the pion and the heavy mesons are either calculable
perturbatively or are suppressed by $\lqcd/m_Q$.

\begin{figure}[t]
\centerline{\includegraphics*[width=0.37\textwidth]{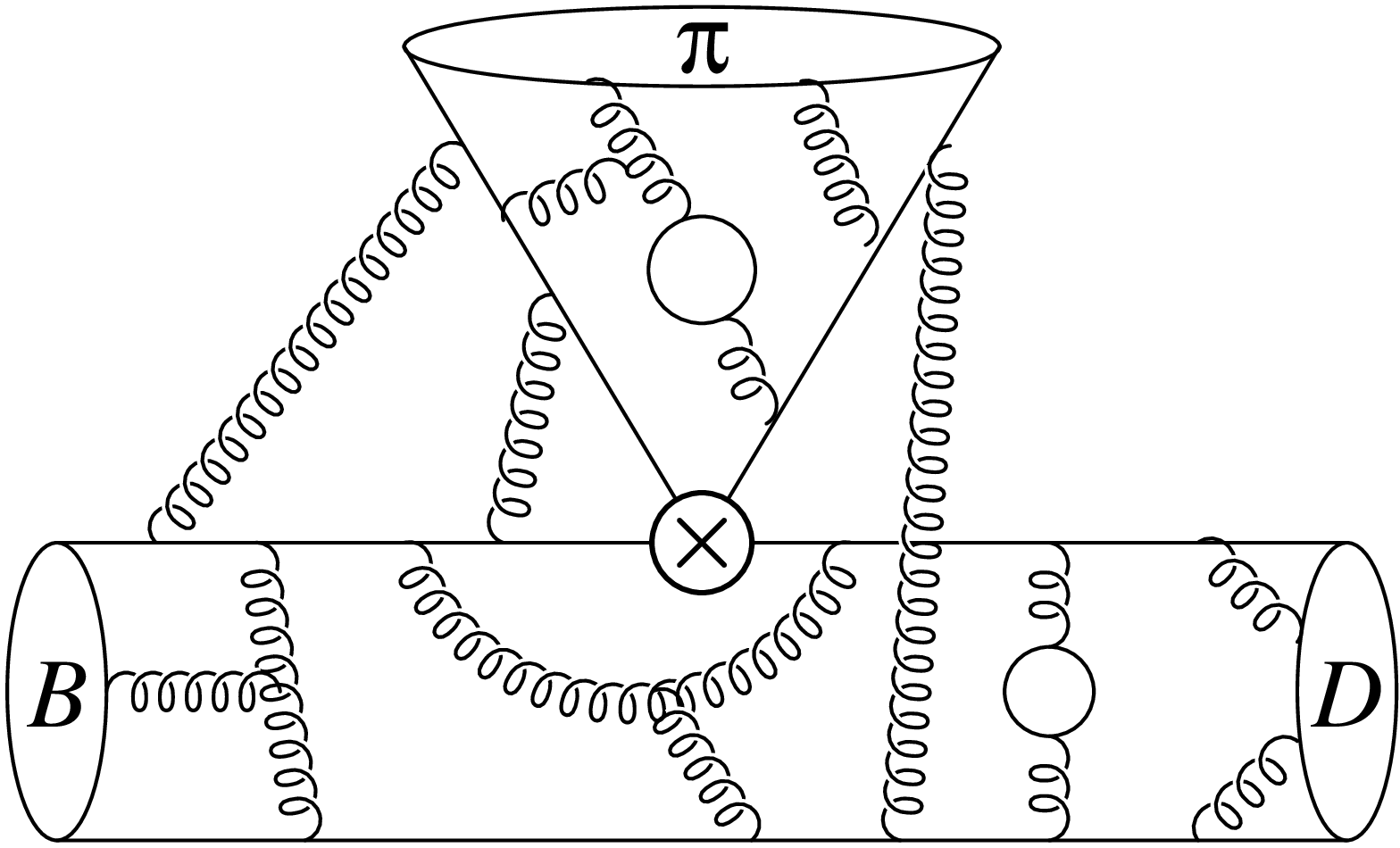}\hspace*{1cm}
\includegraphics*[width=0.37\textwidth]{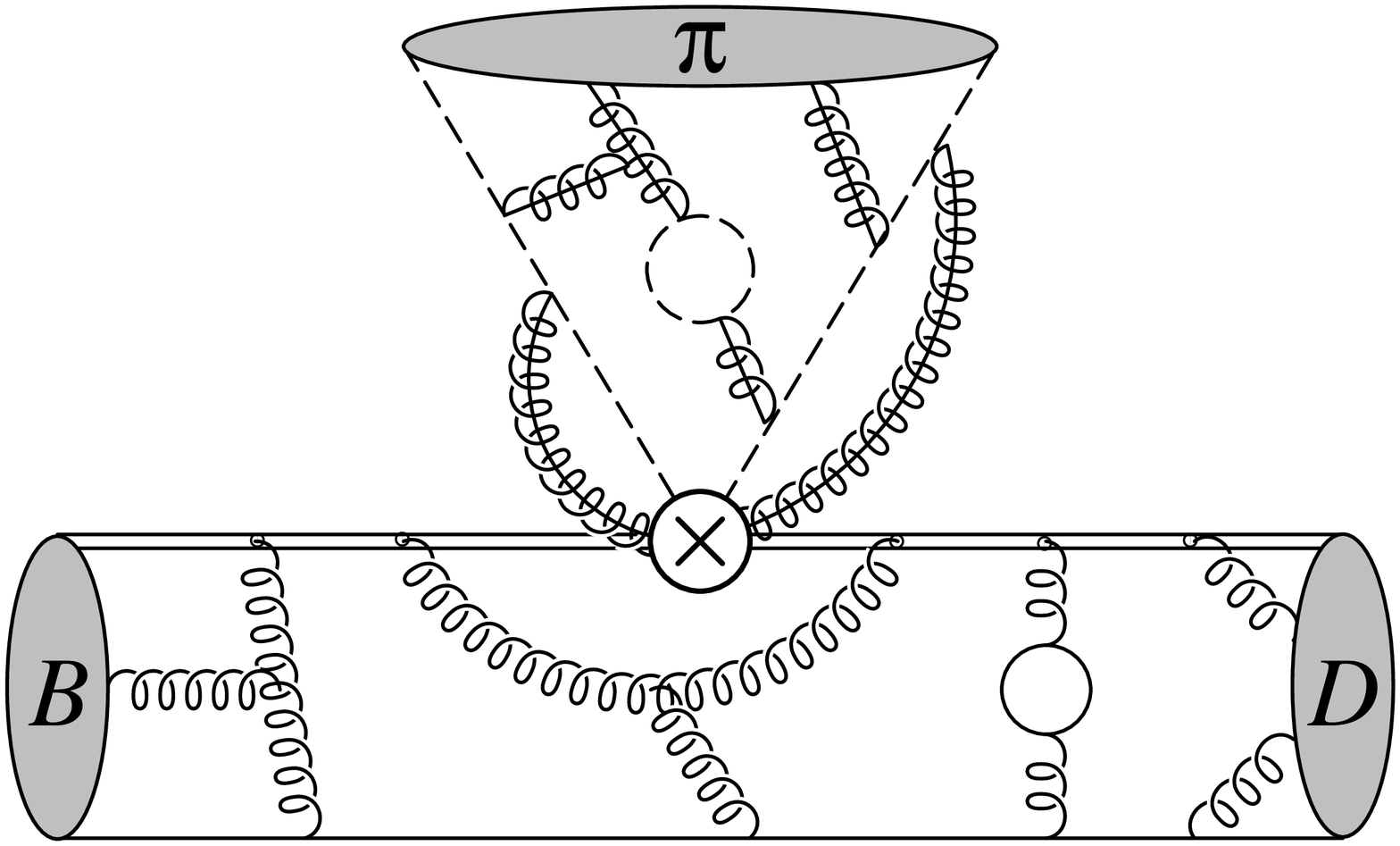}}
\caption{Illustration of factorization in $B\to D\pi$.  Left: typical diagram
in full QCD.  Right: typical diagram in SCET at leading order in $\lqcd/m_Q$. 
The $\otimes$ denotes the weak Hamiltonian, double lines are heavy quarks,
gluons with a line through them are collinear.  
(From Ref.~\protect\citex{BPSdpi}.)}
\label{fig:factplot}
\end{figure}

It has long been known that if $M_1$ is heavy and $M_2$ is light, such as
$\B0bar\to D^{(*)+}\pi^-$, then ``color transparency" may justify
factorization~\cite{BJfact,DGfact,PWfact}.  The physical picture is that the
two quarks forming the $\pi$ must emerge from the weak decay in a small
(compared to $\lqcd^{-1}$) color dipole state rapidly moving away from the $D$
meson.  At the same time the wave function of the brown muck in the heavy
meson only has to change moderately, since the recoil of the $D$ is small. 
While the order $\alpha_s$ corrections were calculated a decade
ago~\cite{PWfact}, it was only shown recently, first to 2-loops~\cite{BBNS} and
then to all orders in perturbation theory,~\cite{BPSdpi} that in such decays
factorization is the leading result in a systematic expansion in powers of
$\alpha_s(m_Q)$ and $\lqcd/m_Q$.  The factorization formula for $\B0bar \to
D^{(*)+}\pi^-$ and $B^- \to D^{(*)0}\pi^-$ decay is~\cite{BBNS,BPSdpi}
\beq\label{BDpifact}
\langle D^{(*)}\pi |\, O_i(\mu_0)\, | B\rangle 
  = iN_{(*)}\, F_{B\to D^{(*)}}\, f_\pi \int_0^1\! \d x\, 
  T(x,\mu_0,\mu)\, \phi_\pi(x,\mu) \,.
\eeq
where $O_{1,2}$ are the color singlet and octet operators in Eq.~(\ref{O12})
that occur in the effective Hamiltonian in Eq.~(\ref{H1}).  Diagrams such as
the one in Fig.~\ref{fig:factplot} on the left give  contributions suppressed
by $\alpha_s$ or $\lqcd/m_b$, and the leading contributions (in $\lqcd/m_b$)
come only from diagrams such as the one in Fig.~\ref{fig:factplot} on the
right.  At leading order, soft gluons decouple from the pion, and collinear
gluons with momenta scaling as $(p^-,p^\perp,p^+) \sim (m_b,\lqcd,\lqcd^2/m_b)$
couple only to the hard vertex [see discussion around Eqs.~(\ref{scetdecomp})
-- (\ref{colfield})], giving rise to the convolution integral.

In Eq.~(\ref{BDpifact}), $N = (m_B^2-m_D^2)/4$ and $N_* = m_{D^*}\,
(\epsilon^*\cdot p_B)/2$ are kinematic factors.  There are several
nonperturbative quantities, $F_{B\to D^{(*)}}$ is the $B\to D^{(*)}$ form
factor at $q^2=m_\pi^2$ measurable in semileptonic $B\to D^{(*)}\ell\bar\nu$
decay, $f_\pi$ is the pion decay constant, and $\phi_\pi$ is the pion
light-cone wave function that describes the probability that one of the quarks
has momentum fraction $x$ in the pion.  The $T(x,\mu_0,\mu)$ is a
perturbatively calculable short distance coefficient.  (Strictly speaking, $T$
depends on a third scale, $\mu'$, that cancels the $\mu'$-dependence of the
Isgur-Wise function, $\xi(w,\mu')$, which determines $F_{B\to D^{(*)}}$.) 
Contrary to naive factorization, which corresponds to setting $\mu_0 = m_b$ and
$T = 1$, Eq.~(\ref{BDpifact}) provides a consistent formulation where the scale
and scheme dependences cancel order by order in $\alpha_s$ between the Wilson
coefficients $C_i(\mu_0)$ and $T(x,\mu_0,\mu)$ in the matrix elements.  

The proof of factorization applies as long as the meson that inherits the brown
muck from the $B$ meson is heavy (e.g., $D^{(*)}$, $D_1$, etc.) and the other
is light (e.g., $\pi$, $\rho$, etc.).  The proof does not apply to decays when
the spectator quark in the $B$ ends up in the pion, such as color suppressed
decays of the type $\B0bar \to D^0\pi^0$, or color allowed decays of the type
$\B0bar \to D_s^- \pi^+$.  Annihilation and hard spectator contributions to all
decays discussed are power suppressed if one assumes that tail ends of the wave
functions behave as $(\lqcd/m_b)^a$ with $a>0$.

While the perturbative corrections in $T(x,\mu_0,\mu)$ are calculable, little
is known from first principles about the correction suppressed by powers of
$\lqcd/m_Q$.  Some possibilities to learn about their size is discussed next.

\subsubsection{\boldmath Tests of factorization}
\label{sec:factest}

It is important to understand quantitatively the accuracy of factorization in
different processes, and the mechanism(s) responsible for factorization and its
violation.  Factorization also holds in the large number of colors limit
($N_c\to \infty$ with  $\alpha_s N_c = \mbox{constant}$) in all $B\to M_1^-
M_2^+$ type decays, with corrections suppressed by $1/N_c^2$, independent of
the final mesons.  If factorization is mostly a consequence of perturbative
QCD, then its accuracy should depend on details of the final state, since the
proof outlined in the previous section relies on $M_2$ being fast ($m/E \ll
1$), whereas the large-$N_c$ argument is independent of this.  It would be nice
to observe deviations that distinguish between these expectations, and to
understand the size of power suppressed effects.

Of the nonperturbative input needed to evaluate Eq.~(\ref{BDpifact}), the $B\to
D^{(*)}$ form factors that enter $F_{B\to D^{(*)}}$ are measured in
semileptonic $B\to D^{(*)}\ell\bar\nu$ decay, and the pion decay constant
$f_\pi$ is also known.  The pion light-cone wave function is $\phi_\pi(x) =
6x(1-x) + \ldots$, where the corrections are not too important since these
decays receive small contributions from $x$ near $0$ or $1$.  Thus, in color
allowed decays, such as $\B0bar\to D^{(*)+}\pi^-$ and $D^{(*)+}\rho^-$,
factorization has been observed to work at the $10\%$ level.  These tests get
really interesting just around this level, since we would like to distinguish
between corrections suppressed by $\lqcd/m_{c,b}$ and/or $1/N_c^2$.

At the level of existing data, factorization also works in $B\to
D_s^{(*)}D^{(*)}$ decays, where both particles are heavy.  It will be 
interesting to check whether there are larger corrections to factorization in
$\B0bar\to D_s^{(*)-} \pi^+$ decay than in $\B0bar\to D^{(*)+} \pi^-$, since
the former is expected to be suppressed in addition to $|V_{ub}/V_{cb}|^2$ by
$\lqcd/m_{c,b}$ as well~\cite{bud,LR}.  For this test, measurement of the $B\to
\pi\ell\bar\nu$ form factor is necessary.  Another test involves decays to
``designer mesons", such as $\B0bar \to D^{(*)+}d^-$ (where $d = a_0, b_1,
\pi_2$, etc.), which vanish in naive factorization, so the order $\alpha_s$ and
$\lqcd/m_{c,b}$ terms are expected to be the leading contributions~\cite{DH}.

One of the simplest detailed tests of factorization is the comparison of the
$\B0bar \to D^{(*)+}\pi^-$ and $B^- \to D^{(*)0}\pi^-$ rates and isospin
amplitudes.  These rates are predicted to be equal in the $m_{c,b} \gg \lqcd$
limit, since they only differ by a power suppressed contribution to $B^- \to
D^{(*)0}\pi^-$ when the spectator in the $B$ ends up in the $\pi$.  Let's work
this out for fun in detail.

\subparagraph{\boldmath $B\to D\pi$ isospin analysis}

The initial $(\B0bar,B^-)$ and final $(D^+, D^0)$ are $I = \frac12$ doublets,
the pions are in an $I=1$ triplet.  So $D\pi$ can be in $I = \frac12$ or
$\frac32$ state, and the decomposition is
\beqa\label{ispinDpi}
|D^0 \pi^0\rangle &=& -\sqrt{\frac13} \, |(D \pi)_{I=1/2}\rangle
  + \sqrt{\frac23} \, |(D \pi)_{I=3/2}\rangle \,, \nn\\
|D^+ \pi^-\rangle &=& \sqrt{\frac23} \, |(D \pi)_{I=1/2}\rangle
  + \sqrt{\frac13} \, |(D \pi)_{I=3/2}\rangle \,, \nn\\
|D^0 \pi^- \rangle &=& |(D \pi)_{I=3/2}\rangle \,.
\eeqa
The $b\to c \bar u d$ Hamiltonian is $|I,I_z\rangle = |1,-1\rangle$.  Similar
to Sec.~\ref{sec:pipiispin}, we need to be careful with the relative
normalization of the $\B0bar$ and $B^-$ decay matrix elements.  The $I =
\frac12$ amplitude only occurs in $\B0bar$ decay, and there is no subtlety. 
The $I = \frac32$ amplitude occurs with different normalization in neutral and
charged $B$ decay: $\langle\B0bar | H | (D\pi)_{I=3/2}\rangle = (1/\sqrt3)
{\cal A}_{3/2}$, while $\langle B^- | H | (D\pi)_{I=3/2}\rangle = {\cal
A}_{3/2}$.  Thus the $A_{3/2} \equiv  {\cal A}_{3/2} / \sqrt3$ amplitude in
$\B0bar$ decay needs to be multiplied by $\sqrt3$ to get the normalization of
the $B^-$ decay amplitude right.  We get
\beqa\label{ADpi}
A^{00} \equiv A(\B0bar\to D^0 \pi^0) &=& - \sqrt{\frac13} \, A_{1/2}
  + \sqrt{\frac23} \, A_{3/2} \,, \nn\\
A^{+-} \equiv A(\B0bar\to D^+ \pi^-) &=& \sqrt{\frac23} \, A_{1/2}
  + \sqrt{\frac13} \, A_{3/2} \,, \nn\\
A^{0-} \equiv A(B^-\to D^0 \pi^-) &=& \sqrt3\, A_{3/2} \,.
\eeqa
This implies the triangle relation:
\beq\label{triDpi}
A^{+-} + \sqrt2\, A^{00} = A^{0-}\,.
\eeq

A prediction of QCD factorization in $B\to D\pi$ decay is that amplitudes
involving the spectator quark in the $B$ going into the $\pi$ should be power
suppressed~\cite{BBNS,BPSdpi}, and therefore,
\beq
{|A^{0-}| \over |A^{+-}|} = 1 + {\cal O}\bigg({\lqcd \over m_{c,b}}\bigg) \,,
\eeq
or in terms of isospin amplitudes, $A_{1/2} = \sqrt2\, A_{3/2}\, [1 + {\cal O}
(\lqcd/m_{c,b})]$.  In this case the triangle in Eq.~(\ref{triDpi}) becomes
squashed, and the strong phase difference between the $A_{1/2}$ and $A_{3/2}$
amplitudes is suppressed, $\delta_{1/2} - \delta_{3/2} = {\cal
O}(\lqcd/m_{c,b})$.  The experimental data are~\cite{BDpiispin,BDpicolsup}
\beqa\label{factratio}
&\ds {{\cal B}(B^-\to D^0 \pi^-) \over {\cal B}(\B0bar\to D^+ \pi^-)}
  = 1.85\pm 0.25\,,& \nn\\[2pt]
& 16.5^\circ < \delta_{1/2} - \delta_{3/2} < 38.1^\circ 
  \quad {\rm (90\%~CL)}\,. &
\eeqa
The ratio of branching ratios is measured to be in the ballpark of $1.8$ also
for $D$ replaced by $D^*$ and $\pi$ replaced by $\rho$.  These deviations from
factorization are usually attributed to ${\cal O}(\lqcd/m_c)$
corrections~\cite{BBNS}, which could be of order $30\%$ in the amplitudes and
twice that in the rates.  One could claim that the strong phase in
Eq.~(\ref{factratio}) should be viewed as small, since $1 - \cos 26^\circ
\simeq 0.1 \ll 1$.  This is open to interpretation, as the answer depends
sensitively on the measure used (and, for example, we think of the CKM angle
$\beta \approx 23.5^\circ$ as order unity).

Studying such two-body channels it is hard to unambiguously identify the source
of the corrections to factorization.  The problem is that the color suppressed
contribution to the $B^-\to D^0 \pi^-$ is formally order $1/N_c$ in the large
$N_c$ limit, and order $\lqcd/m_{c,b}$ in the heavy mass limit, which may be
comparable.  Factorization fails even worse in $D\to K\pi$ decays, however this
does not show model independently that the corrections seen in $B\to D\pi$ are
due to $\lqcd/m_Q$ effects, since the proof of factorization based on the heavy
quark limit does not apply for $D\to K\pi$ to start with.  It does indicate,
however, that the large $N_c$ limit cannot be the full story.

\subparagraph{\boldmath Factorization in $B\to D^{(*)}X$}

Another possibility to study corrections to factorization is to consider $B\to
D^{(*)} X$ decay where $X$ contains two or more hadrons.  The advantage
compared to two-body channels is that the accuracy of factorization can be
studied as a function of kinematics for final states with fixed particle
content, by examining the differential decay rate as a function of the
invariant mass of the light hadronic state $X$~\cite{LLW,ReIs,DGfact}.  If
factorization works primarily due to the large $N_c$ limit, then its accuracy
is not expected to decrease as the invariant mass of $X$, $m_X$, increases. 
However, if factorization is mostly a consequence of perturbative QCD, then the
corrections should grow with $m_X$.  Factorization has also been studied in
inclusive $B\to D^{(*)}X$ decay, and it was suggested that the small velocity
limit ($m_b,\, m_c \gg m_b-m_c \gg \lqcd$) may also play an important role in
factorization~\cite{ucsdguys}.

\begin{figure}[t]
\centerline{\raisebox{-2.5pt}{%
  \includegraphics*[width=.458\textwidth]{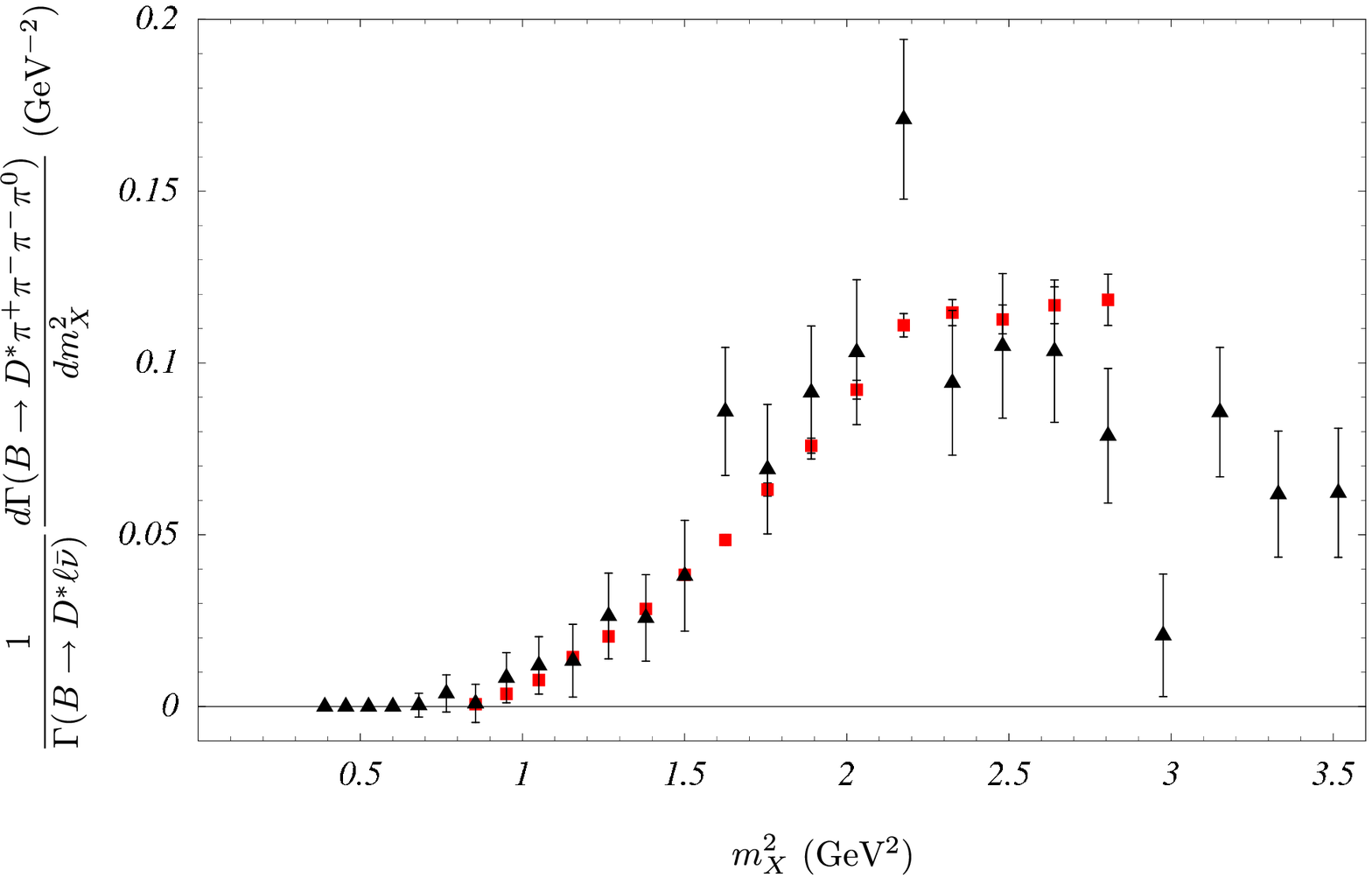}}\hspace{.5cm}
  \includegraphics*[width=.45\textwidth]{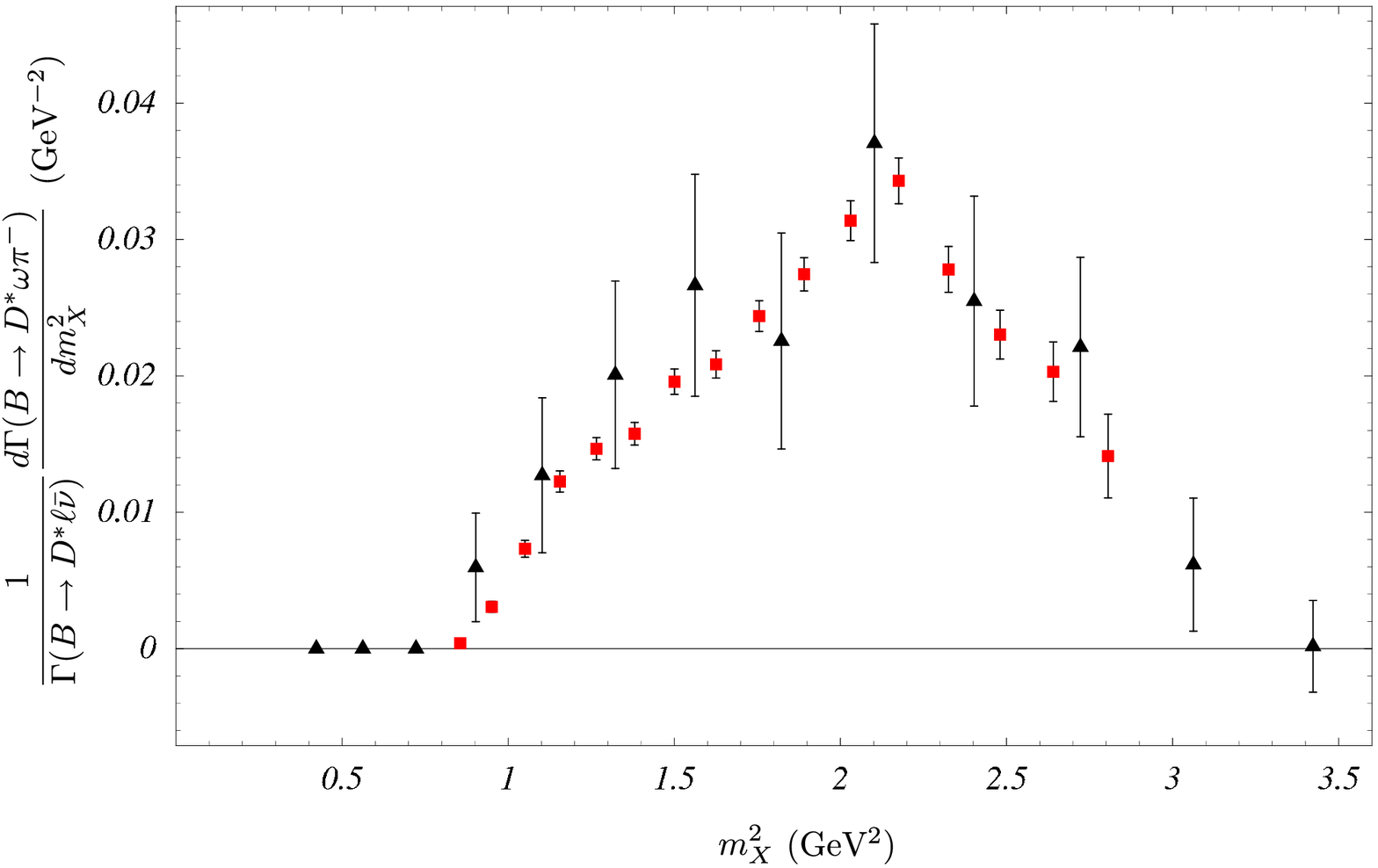}}
\caption{$\d \Gamma(B\to D^*X)/\d m_X^2$, where $X = \pi^+\pi^-\pi^-\pi^0$ 
(left) and $X = \omega\pi$ (right), normalized to the semileptonic width
$\Gamma(B\to D^*\ell\bar\nu)$.  The triangles are $B$ decay
data\protect\footnotemark\ and the squares are the predictions using $\tau$
data.  (From Ref.~\protect\citex{LLW}.)}
\label{fig:BD4pi}
\end{figure}

Combining data for hadronic $\tau$ decay (which effectively measures the
hadronization of a virtual $W$ to $X$) and semileptonic $B$ decay allows such
tests to be made for a variety of final states.  Figure~\ref{fig:BD4pi} shows
the comparison of the $B\to D^*\pi^+\pi^-\pi^-\pi^0$ and $D^*\omega\pi^-$
data~\cite{CLEObd4pi} with the $\tau$ decay~\cite{CLEOtau4pi} data.  The reason
to consider the $4\pi$ final state is because the $2\pi$ and $3\pi$ channels
are dominated by resonances.  The kinematic range accessible in $\tau\to 4\pi$
corresponds to $0.4 \lsim m_{4\pi}/E_{4\pi} \lsim 0.7$ in $B\to 4\pi$ decay.  A
background to these comparisons is that one or more of the pions may arise from
the $\bar c_L \gamma^\mu b_L$ current instead of the $\bar d_L \gamma^\mu u_L$
current.  In the $\omega\pi^-$ mode this is very unlikely to be
significant~\cite{LLW}.  In the $\pi^+\pi^-\pi^-\pi^0$ mode such backgrounds
can be constrained by measuring the $B\to D^*\pi^+\pi^+\pi^-\pi^-$ rate, since
$\pi^+\pi^+\pi^-\pi^-$ cannot come from the $\bar d_L \gamma^\mu u_L$ current. 
CLEO found ${\cal B}(B\to D^*\pi^+\pi^+\pi^-\pi^-) / {\cal B}(B\to
D^*\pi^+\pi^-\pi^-\pi^0) < 0.13$ at 90\%\,CL in the  $m_X^2 < 2.9\,{\rm GeV}^2$
region~\cite{CLEO4piws}, consistent with zero.  When more precise data are
available, observing deviations that grow with $m_X$ would be evidence that
perturbative QCD is an important part of the success of factorization in $B\to
D^*X$.

\footnotetext{In this case the charged and neutral $B$ decay rates do not
differ significantly, ${\cal B}(B^-\to D^{*0} \pi^+ \pi^- \pi^- \pi^0) = (1.80
\pm 0.36)\%$ and ${\cal B}(\B0bar\to D^{*+} \pi^+ \pi^- \pi^- \pi^0) = (1.72
\pm  0.28)\%$~\cite{CLEObd4pi}.  Their ratio is certainly smaller then the
similar ratio in Eq.~(\ref{factratio}), typical for $B\to D^{(*)} \pi$ and
$D^{(*)} \rho$ decays.  In addition, ${\cal B}(\B0bar\to D^{*0} \pi^+ \pi^+
\pi^- \pi^-) = (0.30 \pm 0.09)\%$ is small~\cite{CLEO4piws} and sensitive to
contributions when the spectator in the $B$ does not end up in the $D^*$.}

\subsection[Factorization in charmless $B$ decays]{\boldmath Factorization in
charmless $B$ decays}\label{sec:factnoc}

\begin{figure}[t]
\centerline{\includegraphics*[width=.32\textwidth]{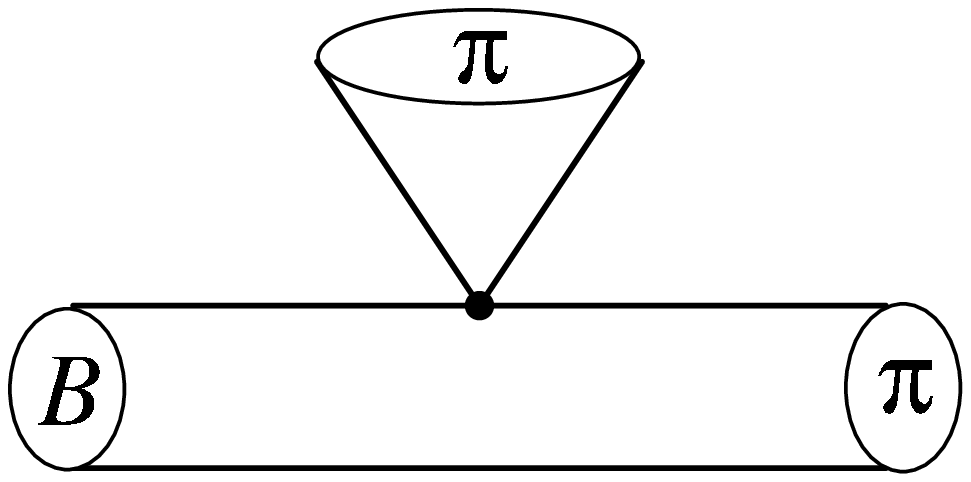}\hspace*{1.5cm}
\includegraphics*[width=.32\textwidth]{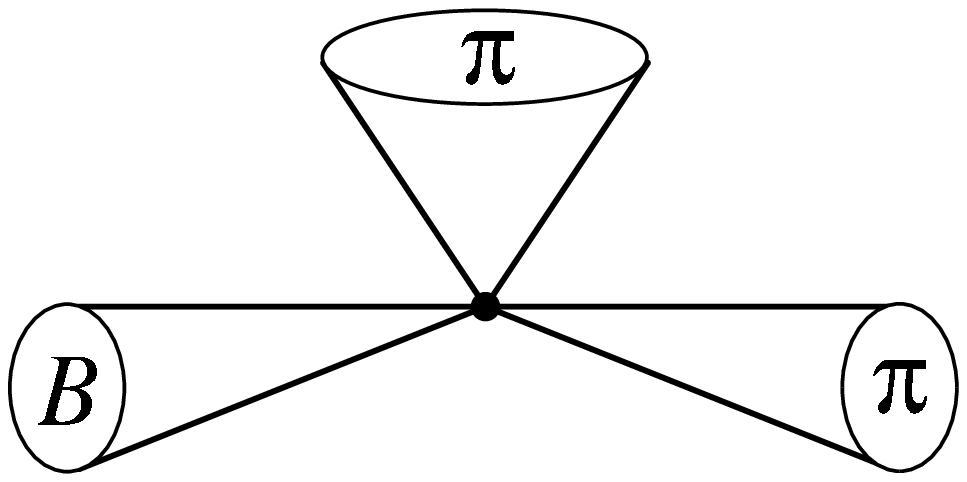}}
\caption{Two contributions to $B\to \pi\pi$ amplitudes (from
Ref.~\protect\citex{iain}).}
\label{fig:charmless}
\end{figure}

Calculating $B$ decay amplitudes to charmless two-body final states is
especially important for the study of $CP$ violation.  There are two
contributions to these decays shown schematically in Fig.~\ref{fig:charmless}.
The first term is analogous to the leading term in $\B0bar\to D^+\pi^-$, while
the second one involves hard spectator interaction.  There are two approaches
to factorization in these decays, which differ even on the question of which of
the two contributions is the leading one in the heavy quark limit.

Beneke {\it et al.} (BBNS)~\cite{bbnslight} proposed a factorization formula
\beq\label{Bpipifact}
\langle \pi\pi | O_i | B\rangle 
= F_{B\to \pi} \int\! \d x\, T^I(x)\, \phi_\pi(x)
  + \int\! \d \xi\, \d x\, \d x'\, T^{II}(\xi,x,x')\, 
  \phi_B(\xi)\, \phi_\pi(x)\, \phi_\pi(x') \,,
\eeq
and showed that it is consistent to first order in $\alpha_s$.  The $T$'s are
calculable short distance coefficient functions, whereas the $\phi$'s are
nonperturbative light-cone distribution functions.  Each of these terms have
additional scale dependences not shown above, similar to those in
Eqs.~(\ref{htlff}) and (\ref{BDpifact}), which are supposed to cancel
order-by-order in physical results.  A major complication of charmless decays
compared $B\to D\pi$ is that understanding the role of endpoint regions of the
light-cone distribution functions is much more involved.  BBNS assume that
Sudakov suppression is not effective at the $B$ mass scale in the endpoint
regions of these distribution functions.  Then the two terms are of the same
order in $\lqcd/m_b$, but the second term is suppressed by $\alpha_s$.

Keum {\it et al.} (KLS)~\cite{keumetal} assume that Sudakov suppression is
effective in suppressing contributions from the tails of the wave functions, $x
\sim \lqcd/m_b$.  Then the first term [in Eq.~(\ref{Bpipifact}) and in
Fig.~\ref{fig:charmless}] is subleading and the second one gives the dominant
contribution.  This issue is related to Sec.~\ref{sec:SCET}, where an open
question was the relative size of the two contributions to the $B\to
\pi\ell\bar\nu$ form factors in Eq.~(\ref{htlff}).  These form factors are
calculable according to KLS (in terms of the poorly known $B$ and $\pi$
light-cone wave functions), whereas they are nonperturbative inputs that can
only be determined from data according to BBNS.

The outstanding open theoretical questions are to prove the factorization
formula to all orders in $\alpha_s$ (this was claimed very recently~\cite{CK}),
to understand the role of Sudakov effects, and to find out which contribution
(if either) is dominant in the heavy mass limit.  Before these questions are
answered, it is not clear that either approach is right.  A complete
formulation of power suppressed corrections is also lacking so far.

Some terms that are formally order $\lqcd/m_b$ in the BBNS approach are known
to be large numerically and must be included to be able to describe the data. 
These are the so-called ``chirally enhanced" terms proportional to $m_K^2/(m_s
m_b)$, which are actually not enhanced by any parameter of QCD in the chiral
limit, they are just ${\cal O}(\lqcd)$, but happen to be large.  The
uncertainty related to weak annihilation contributions also needs to be better
understood.  Note that diagrams usually called annihilation cannot be
distinguished from rescattering.  The $B^0\to D_s K$ data~\cite{BDspiK} seems
to indicate that these are not very strongly suppressed.

\subsubsection[Phenomenology of $B\to \pi\pi,K\pi$]{\boldmath Phenomenology of
$B\to \pi\pi,K\pi$}

While the two approaches discussed above yield different power counting and
sometimes different phenomenological predictions, so far the results from both
groups fit (or could be adjusted to fit) the data on charmless two-body $B$
decays.  It has also been claimed that the effects of charm loops are larger
than predicted by either approach~\cite{charmloops}.  Table~\ref{tab:pipiKpi}
compares theory and data for ratios of certain charmless $B$ decay rates. 
Conclusive tests do not seem easy, and it will take a lot of data to learn
about the accuracy of these predictions.  Predictions for strong phases and
therefore for direct $CP$ violation are typically smaller in BBNS than in the
KLS approach.  More precise experimental data will be crucial.

\begin{table}[t] 
\begin{center}
\begin{tabular}{cccc} \hline\hline
Experimental  &  \multicolumn{2}{c}{Theoretical\, Predictions}  &  World\\[-6pt]
Observable  &  BBNS  &  KLS  &  Average \\ \hline\hline
$\ds {{\cal B}(\pi^+\pi^-) \over {\cal B}(\pi^\mp K^\pm)}$  &
  $0.3 - 1.6$  &  $0.3 - 0.7$  &  $0.28 \pm 0.04$ \\[8pt]
$\ds {{\cal B}(\pi^\mp K^\pm) \over 2\,{\cal B}(\pi^0 K^0)}$  &
  $0.9 - 1.4$  &  $0.8 - 1.05$  &  $1.0 \pm 0.3$ \\[8pt]
$\ds {2\,{\cal B}(\pi^0 K^\pm) \over {\cal B}(\pi^\pm K^0)}$  &
  $0.9 - 1.3$  &  $0.8 - 1.6$  &  $1.3 \pm 0.2$ \\[8pt]
$\ds {\tau_{B^\pm} \over \tau_{B^0}}\, 
  {{\cal B}(\pi^\mp K^\pm) \over {\cal B}(\pi^\pm K^0)}$  &
  $0.6 - 1.0$  &  $0.7 - 1.45$  &  $1.1 \pm 0.1$ \\[8pt]
$\ds {\tau_{B^\pm} \over \tau_{B^0}}\, 
  {{\cal B}(\pi^+ \pi^-) \over 2\,{\cal B}(\pi^\pm \pi^0)}$  &
  $0.6 - 1.1$  &    &  $0.56 \pm 0.14$ \\[6pt] \hline\hline
\end{tabular}
\end{center}
\caption{Experimental data and theoretical predictions/postdictions for ratios
of $B\to \pi\pi,K\pi$ branching ratios (from Ref.~\protect\citex{yossi_ams}).}
\label{tab:pipiKpi}
\end{table}

\begin{figure}[th]
\centerline{\includegraphics*[width=.5\textwidth]{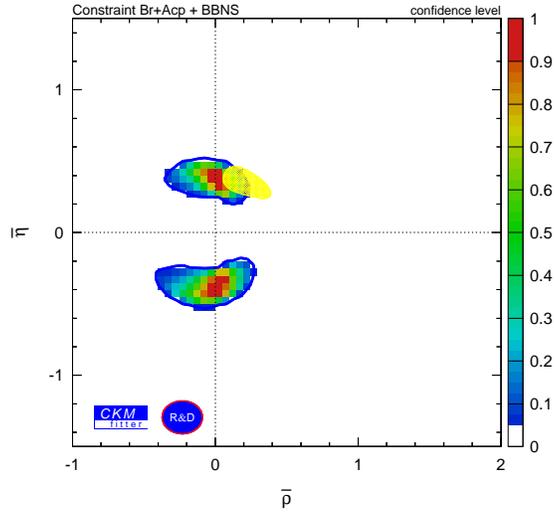}}
\caption{Fit to charmless two-body $B$ decays assuming BBNS (from
Ref.~\protect\citex{ckmfitter2}).}
\label{fig:pipifit}
\end{figure}

A CKM fit assuming BBNS and using the $B\to \pi\pi, K\pi$ rates and direct $CP$
asymmetries is shown in Fig.~\ref{fig:pipifit}.  It yields a $\rho-\eta$ region
consistent with the ``standard" CKM fits, although preferring slightly larger
values of $\gamma$.  Similar results might be obtained using the KLS
predictions as inputs.  A recent analysis including pseudoscalar--vector modes
as well finds an unsatisfactory fit to the data~\cite{saclay}.  Note that if
the lattice results for $\xi^2 \equiv (f_{B_s}^2B_{B_s}) / (f_{B_d}^2 B_{B_d})$
increase when light quark effects are fully understood, the possibility of
which was mentioned in Sec.~(\ref{sec:mix}), and if the $B_s$ mass difference
is near the present limits, that would shift the ``standard" fit to somewhat
larger values of $\gamma$.

Many strategies have been proposed to use $SU(3)$ flavor symmetry to constrain
CKM angles by combining data from several decay modes.  For example, one might
use a combination of $B$ and $B_s$ decays to  $\pi\pi, K \pi, K K$ final states
to gain sensitivity to $\gamma$ without relying on a complete calculation of
the hadronic matrix elements~\cite{fleischer}.  The basic idea is that $B_d \to
\pi^+\pi^-$ and $B_s \to K^+ K^-$ are related by $U$-spin, that exchanges
$d\leftrightarrow s$.  In such analyses one typically still needs some control
over hadronic uncertainties that enter related to, for example, first order
$SU(3)$ breaking effects ($U$-spin breaking is controlled by the same
parameter, $m_s/\Lambda_{\chi SB}$), rescattering effects, etc.  The crucial
question is how experimental data can be used to set bounds on the size of
these uncertainties.  Such analyses will be important and are discussed in more
detail in Frank W\"urthwein's lectures~\cite{fkw}.

\subsubsection*{Summary of factorization}

\begin{itemize}\itemsep 0pt

\item In nonleptonic $B\to D^{(*)} X$ decay, where $X$ is a low mass hadronic
state, factorization is established in the heavy quark limit, at leading order
in $\lqcd/m_Q$.

\item Some of the order $\lqcd/m_c$ corrections are sizable, and there is no
evidence yet of factorization becoming a worse approximation in $B\to D^{(*)}X$
as $m_X$ increases.

\item In charmless nonleptonic decays there are two approaches: BBNS and KLS. 
Different assumptions and power counting, and sometimes different predictions.

\item Progress in understanding charmless semileptonic and rare decay form
factors in small $q^2$ region will help resolve power counting in charmless
nonleptonic decay.

\item New and more precise data will be crucial to test factorization and tell
us about significance of power suppressed contributions in various processes.

\end{itemize}

\subsection{Final remarks}

The recent precise determination of $\sin2\beta$ and other measurements make it
very likely that the CKM contributions to flavor physics and $CP$ violation are
the dominant ones.  The next goal is not simply to measure $\rho$ and $\eta$,
or $\alpha$ and $\gamma$, but to probe the flavor sector of the SM to high
precision by many overconstraining measurements.  Measurements which are
redundant in the SM but sensitive to different short distance physics are very
important, since correlations may give additional information on the possible
new physics encountered (e.g., comparing $\Delta m_s/\Delta m_d$ with ${\cal
B}(B\to X_s\ell^+\ell^-) / {\cal B}(B\to X_d\ell^+\ell^-)$ is not ``just
another way" to measure $|V_{ts}/V_{td}|$).

Hadronic uncertainties are often significant and hard to quantify.  The
sensitivity to new physics and the accuracy with which the SM can be tested
will depend on our ability to disentangle the short distance physics from
nonperturbative effects of hadronization.  While we all want small errors, the
history of $\epsilon'_K$ reminds us to be conservative with theoretical
uncertainties.  One theoretically clean measurement is worth ten dirty ones. 
But what is considered theoretically clean changes with time, and there is
significant progress toward understanding the hadronic physics crucial both
for standard model measurements and for searches for new physics.  For example,
for (i) the determination of $|V_{ub}|$ from inclusive $B$ decay;
(ii)~understanding exclusive rare decay form factors at small $q^2$; and (iii)
establishing factorization in certain nonleptonic decays.

In testing the SM and searching for new physics, our understanding of CKM
parameters and hadronic physics will have to improve in parallel.  Except for a
few clean cases (like $\sin2\beta$) the theoretical uncertainties can be
reduced by doing several measurements, or by gaining confidence about the
accuracy of theoretical assumptions.  Sometimes data may help to constrain or
get rid of nasty things hard to know model independently (e.g., excited state
contributions to certain processes).  

With the recent spectacular start of the $B$ factories an exciting era in
flavor physics has begun.  The precise measurements of $\sin2\beta$ together
with the sides of the unitarity triangle, $|V_{ub}/V_{cb}|$ at the $e^+e^-$ $B$
factories and $|V_{td}/V_{ts}|$ at the Tevatron, will allow us to observe small
deviations from the Standard Model.  The large statistics will allow the study
of rare decays and to improve sensitivity to observables which vanish in the
SM; these measurements have individually the potential to discover physics
beyond the SM.  If new physics is seen, then a broad set of measurements at
both $e^+ e^-$ and hadronic $B$ factories and $K\to \pi \nu\bar\nu$ may allow
to discriminate between classes of models.  It is a vibrant theoretical and
experimental program, the breadth of which is well illustrated by the long list
of important measurements where significant progress is expected in the next
couple of years: 

\begin{itemize}\itemsep 0pt

\item $|V_{td}/V_{ts}|$: the Tevatron should nail it, hopefully soon ---
will all the lattice subtleties be reliably understood by then?

\item $\beta$: reduce error in $\phi K_S$, $\eta' K_S$, and $KKK$ modes ---
will the difference from $S_{\psi K}$ become more significant?

\item $\beta_s$: is $CP$ violation in $B_s\to \psi\phi$ indeed small, as
predicted by the SM?

\item Rare decays: $B\to X_s\gamma$ near theory limited; more precise data on
$q^2$ distribution in $B\to X_s \ell^+\ell^-$ will be interesting.

\item $|V_{ub}|$: reaching $<\!10\%$ would be important.  Need to better
understand $|V_{cb}|$ as well; could be a BABAR/BELLE measurement unmatched by
LHCB/BTeV.

\item $\alpha$: Is the $\pi^+\pi^-\pi^0$ Dalitz plot analysis feasible --- are
there significant resonances in addition to $\rho\pi$?  How small are ${\cal
B}(B\to \pi^0 \pi^0)$ and ${\cal B}(B\to \rho^0 \pi^0)$?

\item $\gamma$: the clean modes are hard --- need to try all.  Start to
understand using data the accuracy of $SU(3)$ relations, factorization, and
related approaches.

\item Search for ``null observables", such as $a_{CP}(b\to s\gamma)$, etc.,
enhancement of $B_{d,s}\to \ell^+\ell^-$, $B\to \ell\nu$, etc.

\end{itemize}

This is surely an incomplete list, and I apologize for all omissions.  Any of
these measurements could have a surprising result that changes the future of
the field.  And it is only after these that LHCB/BTeV and possibly a
super-$B$-factory enter the stage.

\subsubsection{Summary}

\begin{itemize}

\item The CKM picture is predictive and testable --- it passed its first real
test and is probably the dominant source of $CP$ violation in flavor changing
processes.

\item The point is not only to measure the sides and angles of the unitarity
triangle, $(\rho, \eta)$ and $(\alpha, \beta, \gamma)$, but to probe CKM by
overconstraining it in as many ways as possible (large variety of rare decays,
importance of correlations).

\item The program as a whole is a lot more interesting than any single
measurement; all possible clean measurements, both $CP$ violating and
conserving, are important.

\item Many processes can give clean information on short distance physics, and
there is progress toward being able to model independently interpret new
observables.

\end{itemize}

\subsubsection*{Acknowledgements}

It is a pleasure to thank Adam Falk, Sandrine Laplace, Mike Luke, and Yossi Nir
for numerous discussions and insights that influenced these lectures.  I also
thank Gustavo Burdman, Bob Cahn, Yuval Grossman, Uli Nierste, and Iain Stewart
for helpful conversations. This work was supported in part by the Director,
Office of Science, Office of  High Energy and Nuclear Physics, Division of High
Energy Physics, of the  U.S.\ Department of Energy under Contract
DE-AC03-76SF00098 and by a DOE Outstanding Junior Investigator award.



\addcontentsline{toc}{section}{References}

\begin{thebibliography}{99}


\bibitem{book1}
A.V.~Manohar and M.B.~Wise, {\it Heavy Quark Physics,}
Cambridge Monogr.\ Part.\ Phys.\ Nucl.\ Phys.\ Cosmol.\ 10 (2000).

\bibitem{book2}
I.I.~Bigi and A.I.~Sanda, {\it $CP$ Violation,}
Cambridge University Press, New York (2000).

\bibitem{book3}
G.C.~Branco, L.~Lavoura and J.P.~Silva, {\it $CP$ Violation,}
Clarendon Press, Oxford (1999).

\bibitem{babook}
P.F.~Harrison and H.R.~Quinn (eds),
{\it The BABAR Physics Book: Physics at an Asymmetric B Factory}, SLAC-R-0504.

\bibitem{tevbook}
K.~Anikeev {\it et al.},
{\it B physics at the Tevatron: Run II and beyond,} hep-ph/0201071.

\bibitem{lhcrep}
P.~Ball {\it et al.}, {\it B decays at the LHC,} hep-ph/0003238.

\bibitem{yossi}
Y.~Nir, Lectures at 55th Scottish Universities Summer School, hep-ph/0109090.

\bibitem{adam}
A.~Falk, Lectures at TASI 2000, hep-ph/0007339.

\bibitem{sanda}
T.~Sanda, these proceedings, http://www-conf.slac.stanford.edu/ssi/2002.

\bibitem{lattice}
P.~Lepage, these proceedings, http://www-conf.slac.stanford.edu/ssi/2002.

\bibitem{np}
A.~Kagan, these proceedings, http://www-conf.slac.stanford.edu/ssi/2002.

\bibitem{tb}
T.~Browder, these proceedings, http://www-conf.slac.stanford.edu/ssi/2002.

\bibitem{dbm}
D.~MacFarlane, these proceedings, http://www-conf.slac.stanford.edu/ssi/2002.

\bibitem{is}
I.~Shipsey, these proceedings, http://www-conf.slac.stanford.edu/ssi/2002.

\bibitem{rt}
R~Tschirhart, these proceedings, http://www-conf.slac.stanford.edu/ssi/2002.

\bibitem{fkw}
F.~W\"urthwein, these proceedings, http://www-conf.slac.stanford.edu/ssi/2002.

\bibitem{strongCP}
H.~Quinn, Dirac Medal Lecture, hep-ph/0110050;\\
M.~Dine, Lectures at TASI 2000, hep-ph/0011376.

\bibitem{KM}
M.~Kobayashi and T.~Maskawa, Prog.\ Theor.\ Phys.\  {\bf 49} (1973) 652.

\bibitem{C}
N.~Cabibbo, Phys.\ Rev.\ Lett.\  {\bf 10} (1963) 531.

\bibitem{sakharov}
A.D.~Sakharov, Pisma Zh.\ Eksp.\ Teor.\ Fiz.\  5 (1967) 32
[JETP Lett.\  {\bf 5} (1967) 24].

\bibitem{baryogen}
For reviews, see: A.G.~Cohen, D.B.~Kaplan and A.E.~Nelson,
Ann.\ Rev.\ Nucl.\ Part.\ Sci.\  {\bf 43} (1993) 27 [hep-ph/9302210];
A.~Riotto and M.~Trodden,
Ann.\ Rev.\ Nucl.\ Part.\ Sci.\  {\bf 49} (1999) 35 [hep-ph/9901362].

\bibitem{leptogen}
M.~Fukugita and T.~Yanagida, Phys.\ Lett.\ B {\bf 174} (1986) 45;
For a review, see: W.~Buchmuller and M.~Plumacher,
Int.\ J.\ Mod.\ Phys.\ A {\bf 15} (2000) 5047 [hep-ph/0007176].

\bibitem{Branco}
G.C.~Branco, T.~Morozumi, B.M.~Nobre and M.N.~Rebelo,
Nucl.\ Phys.\ B {\bf 617} (2001) 475 [hep-ph/0107164].

\bibitem{Kcpv}
J.H.~Christenson, J.W.~Cronin, V.L.~Fitch and R.~Turlay,
Phys.\ Rev.\ Lett.\  {\bf 13} (1964) 138.

\bibitem{superweak}
L.~Wolfenstein, Phys.\ Rev.\ Lett.\  {\bf 13} (1964) 562.

\bibitem{pdg}
K.~Hagiwara {\it et al.}, Particle Data Group, 
Phys.\ Rev.\ D {\bf 66} (2002) 010001.

\bibitem{gino}
G.~Isidori, Talk at Lepton-Photon 2001, hep-ph/0110255.

\bibitem{GIM}
S.L.~Glashow, J.~Iliopoulos and L.~Maiani, Phys.\ Rev.\ D {\bf 2} (1970) 1285.

\bibitem{bnl} 
S.~Adler {\it et al.}, E787 Collaboration, 
Phys.\ Rev.\ Lett.\  {\bf 88} (2002) 041803 [hep-ex/0111091].

\bibitem{Littenberg}
L.S.~Littenberg, Phys.\ Rev.\ D {\bf 39} (1989) 3322.

\bibitem{ckmfitter}
A.~Hocker, H.~Lacker, S.~Laplace and F.~Le Diberder,
Eur.\ Phys.\ J.\ C {\bf 21} (2001) 225 [hep-ph/0104062]; and updates at
http://ckmfitter.in2p3.fr/.

\bibitem{Inami-Lim}
T.~Inami and C.S.~Lim, Prog.\ Theor.\ Phys.\ {\bf 65} (1981) 297
[Erratum-ibid.\  {\bf 65} (1981) 1772].

\bibitem{lepbosc}
$B$ oscillations working group, http://lepbosc.web.cern.ch/.

\bibitem{latticerev}
For a review, see: L.~Lellouch, Plenary talk at ICHEP 2002,
hep-ph/0211359.

\bibitem{Bslogs}
B.~Grinstein {\it et al.}, 
Nucl.\ Phys.\ B {\bf 380} (1992) 369 [hep-ph/9204207].

\bibitem{chiralsubl}
A.F.~Falk, Phys.\ Lett.\ B {\bf 305} (1993) 268 [hep-ph/9302265]; \\
L.~Randall and E.~Sather, Phys.\ Rev.\ D {\bf 49} (1994) 6236 [hep-ph/9211268].

\bibitem{gbml}
I thank G.~Burdman and M.~Luke for discussions related to this point.

\bibitem{cplear}
A.~Angelopoulos {\it et al.}, CPLEAR Collaboration,
Phys.\ Lett.\ B {\bf 444} (1998) 43.

\bibitem{llnp}
S.~Laplace, Z.~Ligeti, Y.~Nir and G.~Perez,
Phys.\ Rev.\ D {\bf 65} (2002) 094040 [hep-ph/0202010].

\bibitem{rf}
R.~Fleischer, Phys.\ Rept.\  {\bf 370} (2002) 537 [hep-ph/0207108].

\bibitem{babelle}
B.~Aubert {\it et al.}, BABAR Collaboration,
Phys.\ Rev.\ Lett.\ {\bf 89} (2002) 201802 [hep-ex/0207042];
Phys.\ Rev.\ Lett.\ {\bf 87} (2001) 091801 [hep-ex/0107013];\\
K.~Abe {\it et al.}, BELLE Collaboration,
Phys.\ Rev.\ D {\bf 66} (2002) 071102 [hep-ex/0208025];
Phys.\ Rev.\ Lett.\ {\bf 87} (2001) 091802 [hep-ex/0107061].

\bibitem{yossi_ams}
Y.~Nir, Plenary talk at ICHEP 2002, hep-ph/0208080.

\bibitem{babarphik}
B.~Aubert {\it et al.}, BABAR Collaboration, hep-ex/0207070.

\bibitem{bellephik}
K.~Abe {\it et al.}, BELLE Collaboration, hep-ex/0212062.



\bibitem{AP}
T.~Appelquist and H.D.~Politzer, Phys.\ Rev.\ Lett.\  {\bf 34} (1975) 43.

\bibitem{HQS}
N.~Isgur and M.B.~Wise, Phys.\ Lett.\ B {\bf 232} (1989) 113; 
Phys.\ Lett.\ B {\bf 237} (1990) 527.

\bibitem{Geor}
H.~Georgi, Phys.\ Lett.\ B {\bf 240} (1990) 447.

\bibitem{IWprl}
N.~Isgur and M.B.~Wise, Phys.\ Rev.\ Lett.\  {\bf 66} (1991) 1130.

\bibitem{cleobroad}
S.~Anderson {\it et al.}, CLEO Collaboration,
Nucl.\ Phys.\ A {\bf 663} (2000) 647 [hep-ex/9908009].

\bibitem{FM}
A.F.~Falk and T.~Mehen, Phys.\ Rev.\ D {\bf 53} (1996) 231 [hep-ph/9507311].

\bibitem{Czar}
A.~Czarnecki, Phys.\ Rev.\ Lett.\  {\bf 76} (1996) 4124 [hep-ph/9603261];\\
A.~Czarnecki and K.~Melnikov, Nucl.\ Phys.\ B {\bf 505} (1997) 65
[hep-ph/9703277].

\bibitem{Luke}
M.E.~Luke, Phys.\ Lett.\ B {\bf 252} (1990) 447.

\bibitem{LNN}
M.~Neubert, Z.~Ligeti and Y.~Nir, Phys.\ Lett.\ B {\bf 301} (1993) 101 
[hep-ph/9209271]; Phys.\ Rev.\ D {\bf 47} (1993) 5060 [hep-ph/9212266];
Z.~Ligeti, Y.~Nir and M.~Neubert, Phys.\ Rev.\ D {\bf 49} (1994) 1302
[hep-ph/9305304].

\bibitem{latticeD}
S.~Hashimoto {\it et al.}, Phys.\ Rev.\ D {\bf 61} (2000) 014502 
[hep-ph/9906376].

\bibitem{latticeDs}
S.~Hashimoto {\it et al.}, Phys.\ Rev.\ D {\bf 66} (2002) 014503 
[hep-ph/0110253].

\bibitem{BGL}
C.G.~Boyd, B.~Grinstein, R.F.~Lebed, Phys.\ Lett.\ B {\bf 353} (1995) 306
[hep-ph/9504235]; Nucl.\ Phys.\ B {\bf 461} (1996) 493 [hep-ph/9508211];
Phys.\ Rev.\ D {\bf 56} (1997) 6895 [hep-ph/9705252];
I.~Caprini, L.~Lellouch and M.~Neubert, Nucl.\ Phys.\ B {\bf 530} (1998) 153
[hep-ph/9712417].

\bibitem{BGZL}
B.~Grinstein and Z.~Ligeti, 
Phys.\ Lett.\ B {\bf 526} (2002) 345 [hep-ph/0111392].

\bibitem{IsWi}
N.~Isgur and M.B.~Wise, Phys.\ Rev.\ D {\bf 42} (1990) 2388.

\bibitem{lw}
Z.~Ligeti and M.B.~Wise, Phys.\ Rev.\ D {\bf 53} (1996) 4937 [hep-ph/9512225];
Z.~Ligeti, I.W.~Stewart and M.B.~Wise,
Phys.\ Lett.\ B {\bf 420} (1998) 359 [hep-ph/9711248].

\bibitem{Gtdr}
B.~Grinstein, Phys.\ Rev.\ Lett.\  {\bf 71} (1993) 3067 [hep-ph/9308226].

\bibitem{Szczepaniak}
A.~Szczepaniak, E.M.~Henley and S.J.~Brodsky, 
Phys.\ Lett.\ B {\bf 243} (1990) 287.

\bibitem{hmmm}
R.~Akhoury, G.~Sterman and Y.P.~Yao,
Phys.\ Rev.\ D {\bf 50} (1994) 358;\\
H.n.~Li and H.L.~Yu, Phys.\ Rev.\ D {\bf 53} (1996) 2480 [hep-ph/9411308];\\
A.~Szczepaniak, Phys.\ Rev.\ D {\bf 54} (1996) 1167.

\bibitem{charles}
J.~Charles {\it et al.}, Phys.\ Rev.\ D {\bf 60} (1999) 014001 [hep-ph/9812358].

\bibitem{iain}
I.~Stewart, talk at BEACH'02 (http://beach2002.physics.ubc.ca), 
hep-ph/0209159.

\bibitem{SCET0}
C.W.~Bauer, S.~Fleming and M.E.~Luke, Phys.\ Rev.\ D {\bf 63} (2001) 014006
[hep-ph/0005275].

\bibitem{SCET1}
C.W.~Bauer, S.~Fleming, D.~Pirjol and I.W.~Stewart,
Phys.\ Rev.\ D {\bf 63} (2001) 114020 [hep-ph/0011336].

\bibitem{SCET2}
C.W.~Bauer and I.W.~Stewart, 
Phys.\ Lett.\ B {\bf 516}, 134 (2001) [hep-ph/0107001];
C.W.~Bauer, D.~Pirjol and I.W.~Stewart, 
Phys.\ Rev.\ D {\bf 65} (2002) 054022 [hep-ph/0109045].

\bibitem{BF}
M.~Beneke and T.~Feldmann, Nucl.\ Phys.\ B {\bf 592} (2001) 3 [hep-ph/0008255].

\bibitem{Lirecent}
T.~Kurimoto, H.n.~Li and A.I.~Sanda,
Phys.\ Rev.\ D {\bf 65} (2002) 014007 [arXiv:hep-ph/0105003];
H.n.~Li, Phys.\ Rev.\ D {\bf 66} (2002) 094010 [arXiv:hep-ph/0102013].

\bibitem{BPSsl}
C.W.~Bauer, D.~Pirjol and I.W.~Stewart, hep-ph/0211069;\\
D.~Pirjol and I.W.~Stewart, hep-ph/0211251.

\bibitem{BH}
G.~Burdman and G.~Hiller, Phys.\ Rev.\ D {\bf 63} (2001) 113008 
[hep-ph/0011266].

\bibitem{OPE}
J.~Chay, H.~Georgi and B.~Grinstein, Phys.\ Lett.\ B {\bf 247} (1990) 399;\\
M.A.~Shifman and M.B.~Voloshin, Sov.\ J.\ Nucl.\ Phys.\  {\bf 41} (1985) 120;
I.I.~Bigi, N.G.~Uraltsev and A.I.~Vainshtein,
Phys.\ Lett.\ B {\bf 293} (1992) 430 [Erratum-ibid.\ B {\bf 297} (1992) 477]
[hep-ph/9207214];
I.I.~Bigi, M.A.~Shifman, N.G.~Uraltsev and A.I.~Vainshtein,
Phys.\ Rev.\ Lett.\ {\bf 71} (1993) 496 [hep-ph/9304225];\\
A.V.~Manohar and M.B.~Wise, Phys.\ Rev.\ D {\bf 49} (1994) 1310 
[hep-ph/9308246].

\bibitem{PQW}
E.C.~Poggio, H.R.~Quinn and S.~Weinberg, Phys.\ Rev.\ D {\bf 13} (1976) 1958.

\bibitem{NIdual}
N.~Isgur, Phys.\ Lett.\ B {\bf 448} (1999) 111 [hep-ph/9811377];
Phys.\ Rev.\ D {\bf 60} (1999) 074030 [hep-ph/9904398].

\bibitem{upsexp}
A.H.~Hoang, Z.~Ligeti and A.V.~Manohar, Phys.\ Rev.\ Lett.\ {\bf 82} (1999) 277
[hep-ph/9809423]; Phys.\ Rev.\ D {\bf 59} (1999) 074017 [hep-ph/9811239].

\bibitem{bmasses}
A.H.~Hoang, Phys.\ Rev.\ D {\bf 61} (2000) 034005  [hep-ph/9905550];
M.~Beneke and A.~Signer, Phys.\ Lett.\ B {\bf 471} (1999) 233 [hep-ph/9906475];
K.~Melnikov and A.~Yelkhovsky, Phys.\ Rev.\ D {\bf 59} (1999) 114009
[hep-ph/9805270].

\bibitem{BSUreview}
I.I.~Bigi, M.A.~Shifman and N.~Uraltsev,
Ann.\ Rev.\ Nucl.\ Part.\ Sci.\ {\bf 47} (1997) 591 [hep-ph/9703290].

\bibitem{shape}
C.W.~Bauer, Z.~Ligeti, M.~Luke and A.V.~Manohar, hep-ph/0210027;\\
M.~Battaglia {\it et al.}, hep-ph/0210319.

\bibitem{volo}
M.B.~Voloshin, Phys.\ Rev.\ D51 (1995) 4934 [hep-ph/9411296].

\bibitem{gremmetal}
M.~Gremm, A.~Kapustin, Z.~Ligeti and M.B.~Wise,
Phys.\ Rev.\ Lett.\ {\bf 77} (1996) 20 [hep-ph/9603314].

\bibitem{GK}
M.~Gremm and A.~Kapustin, Phys.\ Rev.\ D {\bf 55} (1997) 6924 [hep-ph/9603448].

\bibitem{GS}
M.~Gremm and I.~Stewart, Phys.\ Rev.\ D {\bf 55} (1997) 1226 [hep-ph/9609341].

\bibitem{FLSmass}
A.F.~Falk, M.~Luke, and M.J.~Savage, Phys.\ Rev.\ D {\bf 53} (1996) 2491 
[hep-ph/9507284]; Phys.\ Rev.\ D {\bf 53} (1996) 6316 [hep-ph/9511454];\\
A.F.~Falk and M.~Luke, Phys.\ Rev.\ D {\bf 57} (1998) 424 [hep-ph/9708327].

\bibitem{kl}
A.~Kapustin and Z.~Ligeti, Phys.\ Lett.\ B {\bf 355} (1995) 318 
[hep-ph/9506201];\\
R.D.~Dikeman, M.A.~Shifman and N.G.~Uraltsev,
Int.\ J.\ Mod.\ Phys.\ A {\bf 11} (1996) 571 [hep-ph/9505397].

\bibitem{llmw}
Z.~Ligeti, M.~Luke, A.V.~Manohar and M.B.~Wise, Phys.\ Rev.\ D {\bf 60} (1999)
034019 [hep-ph/9903305].

\bibitem{bauer}
C.~Bauer, Phys.\ Rev.\ D {\bf 57} (1998) 5611
[Erratum-ibid.\ D {\bf 60} (1999) 099907] [hep-ph/9710513].

\bibitem{momentsdata}
D.~Cronin-Hennessy {\it et al.}, CLEO Collaboration,
Phys.\ Rev.\ Lett.\ {\bf 87} (2001) 251808 [hep-ex/0108033];
R.~Briere {\it et al.}, CLEO Collaboration, hep-ex/0209024;
B.~Aubert {\it et al.}, BABAR Collaboration, hep-ex/0207084;
DELPHI Collaboration, 2002-071-CONF-604; 2002-070-CONF-605.

\bibitem{CLEObsgmom}
S.~Chen {\it et al.}, CLEO Collaboration,
Phys.\ Rev.\ Lett.\ {\bf 87} (2001) 251807 [hep-ex/0108032].

\bibitem{mass}
V.D.~Barger, C.S.~Kim and R.J.~Phillips, Phys.\ Lett.\ B {\bf 251} (1990) 629;\\
J.~Dai, Phys.\ Lett.\ B {\bf 333} (1994) 212 [hep-ph/9405270].

\bibitem{FLW}
A.F.~Falk, Z.~Ligeti and M.B.~Wise, Phys.\ Lett.\ B {\bf 406} (1997) 225
[hep-ph/9705235].

\bibitem{BDU}
I.~Bigi, R.D.~Dikeman and N.~Uraltsev, Eur.\ Phys.\ J.\ C {\bf 4} (1998) 453
[hep-ph/9706520].

\bibitem{llrhadron}
A.K.~Leibovich, I.~Low and I.Z.~Rothstein,
Phys.\ Lett.\ B {\bf 486} (2000) 86 [hep-ph/0005124]; 
Phys.\ Rev.\ D {\bf 62} (2000) 014010 [hep-ph/0001028].

\bibitem{energy}
A.O.~Bouzas and D.~Zappala, 
Phys.\ Lett.\ B {\bf 333} (1994) 215 [hep-ph/9403313];\\
C.~Greub and S.-J.~Rey, Phys.\ Rev.\ D {\bf 56} (1997) 4250 [hep-ph/9608247].

\bibitem{BLL1}
C.W.~Bauer, Z.~Ligeti and M.~Luke, Phys.\ Lett.\ B {\bf 479} (2000) 395
[hep-ph/0002161]; see also: hep-ph/0007054.

\bibitem{BLL2}
C.W.~Bauer, Z.~Ligeti and M.~Luke, Phys.\ Rev.\ D {\bf 64} (2001) 113004
[hep-ph/0107074].

\bibitem{KoMe}
R.V.~Kowalewski and S.~Menke,
Phys.\ Lett.\ B {\bf 541} (2002) 29 [hep-ex/0205038].

\bibitem{ugo}
U.~Aglietti, M.~Ciuchini and P.~Gambino,
Nucl.\ Phys.\ B {\bf 637} (2002) 427 [hep-ph/0204140].

\bibitem{structure}
M.~Neubert, Phys.\ Rev.\ D {\bf 49} (1994) 4623 [hep-ph/9312311];\\
I.I.~Bigi, M.A.~Shifman, N.G.~Uraltsev and A.I.~Vainshtein,
Int.\ J.\ Mod.\ Phys.\ A {\bf 9} (1994) 2467 [hep-ph/9312359].

\bibitem{extractshape}
A.K.~Leibovich and I.Z.~Rothstein, Phys.\ Rev.\ D {\bf 61} (2000) 074006
[hep-ph/9907391];
A.K.~Leibovich, I.~Low and I.Z.~Rothstein, Phys.\ Rev.\ D {\bf 61} (2000) 
053006 [hep-ph/9909404].

\bibitem{notO7}
M.~Neubert, Phys.\ Lett.\ B {\bf 513} (2001) 88 [hep-ph/0104280];
A.K.~Leibovich, I.~Low and I.Z.~Rothstein, Phys.\ Lett.\ B {\bf 513} (2001) 83
[hep-ph/0105066].

\bibitem{cleoVub}
A.~Bornheim {\it et al.}, CLEO Collaboration, 
Phys.\ Rev.\ Lett.\ {\bf 88} (2002) 231803 [hep-ex/0202019].

\bibitem{subltwistus}
A.K.~Leibovich, Z.~Ligeti, M.B.~Wise, 
Phys.\ Lett.\ B {\bf 539} (2002) 242 [hep-ph/0205148].

\bibitem{FLS}
A.F.~Falk, M.~Luke, and M.J.~Savage, Phys.\ Rev.\ D {\bf 49} (1994) 3367
[hep-ph/9308288].

\bibitem{subltwist}
C.W.~Bauer, M.~Luke and T.~Mannel, 
Phys.\ Lett.\ B {\bf 543} (2002) 261 [hep-ph/0205150]; hep-ph/0102089.

\bibitem{subltwistmn}
M.~Neubert, Phys.\ Lett.\ B {\bf 543} (2002) 269 [hep-ph/0207002].

\bibitem{Voloshin}
I.I.~Bigi and N.G.~Uraltsev, 
Nucl.\ Phys.\ B {\bf 423} (1994) 33 [hep-ph/9310285];\\
M.B.~Voloshin, Phys.\ Lett.\ B {\bf 515} (2001) 74 [hep-ph/0106040].

\bibitem{neubertq2}
M.~Neubert, JHEP {\bf 0007} (2000) 022 [hep-ph/0006068];\\
M.~Neubert and T.~Becher, Phys.\ Lett.\ B {\bf 535} (2002) 127 [hep-ph/0105217].

\bibitem{dss}
A.K.~Leibovich, Z.~Ligeti, I.W.~Stewart and M.B.~Wise,
Phys.\ Rev.\ Lett.\  {\bf 78} (1997) 3995 [hep-ph/9703213];
Phys.\ Rev.\ D {\bf 57} (1998) 308 [hep-ph/9705467].

\bibitem{IWsr}
N.~Isgur and M.B.~Wise, Phys.\ Rev.\ D {\bf 43} (1991) 319.

\bibitem{orsay}
A.~Le Yaouanc {\it et al.}, Phys.\ Lett.\ B {\bf 520} (2001) 59 
[hep-ph/0107047]; Phys.\ Lett.\ B {\bf 520} (2001) 25 [hep-ph/0105247].

\bibitem{belleDss}
K.~Abe {\it et al.}, BELLE Collaboration, BELLE-CONF-0235 (ABS724).

\bibitem{cleoDss}
J.~Gronberg {\it et al.}, CLEO Collaboration, 
CLEO CONF 96-25, ICHEP96 PA05-069.

\bibitem{mndss}
M.~Neubert, Phys.\ Lett.\ B {\bf 418} (1998) 173 [hep-ph/9709327].

\bibitem{burdmanAfb}
G.~Burdman, Phys.\ Rev.\ D {\bf 57} (1998) 4254 [hep-ph/9710550].

\bibitem{BFS}
M.~Beneke, T.~Feldmann and D.~Seidel, Nucl.\ Phys.\ B {\bf 612} (2001) 25
[hep-ph/0106067].

\bibitem{BoBu}
S.W.~Bosch and G.~Buchalla, 
Nucl.\ Phys.\ B {\bf 621} (2002) 459 [hep-ph/0106081].

\bibitem{KN}
A.L.~Kagan and M.~Neubert, 
Phys.\ Lett.\ B {\bf 539} (2002) 227 [hep-ph/0110078].

\bibitem{GLN}
Y.~Grossman, Z.~Ligeti and E.~Nardi,
Nucl.\ Phys.\ B {\bf 465} (1996) 369
[Erratum-ibid.\ B {\bf 480} (1996) 753] [hep-ph/9510378].

\bibitem{babellebsg}
B.~Aubert {\it et al.}, BABAR Collaboration, 
hep-ex/0207076; hep-ex/0207074;\\
K.~Abe {\it et al.}, BELLE Collaboration, 
Phys.\ Lett.\ B {\bf 511} (2001) 151 [hep-ex/0103042].

\bibitem{sll}
J.~Kaneko {\it et al.}, BELLE Collaboration, hep-ex/0208029.

\bibitem{Kll}
K.~Abe {\it et al.}, BELLE Collaboration, 
Phys.\ Rev.\ Lett.\  {\bf 88} (2002) 021801 [hep-ex/0109026];\\
B.~Aubert {\it et al.}, BABAR Collaboration, hep-ex/0207082.

\bibitem{acp_bsg}
T.E.~Coan {\it et al.}, CLEO Collaboration, 
Phys.\ Rev.\ Lett.\  {\bf 86} (2001) 5661 [hep-ex/0010075].

\bibitem{acp_bKg}
K.~Abe {\it et al.}, BELLE Collaboration, BELLE-CONF-0239 (ABS728);\\
B.~Aubert {\it et al.}, BABAR Collaboration, 
Phys.\ Rev.\ Lett.\  {\bf 88} (2002) 101805 [hep-ex/0110065].

\bibitem{Hurth}
T.~Hurth, hep-ph/0212304.



\bibitem{babelleaeff}
B.~Aubert {\it et al.}, BABAR Collaboration, hep-ex/0207055;
Phys.\ Rev.\ D {\bf 65} (2002) 051502 [hep-ex/0110062];\\
K.~Abe {\it et al.}, BELLE Collaboration, hep-ex/0301032; hep-ex/0204002.

\bibitem{pipi}
M.~Gronau and D.~London,
Phys.\ Rev.\ Lett.\  {\bf 65} (1990) 3381.

\bibitem{rhopi}
H.J.~Lipkin, Y.~Nir, H.R.~Quinn and A.~Snyder,
Phys.\ Rev.\ D {\bf 44} (1991) 1454.

\bibitem{rhopidalitz}
A.E.~Snyder and H.R.~Quinn,
Phys.\ Rev.\ D {\bf 48} (1993) 2139.

\bibitem{ADK}
R.~Aleksan, I.~Dunietz and B.~Kayser, Z.\ Phys.\ C {\bf 54} (1992) 653.

\bibitem{GW}
M.~Gronau and D.~Wyler, Phys.\ Lett.\ B {\bf 265}, 172 (1991);\\
M.~Gronau and D.~London, Phys.\ Lett.\ B {\bf 253}, 483 (1991).

\bibitem{DCP-K}
K.~Abe {\it et al.}, BELLE Collaboration, hep-ex/0207012;\\
B.~Aubert {\it et al.}, BABAR Collaboration, hep-ex/0207087.

\bibitem{ADS}
D.~Atwood, I.~Dunietz and A.~Soni,
Phys.\ Rev.\ Lett.\  {\bf 78} (1997) 3257 [hep-ph/9612433];
Phys.\ Rev.\ D {\bf 63} (2001) 036005 [hep-ph/0008090].

\bibitem{GLS}
Y.~Grossman, Z.~Ligeti, A.~Soffer, hep-ph/0210433.

\bibitem{BJfact}
J.D.~Bjorken, Nucl.\ Phys.\ Proc.\ Suppl.\ {\bf 11} (1989) 325;

\bibitem{DGfact}
M.J.~Dugan and B.~Grinstein, Phys.\ Lett.\ B {\bf 255} (1991) 583.

\bibitem{PWfact}
H.D.~Politzer and M.B.~Wise, Phys.\ Lett.\ B {\bf 257} (1991) 399.

\bibitem{BBNS}
M.~Beneke, G.~Buchalla, M.~Neubert and C.T.~Sachrajda,
Nucl.\ Phys.\ B {\bf 591} (2000) 313 [hep-ph/0006124].

\bibitem{BPSdpi}
C.W.~Bauer, D.~Pirjol and I.W.~Stewart, 
Phys.\ Rev.\ Lett.\ {\bf 87} (2001) 201806 [hep-ph/0107002]; 
and ref.~\citex{SCET2}.

\bibitem{bud}
Z.~Ligeti, Plenary talk at EPS-HEP 2001, hep-ph/0112089.

\bibitem{LR}
Z.~Luo and J.~L.~Rosner,
Phys.\ Rev.\ D {\bf 64} (2001) 094001 [hep-ph/0101089];\\
C.W.~Chiang, Z.~Luo and J.L.~Rosner,
Phys.\ Rev.\ D {\bf 66} (2002) 057503 [hep-ph/0206006].

\bibitem{DH}
M.~Diehl and G.~Hiller, JHEP {\bf 0106} (2001) 067 [hep-ph/0105194].

\bibitem{BDpiispin}
S.~Ahmed {\it et al.}, CLEO Collaboration,
Phys.\ Rev.\ D {\bf 66} (2002) 031101 [hep-ex/0206030].

\bibitem{BDpicolsup}
K.~Abe {\it et al.}, BELLE Collaboration, 
Phys.\ Rev.\ Lett.\  {\bf 88} (2002) 052002 [hep-ex/0109021];
T.E.~Coan {\it et al.}, CLEO Collaboration, 
Phys.\ Rev.\ Lett.\  {\bf 88} (2002) 062001 [hep-ex/0110055];
B.~Aubert {\it et al.}, BABAR Collaboration, hep-ex/0207092.

\bibitem{LLW}
Z.~Ligeti, M.E.~Luke and M.B.~Wise,
Phys.\ Lett.\ B {\bf 507}, 142 (2001) [hep-ph/0103020].

\bibitem{ReIs}
C.~Reader and N.~Isgur, Phys.\ Rev.\ D {\bf 47} (1993) 1007.

\bibitem{ucsdguys}
C.W.~Bauer, B.~Grinstein, D.~Pirjol and I.W.~Stewart, hep-ph/0208034.

\bibitem{CLEObd4pi}
J.P.~Alexander {\it et al.}, CLEO Collaboration, 
Phys.\ Rev.\ D {\bf 64} (2001) 092001 [hep-ex/0103021].

\bibitem{CLEOtau4pi}
K.W.~Edwards {\it et al.}, CLEO Collaboration, 
Phys.\ Rev.\ D {\bf 61} (2000) 072003 [hep-ex/9908024].

\bibitem{CLEO4piws}
K.W.~Edwards {\it et al.}, CLEO Collaboration, 
Phys.\ Rev.\ D {\bf 65} (2002) 012002 [hep-ex/0105071].

\bibitem{bbnslight}
M.~Beneke, G.~Buchalla, M.~Neubert and C.T.~Sachrajda,
Phys.\ Rev.\ Lett.\  {\bf 83} (1999) 1914 [hep-ph/9905312]; 
Nucl.\ Phys.\ B {\bf 606} (2001) 245 [hep-ph/0104110].

\bibitem{keumetal}
Y.Y.~Keum, H--n.~Li and A.I.~Sanda, Phys.\ Lett.\ B {\bf 504} (2001) 6;
Phys.\ Rev.\ D {\bf 63} (2001) 054008;
Y.Y.~Keum and H--n.~Li, Phys.\ Rev.\ D {\bf 63} (2001) 074006.

\bibitem{CK}
J.~Chay and C.~Kim, hep-ph/0301055; hep-ph/0301262.

\bibitem{BDspiK}
P.~Krokovny {\it et al.}, BELLE Collaboration,
Phys.\ Rev.\ Lett.\  {\bf 89} (2002) 231804 [hep-ex/0207077];\\
B.~Aubert {\it et al.}, BABAR Collaboration, hep-ex/0211053.

\bibitem{charmloops}
M.~Ciuchini {\it et al.}, Phys.\ Lett.\ B {\bf 515} (2001) 33 
[hep-ph/0104126];\\
S.J.~Brodsky and S.~Gardner, 
Phys.\ Rev.\ D {\bf 65} (2002) 054016 [hep-ph/0108121].

\bibitem{ckmfitter2}
A.~Hocker, H.~Lacker, S.~Laplace and F.~Le Diberder,
AIP Conf.\ Proc.\  {\bf 618} (2002) 27 [hep-ph/0112295];
S.~Laplace hep-ph/0209188; and Ref.~\citex{ckmfitter}.

\bibitem{saclay}
R.~Aleksan {\it et al.}, hep-ph/0301165.

\bibitem{fleischer}
R.~Fleischer, Phys.\ Lett.\ B {\bf 459}, 306 (1999) [hep-ph/9903456].

\end{thebibliography}
\end{document}